\documentclass[usenatbib]{mnras}

\usepackage{newtxtext,newtxmath}

\usepackage[T1]{fontenc}
\usepackage{ae,aecompl}

\usepackage{graphicx}	
\usepackage{amsmath}	
\usepackage{amssymb}	

\usepackage{longtable}  
\usepackage{environ}  
\usepackage{pdflscape}  
\usepackage{threeparttable}  
\usepackage{threeparttablex}  

\title[Denuded Dwarfs Demystified]
{Denuded Dwarfs Demystified: \break
{\color{black}  Gas Loss from dSph Progenitors and Implications for the Minimum Mass of Galaxies
}
}

\author[N. Ivkovich and M. L. McCall]{
Nina Ivkovich
and Marshall L. McCall\thanks{E-mail: mccall@yorku.ca}
\\
~Department of Physics and Astronomy, York University, Toronto, Ontario, Canada M3J 1P3\\
}

\date{Accepted XXX. Received YYY; in original form ZZZ}

\pubyear{2018}

\begin{document}
\label{firstpage}
\pagerange{\pageref{firstpage}--\pageref{lastpage}}
\maketitle

\begin{abstract}  
{\color{black} The placement of early-type dwarf galaxies (dSphs and dEs) with respect to the Potential Plane defined by pressure-supported late-type dwarf galaxies (dIs and BCDs) has been determined from surface brightness profiles.}  dEs and the most luminous dSphs lie on the Plane, suggesting that they emerged from late-type dwarfs that converted most of their gas into stars.  
However, there is a critical value of the potential at which dSphs start to fall systematically below the Plane, with the deviation growing as the potential becomes shallower.   The displacements are attributed to depletion of baryons through gas loss, smaller galaxies having lost proportionately more gas.  The critical potential corresponds to an escape velocity of 
$50 \pm 8 \, \rm km \, s^{-1}$,  
which is what is expected for gas with a temperature of 
$13,000 \pm 4,000 \, \rm K$,  
typical of a low-metallicity HII region.
This suggests that photoionization was responsible for instigating the loss of gas by galaxies with potentials shallower than the critical value, with evacuation occurring over a few tens of millions of years.  Extreme ratios of dynamical to luminous masses observed for the smallest dSphs are an artifact of mass loss.   Because the efficiency with which gas was converted into stars was lower for dSphs with shallower potentials, there should be a minimum baryonic mass for a galaxy below which the stellar mass is negligible.  
Gross extrapolation of the trend of inferred gas masses with stellar masses suggests a value 
between 500 and $10,000 \, \rm \mathcal{M}_\odot$.
{\color{black} The corresponding dynamical mass is below 
$10^6 \, \rm \mathcal{M}_\odot$}.  
\end{abstract}


\section{
Introduction  
} 
\label{intro}

{\color{black}
\subsection{
Definition of ``Dwarf''
}
}
\label{nomenclature}

{\color{black}
The term "dwarf" is loosely used to refer to a galaxy with an elliptical or irregular morphology that has a luminosity that is very much smaller than that of the Milky Way.  How much smaller is generally not quantified.  Since this paper draws upon a sample of ``dwarfs'' to discern the origin of differences between different kinds, it is important first to establish a quantitative criterion to pinpoint what ``small'' means.  {\color{black}  In that regard, it should be noted that throughout this paper distances and quantities derived from them are anchored to the scale of \citet{mcc14a}, on which the Virgo Cluster has a distance modulus of 30.98.}

\citet{lel14a} felt that any criterion defining a dwarf should eliminate gas-rich disk galaxies with bulges.  Therefore, they defined a dwarf galaxy to be any galaxy for which the square root of the kinetic energy per unit mass equates to an asymptotic rotational velocity at or below $100 \, \rm km \, s^{-1}$.  The corresponding limit to the absolute magnitude in $K_s$ is about $-21$.  By this definition, the most luminous gas-rich dwarfs are rotationally-supported and lie on the Tully-Fisher relation.  Morphologically, though, they are spiral galaxies.
}

{\color{black}
There is another way to define dwarfs energetically, one that is consistent with morphological typing.  Figure~\ref{fig_mbar_vs_vrot} displays the baryonic Tully-Fisher relation for a sample of dwarf irregular galaxies and blue compact dwarfs \citep{mcc12b} and a sample of generally more luminous but especially gas-rich galaxies that are rotating \citep{sta09a}, all converted to the distance scale adopted in this paper (see \S\ref{distances}).  
\begin{figure}
\includegraphics[keepaspectratio=true,width=8.5cm]{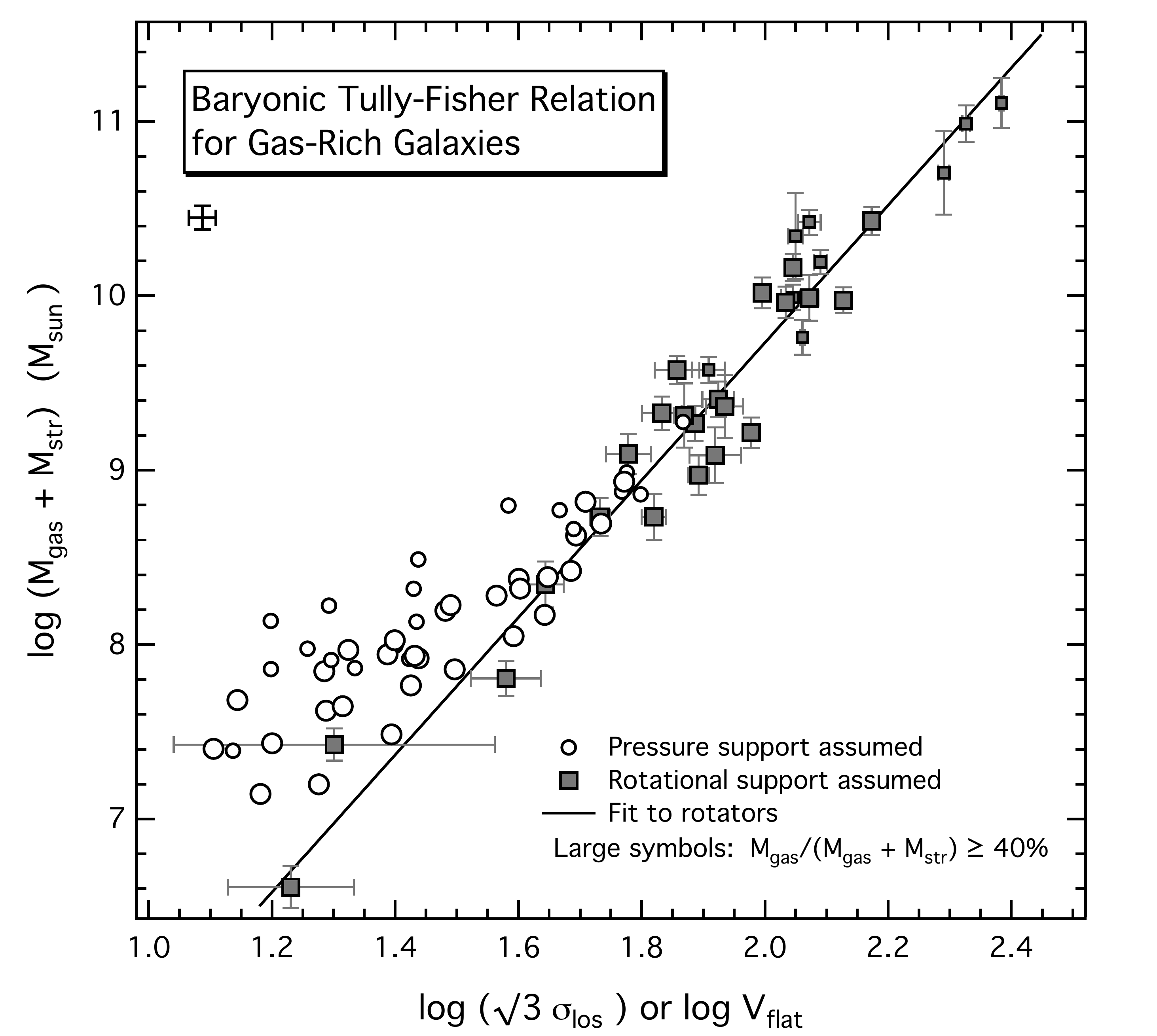}
\caption{\label{fig_mbar_vs_vrot}
{\color{black} The baryonic Tully-Fisher relation for gas-rich galaxies.  For rotating galaxies (filled squares), the logarithm of the baryonic mass is plotted versus the logarithm of the deprojected rotational velocity $V_\textit{flat}$ in the flat part of the rotation curve.  For dwarf irregular galaxies and blue compact dwarfs that are believed to be predominantly supported by pressure (open circles), the logarithm of the baryonic mass is plotted against the logarithm of the rotational velocity equivalent energetically to the line-of-sight velocity dispersion $\sigma_\textit{los}$.  The mean uncertainties for these galaxies are indicated by the error cross.  Galaxies with gas fractions that are 40\% or higher, for which uncertainties in the mass-to-light ratio of stars least affect the baryonic masses, are highlighted by large symbols.
}
}
\end{figure}
Line-of-sight velocity dispersions measured for systems believed to be pressure-supported have been converted to equivalent rotational velocities by multiplying by $\sqrt 3$.  The majority of galaxies in each sample have gas fractions that are 40\% or greater, so uncertainties in mass-to-light ratios adopted for stars have only a minor effect on baryonic masses.   There is a break in the slope of the relationship beginning at an energy equating to a rotational velocity of  $\sim 50\, \rm km \, s^{-1}$ (\citealt{mcc12a}), which for pressure-supported objects is a velocity dispersion of $\sim 30\, \rm km \, s^{-1}$.   The corresponding absolute magnitude in $K_s$ is about $-18.0$ ($-15.5$ in $V$).  The correlation for galaxies with less energy (lower luminosity and mass) is half as steep, and the scatter far exceeds observational errors \citep{mcc12a}.  Notable also is the observation by \citet{lee07a} that the distribution of H$\alpha$ equivalent widths for galaxies widens by a factor of two at absolute magnitudes fainter than about $-16$ in $V$, implying that star formation becomes regulated differently than in more luminous galaxies.   Such galaxies are believed to be predominantly supported by pressure \citep{mcc12b}, and are classified as irregular galaxies, not spirals.  

It seems somewhat arbitrary to regard the absence of a bulge as a criterion to define a dwarf because the Tully-Fisher relation is independent of bulge strength.   It is more logical to define a dwarf on the basis of dynamical support, especially considering the mesh with morphology.   Pressure-supported galaxies are the primary subject of this paper, and the term "dwarf" is used to refer to a galaxy with a velocity dispersion of around $30 \, \rm km \, s^{-1}$ or less.

{\color{black} Although the definition adopted for a dwarf is based upon the properties of gas-rich galaxies, in this paper it is extended to include gas-poor ones.  In so doing, many galaxies traditionally regarded as dwarf ellipticals on the basis of their surface brightness profiles \citet{san84a} can no longer be viewed as ``true'' dwarfs (see, e.g., \citealt{tol14a}).  Much of our knowledge about them is based upon studies of the Virgo Cluster, in which galaxies labeled as dEs exhibit diverse properties, including in many cases rotation.   However, \citet{rys13a} point out that some rotators may be stripped or harassed lenticulars or spirals that retain at least a fraction of the angular momentum of their progenitors.  From an evolutionary standpoint, then, they aren't really dwarfs.  Nevertheless, there exist low-luminosity pressure-supported galaxies, such as VCC~1261, that do not qualify as dwarfs based on the criterion laid down above.}
}

{\color{black}
\subsection{
Dwarf Nomenclature and Characteristics
}
\label{nomenclature}
}

{\color{black} Fundamentally, there are two kinds of pressure-supported dwarfs:}  gas-rich and gas-poor.  {\color{black} Gas-rich versions} fall into two categories:  dwarf irregular galaxies (dIs) and blue compact dwarfs (BCDs).  Gas-rich galaxies display surface brightness profiles that are exponential at large radii, but towards their centres profiles of dIs flatten whereas those of {\color{black} many} BCDs continue to brighten {\color{black} \mbox{\citep{vad05a, cai15a}}}.
Gas-poor dwarfs also have been classified into two groups:  dwarf spheroidal galaxies (dSphs) and dwarf elliptical galaxies (dEs).  The former, which are the least luminous, display surface brightness profiles like those of dIs {\color{black} \citep{kor85a, vad08a}}.  {\color{black} The latter, originally recognized in the Virgo cluster, are defined to be objects with surface brightness profiles flatter than a de Vaucouleurs law, which distinguishes them from giant ellipticals \citep{san84a}.}  Core profiles often rise steeply (with many being nucleated), but the behaviour at large radii can be close to exponential \citep{tol14a, jan14a}.  The various structural similarities amongst all of these galaxies suggest evolutionary connections, and admit the possibility that dSphs and/or dEs are dIs and/or BCDs that have lost their gas.

It is useful to draw upon terminology describing positioning in the Hubble sequence to distinguish between gas-rich and gas-poor dwarfs.  In this paper, dIs and BCDs together will be called late-type dwarfs (LTDs) and dSphs and dEs as a group will be termed early-type dwarfs (ETDs).  

Among LTDs, the {\color{black} photometric} distinction between dIs and BCDs can be ascribed to star formation activity.  By regarding the enhancement of surface brightness in the core of a BCD to be merely the consequence of a burst of star formation, \citet{vad05a} demonstrated {\color{black} through surface photometry in $K_s$} that the two kinds of galaxies are structurally indistinguishable.   

{\color{black} \citet{lel14a} showed that BCDs for which rotational motions are evident (the most luminous) generally have steeper inner velocity gradients than dIs displaying rotation, suggesting that their starburst is linked to the inner shape of the potential well.  However, there are many such galaxies that are morphologically distorted, even with multiple starburst sites, possibly as a consequence of an interaction \citep{cai15a}.  Nevertheless, \citet{lel14a} recognized compact dIs with inner velocity gradients similar to those seen in BCDs, and concluded that BCDs must evolve into such galaxies provided that mass is not redistributed.}

The distinction between dSphs and dEs is more complex, but appears to be anchored to mass.  \citet{gre01a} considered dSphs to be fainter than 
$M_V \sim -14$.  
{\color{black} In keeping with \citet{san84a}, she recognized dEs to be fainter than
$M_V \sim -17$, consistent with what is expected using the modern distance modulus for the Virgo Cluster adopted by \citet{mcc14a}.}   
However, an apparent two-magnitude gap between the brightest dSphs and faintest dEs has been shown to be a selection effect.  \citet{for11a} identified faint dEs in the Virgo cluster with structural scaling relations extending close to the realm of dSphs.  {\color{black} As a function of increasing luminosity, the Sersic index for dSphs stays more or less constant around unity and then seamlessly transitions to a slowly rising trend for dEs \citep{mis08a,mis09a,lie12a}.}  {\color{black} However}, velocity dispersions show that dark matter predominates over baryons in the inner regions of dSphs \citep{wol10a}, but that stars make up half of the mass {\color{black} within the half-light radius} of dEs \citep{tol14a}.  {\color{black} Furthermore},  \citet{lie12a} presented evidence for a break in the slope of the colour-magnitude relation for dEs in the Virgo cluster at 
$M_\textit{V} \sim -14$,  
with fainter galaxies uniformly blue and brighter galaxies trending redward with increasing luminosity.  They proposed that the break signifies a critical mass required for a galaxy to retain gas, and explained the blue colours of dEs with lower masses to be a consequence of less efficient enrichment.  {\color{black} Such an explanation is attractive because it simultaneously allows structural trends for dSphs and dEs to join.}  Perhaps dEs fainter than the break luminosity should be regarded as the dSphs of the cluster.  From an evolutionary standpoint then, the logical luminosity in the near-infrared at which to distinguish dSphs from dEs is 
$M_\textit{Ks} \sim -16.5$   
(see also \citealt{for11a}), since the mean $V-K_s$ colour of dEs in Virgo is 
2.5  
(see \S\ref{photometry_dsph}).

Often the classification ``dSph'' is applied to ETDs located in the Local Group, whereas ``dE'' is applied to similar galaxies beyond, most commonly in clusters.    
The environmental distinction is highlighted in this paper, in part because of the different approaches to the analysis of the two categories of galaxies.  All but three of the ``dSphs'' in the Local Group sample to be discussed satisfy the luminosity condition, too, having absolute magnitudes fainter than 
$-16.5$ in $K_s$.  


\subsection{
Relationship of Early-Type Dwarfs to Late-Type Dwarfs
}  

It has long been known that there exists a Fundamental Plane connecting measurements of radii, velocity dispersions and surface brightnesses of luminous early-type galaxies \citep{djo87a, dre87a}.  ETDs connect to it in a two-dimensional Fundamental Manifold \citep{zar06a, zar06b}.  \citet{tol11a} even showed that {\color{black} pressure-supported} bodies spanning eight orders of magnitude in luminosity scatter around a Fundamental Curve in mass-radius-luminosity space.

Because they are structurally similar to dIs, it has been proposed that ETDs are LTDs that are deficient in gas 
\citep{kor85a}.  
Further support for this idea came from the recognition of a Fundamental Plane for {\color{black} pressure-supported} LTDs, which attributes scatter in the relationship between the near-infrared luminosity and internal motions to variations in surface brightness 
\citep{vad05a}.  
It revealed that the correlation between dynamical and structural properties for dSphs is similar to that for {\color{black} pressure-supported} dIs and BCDs \citep{vad08a}.  Also, in the Virgo Cluster, structurally compact ETDs display SDSS light profiles similar to the low surface brightness component of BCDs located there, and the more diffuse ETDs have properties akin to irregulars \citep{mey14a}.  In another near-infrared study, \citet{you14a} confirmed that underlying structural properties of dIs in the Local Volume overlap with those of ETDs in both the Local Volume and the Virgo Cluster.  {\color{black} Furthermore, \citet{lel14a} showed that rotating ETDs in the Virgo Cluster have inner velocity gradients that correlate with central surface brightnesses similarly to LTDs, and that the gradients themselves are comparable to those of rotating BCDs and compact dIs.}  However, correlations with luminosity observed {\color{black} by \citet{you14a}} for Local Volume galaxies and Virgo Cluster dEs showed significant differences in both slope and scatter, suggestive of environmental regulation.  A principal components analysis by \citet{woo08a} also showed that the relationships between stellar masses {\color{black} (derived using stellar population models from observations in $V$),} radii, and internal motions for ETDs and LTDs overlap.  Since the dispersion in properties along one principal axis was eight or more times that along the other two, \citet{woo08a} felt that dwarfs might better be regarded as populating a "Fundamental Line" in their parameter space, {\color{black} a suggestion previously made by \citet{dek03a} for galaxies with masses between $6 \times 10^5$ and $3 \times 10^{10} \, \rm M_\odot$.}

\citet{kor16a} examined the placement of dIs and dSphs with respect to scaling relations for dark matter in rotationally-supported spiral and irregular galaxies (i.e. disk galaxies with weak or non-existent bulges).  Specifically, they focussed on connections among the central density, core radius, velocity dispersion, and absolute magnitude in $B$.  From the standpoint of continuity, they contended that the relationships observed for rotating systems should extend into the realm of pressure-supported systems, and sought a path for unification.  Particularly revealing was the relationship between the central density of the dark matter halo and the absolute magnitude.  Relative to the extrapolated trend defined by the bright galaxies, absolute magnitudes of dIs and dSphs proved to be fainter than expected for their densities by
$\sim 3.5 \, \rm mag$  
and 
$\sim 4.6 \, \rm mag$,  
respectively.
The implication is that the dwarfs have proportionately higher surface mass densities of dark matter relative to baryons compared to {\color{black} galaxies that \citet{kor16a} regard as rotationally-supported} systems.  The offsets were attributed to baryon depletion, which could have been a consequence of proportionately less efficient capture of gas in the early universe or proportionately greater loss of gas later on, say by supernovae-driven winds.    This finding overlaps that of \citet{tol11a}, who argued that the weak correlation of mass with luminosity that they observed for the faintest dSphs was evidence that growth of small galaxies is limited by the shallowness of their potential wells.

With the presumption that trends for rotationally-supported galaxies are extensions of trends for dIs and dSphs, \citet{kor16a} utilized the observed offsets to gain insights into relative baryon depletion factors.  They concluded that the ratio of the surface density of baryonic matter to dark matter decreases with decreasing luminosity for galaxies fainter than $-18$ in $V$.

Although \citet{kor16a} found dSphs to be a magnitude fainter than dIs with the same central density, a comparison of baryonic masses does not easily follow.  In $B$, star formation enhances the brightess of dIs with respect to dSphs, possibly explaining much of the difference.  Furthermore, the authors did not account for gas, which augments the baryonic total for dIs substantially.  Because of the small samples, the comparison could be biased by selection effects, too.  Nevertheless, within the scatter, scaling relations for the two types of galaxies largely overlap, verifying at least their structural similarities, especially from the standpoint of dark matter.  In fact, \citet{kor16a} consider the presence or absence of gas to be a secondary issue.

Using measurements of [Fe/H] in individual stars, \citet{kir13a} showed that LTDs and dSphs follow the same relation between metallicity and overall stellar mass.  However, the {\color{black} metallicity distributions of stars within} dSphs display a sharp cutoff at high metallicity, leading \citet{kir13a} to conclude that star formation was curtailed, presumably through the loss of gas.  Many dEs have distinct nuclei and/or enhanced cores, and the chemical properties of dSphs indicate that much of the star formation occurred in a burst \citep{ric95a}.   Thus, ETDs might more appropriately be linked to BCDs.  Yet, \citet{wei11a} concluded that dwarfs of all types formed most of their stars before $z \sim 1$, and that the mean histories of star formation were comparable up until a few Gyr ago.

ETDs {\color{black} (and even LTDs with gas fractions below 0.4 -- see \citealt{geh06b})}  are generally found in environments where the mean density of matter is high, either {\color{black} in the vicinity of} giant hosts or in clusters.  In this sense, many may be considered to be the visible manifestation of subhaloes.  This has led many researchers to suggest that gas deficiencies stem from tidal effects \citep{may01a} or ram-pressure stripping \citep{ein74a,lin83a}.  Indeed, star formation histories elucidated from studies of resolved stellar populations suggest that gas loss was stimulated by an external mechanism \citep{wei11a}.  Furthermore, there are known to be rotationally-supported dE galaxies in the Virgo Cluster that fit the Tully-Fisher relation, suggesting that they are stripped versions of rotating star-forming galaxies \citep{tol11b}.  However, measurements of proper motions show that some dSphs around the Milky Way, such as Fornax and Leo I, are on orbits with pericentres substantially greater than those of the LMC and the SMC \citep{lux10a,soh13a}, both of which still host a lot of gas.  Furthermore, no neutral hydrogen has been found around dSphs in the Local Group \citep{bai07a,spe14a}.  Although it is clear that environment plays a role in distinguishing the gaseous content of LTDs from ETDs, stripping need not be the only way.  Conceivably, the conditions under which ETDs developed led to more rapid star formation, in which case gas loss could have been promoted through faster consumption and/or more vigourous winds \citep{dek86a,dek03a,rob16a}.

{\color{black} Based upon their internal motions, dSphs have dynamical masses that are extremely large compared to their luminosities 
\citep{wol10a}.   
In fact, mass-to-light ratios for ultra-faint dSphs are larger than observed for any other kind of galaxy, including ultra-diffuse galaxies 
\citep{wol10a,dok16a,tol18a}.}
At first glance, this distinction might argue against a connection with LTDs.  However, dynamical mass-to-light ratios are misleading, because the light only accounts for baryons in stellar form.  What really needs to be examined is the ratio of the dynamical mass relative to the total baryonic mass prior to gas loss.  If the stellar mass in a dSph today represents a small fraction of the baryonic mass originally present, then a large mass-to-light ratio would be an artifact of inefficient conversion of gas into stars. 

\subsection{
A Path Forward:  The Potential Plane
}

The Fundamental Plane for LTDs suffers from significant scatter over and above what can be attributed to observational errors 
\citep{vad08a}.  
\citet{mcc12b} 
surmised that if LTDs were virialized, there might be an observable relationship between the kinetic energy per unit mass conveyed by the velocity dispersion (as conveyed by the HI line width) and the potential energy per unit mass indicated by baryons, if the baryonic mass scaled with the mass of dark matter.  This is the precept of the Fundamental Manifold envisioned by 
\citet{zar11a}.  
Indeed, a strong relationship was found, although not that expected for virialized bodies.  The potential as judged from the total baryonic mass correlates with a linear combination of the velocity dispersion and near-infrared surface brightness (corrected for projection), so tightly in fact that scatter can be entirely attributed to measurement uncertainties.  This relationship is referred to as  the ``Potential Plane''.


By virtue of being founded upon baryonic masses, the Potential Plane opens up a possible avenue for recognizing how much gas ETDs once had.  If it is hypothesized that ETDs were once dIs or BCDs and that dark matter dominates the mass of all dwarfs, then the primary visible effect of expelling gas would be a reduction in the baryonic mass.  Thus, ETDs would be expected to display baryonic potentials systematically below those of LTDs with similar structural and dynamical properties.   How much gas they lost could be gauged from the offset.  The constraint would be invaluable for determining how the galaxies evolved prior to losing their gas and how much of an impact they have had upon the intergalactic medium.

In this paper, light profiles of ETDs with a wide range of luminosities are modeled in an identical way to LTDs in order to identify the placement of the galaxies with respect to the Potential Plane defined by LTDs.  {\color{black} Unlike \citet{lel14a}, the primary focus is on faint dwarfs, i.e., those believed to be} predominantly supported by pressure, and the intent is to gain insights into why galaxies that are so structurally similar have come to have such different gaseous properties.  Unlike \citet{kor16a}, the paper emphasizes photometric comparisons in $K_s$, thereby overcoming any biases that might come from variations in star formation activity in LTDs.  In comparing baryonic properties, gas is accommodated.  Furthermore, the paper extends the analysis to dEs, i.e. more luminous gas-poor dwarfs than are typically found locally.  

In \S\ref{sample}, the sample of ETDs to be studied is constructed.  In \S\ref{measurements}, surface brightness profiles for sample galaxies are fitted in the same way as for LTDs.  In addition, distances and absolute magnitudes for sample galaxies are evaluated homogeneously.
In \S\ref{analysis}, stellar masses are estimated, and the location of ETDs with respect to the Potential Plane is pinpointed.  Implications for evolution follow in \S\ref{implications}, where the mass of gas lost by dSphs is quantified, the effect of gas loss on dynamical mass-to-light ratios is analyzed, and the energy at which ETDs diverge from the Potential Plane is utilized to constrain the cause of gas loss.  As well, an estimate for the minimum baryonic mass of a galaxy is extrapolated.  {\color{black} Results are discussed in the context of past research in \S\ref{discussion}, and} conclusions are presented in \S\ref{conclusions}.

\section{
Sample
}
\label{sample}

In order to locate an ETD  with respect to the Potential Plane, it is necessary to quantify its surface brightness, size, luminosity, and internal motions.  Thus, candidates for study were restricted to those galaxies for which both the surface brightness profile and the velocity dispersion of stars {\color{black} were published}, and for which the distance could be reliably determined from published constraints {\color{black} (such as the tip of the red giant branch, RR Lyrae stars, or membership in a galaxy group or cluster)}. 

In all, 
23 dSphs  
were identified for study, of which 
five  
are classed as ultra-{\color{black}faint} dwarfs.  
{\color{black} 
The galaxies are} identified in Table~\ref{tab_obsdsph}.
{\color{black}
None  
have a velocity dispersion more than one sigma higher than $30 \, \rm km \, s^{-1}$.}  Extant data on stellar constituents enables the determination of distances from either the tip of the red giant branch (TRGB) or RR Lyrae stars (see \S\ref{distances}).  The three most luminous galaxies, namely NGC 147, 185, and 205, might be considered to be dEs on the basis of their {\color{black} absolute magnitudes} and light profiles (see \S\ref{nomenclature}), but they are grouped together with fainter dSphs because of their location in the Local Group and the nature of the data available for them.  {\color{black}  All of them exhibit rotation \citep{geh10a},
but the associated energy in all cases is less than 40\% of that in random motions.}

Given the difficulty of determining accurate distances individually to unresolved dwarfs, it was advantageous to restrict the dE sample to clusters or groups.  Recently, high-quality velocity dispersions and Sersic models of deep $H$-band surface photometry were presented for a large collection of dEs in the Virgo Cluster \citep{tol14a,jan14a}, so they are the primary focus of attention here.  The simplest objects, namely those classified as dE(N) and dE(nN) (nucleated and non-nucleated featureless dwarfs, respectively; see \citealt{lis07a}) were selected for study.  The former are distributed like E/S0 galaxies in Virgo, while the latter appear to be part of the unrelaxed population of spirals and star-forming dwarfs \citep{lis07a}.

Of the dEs selected, the faintest are substantially brighter than the bulk of the dSphs.  To extend the sample to lower luminosities, 
three  
pressure-supported dEs studied by \citet{for11a} were added 
(PGC~032348, VCC~846, and VCC~1826).   
{\color{black} All are nucleated \citep{tre02a, for11a}}.
Two  
of these are in the Virgo Cluster, while the third is a member of the Leo I Group \citep{pat03a}. 

{\color{black}
In all,
21  
dE galaxies were selected for study, 
four  
of which are non-nucleated.  
They are identified in Table~\ref{tab_obsde}.
Nine  
have a velocity dispersion exceeding $30 \, \rm km \ s^{-1}$ by more than one sigma.  {\color{black} For 16 of the 18 galaxies for which rotational motions have been constrained \citep{tol14a}, the energy exhibited by rotation is only 20\% or less of that associated with random motions, and in that sense the dEs can be regarded as predominantly supported by pressure.  For the other two, the energy in rotation is 50\% of that in random motions.}
}

\section{
Measurements  
}
\label{measurements}

\subsection{Kinematics}
With the exception of NGC 147, 185, and 205, for which internal motions have been gauged from spectroscopy of integrated light, velocity dispersions for sample dSphs come from measurements of the radial velocities of individual stars.  Adopted values {\color{black} and uncertainties} along with sources are provided in Table~\ref{tab_obsdsph}.  For dEs, velocity dispersions come exclusively from {\color{black} long-slit} spectroscopy of the integrated light \citep{tol14a,for11a}.   {\color{black}Regardless of technique, measurements are reflective of motions within the half-light radius, which is where velocity dispersion profiles are relatively flat.} Adopted values of velocity dispersions {\color{black} and their uncertainties and sources} are provided in Table~\ref{tab_obsde}. 

{\color{black} 
Because two kinds of measurements of velocity dispersions are employed, it is important to be aware of factors that might bias comparisons.  First, mass segregation (if it exists) should not be a factor because both discrete and integrated measurements preferentially sample the giant population.  Stellar binarity inflates velocity dispersions measured by any method, but probably by less than 
30\%  
for galaxies with velocity dispersions between 
4 and $10 \, \rm km \, s^{-1}$  
\citep{min10a}.  Although it is conceivable that dEs host central black holes whereas dSphs do not, measurements of velocity dispersions were made in such a way that they are unlikely to be inflated significantly by associated dynamical effects.   For globular clusters, measurements of velocity dispersions derived from integrated light can be biased by ``shot noise'' from bright giants if a few contribute a significant fraction of the light in the sampling aperture \citep{dub97a,lan13a,lut15a}.  That is not expected to be a problem here because the galaxies for which there are such measurements are generally unresolved, so sampling apertures encompass many more stars \citep{dub97a}.   \citet{lut15a} showed that in the extremely dense core of the massive globular cluster NGC 6388 discrete measurements of velocities from observations acquired with adaptive optics are biased towards the mean cluster velocity as a result of contamination by the wings of the point spread function for other stars and by the unresolved background, thereby lowering the velocity dispersion.  Velocity dispersions adopted here for dwarf spheroidal galaxies were not measured using adaptive optics, and in any case such galaxies are nowhere near as concentrated as NGC 6388.  What matters most is that velocity dispersions derived by integrated and discrete measurements agree in globular cluster fields where the aforementioned problems are minimal or appropriately corrected \citep{lut15a}.
}

In establishing the Potential Plane, motions in LTDs were quantified by {\color{black} $W_{20}$, the width of the HI line profile at 20\% of the peak.  HI line profiles for most of the galaxies in the LTD sample are Gaussian (see, for example, \citealt{huc03a}).}  Therefore, to examine the placement of ETDs with respect to the Potential Plane, a Gaussian velocity distribution was assumed {\color{black} in order} to estimate $W_{20}$ from the measured line-of-sight stellar velocity dispersion $\sigma_\textit{los}$ {\color{black} \citep{ver01a}}:
\begin{equation}
\label{w20_etd}
W_{20} = (8 \ln 5)^{1/2} \sigma_\textit{los}
\end{equation}  

{\color{black} For three pressure-supported LTDs (Leo A, DDO 210, and Sag DIG), 
\citet{kir17a} compared stellar velocity dispersions determined from spectroscopy of individual stars to intensity-weighted velocity dispersions derived from HI maps.  Considering uncertainties, they concluded that the two methods deliver consistent results.  Thus, it is not expected that positioning of ETDs with respect to the Potential Plane of LTDs would be biased significantly by {\color{black} the use of a different tracer of kinematics}.}

\subsection{Photometry}

\subsubsection{Dwarf Spheroidal Galaxies}
\label{photometry_dsph}

In order to reduce the impact of young stars on light profiles, the Potential Plane for LTDs was defined using surface photometry in the $K_s$ band \citep{vad05a}.  However, in the dSph sample, useful images in this band were available for only three galaxies (NGC~147, NGC~185 and NGC~205).  For the rest, it was necessary to fit data in $V$ and transform photometric parameters to $K_s$.  Fortunately, dSphs are not forming stars, so optical profiles should resemble those in the near-infrared.  

To estimate $V-K_s$, Goldmine data for 40 dEs in the Virgo cluster \citep{gav03a} were examined.  Goldmine $K$ magnitudes were transformed from the ARNICA to the 2MASS system using equations provided by \citet{cut08a},  and both $V$ and $K$ magnitudes were corrected for extinction and redshift using the York Extinction Solver (see \S\ref{extinctions}).  The 
32  
galaxies with absolute magnitudes fainter than 
$-17$  
in $V$ display no clear trend of $V-K_s$ with luminosity.  For them, the mean reddening-free $V - K_s$ colour is
$2.49 \pm 0.27$.   
This value was adopted to convert dSph photometry in $V$ to $K_s$.  The quoted uncertainty is the standard deviation, which is relevant if colour variations are exclusively cosmic in origin.  However, some of the dispersion must be due to observational errors, so for any single galaxy it is best to regard any error estimate stemming from it as an upper limit.

For many galaxies, surface brightness profiles have been conveyed via star counts \citep{irw95a,mun10a,oka12a}.   To convert $\hbox{counts} / \hbox{arcmin}^2$ to $\hbox{mag} / \hbox{arcsec}^2$, the published central surface brightness or apparent total magnitude (with extinction added) was compared with the counts expected at the centre or overall from the published fit of a King law \citep{kin62a} to the stellar densities.

{\color{black}
The Sersic function \citep{ser63a} has proved to be attractive for modeling light profiles of dwarfs because the extra degree of freedom provided by the Sersic index offers the means to fit a wide range of shapes.  For example, \citet{you14a} chose to model dI profiles with a Sersic function, mainly because of its success in modeling dEs, and partly because of interest in studying variations in the Sersic index itself.  However, this freedom has the potential to obscure underlying simplicity.  BCDs in the field are a case in point.  Their light profiles are typically very different from those of dIs in their cores but similar at large radii.  Rather than fitting the two kinds of galaxies with different Sersic functions, it is simpler and perhaps more logical to regard a BCD as a dI with a starburst on top of it.   In $K_s$, the surface brightness profile of the diffuse component of a dI is approximated well by a hyperbolic secant function 
\citep{vad05a}.  
Modeling of the light profile of a BCD can be accomplished by superimposing a Gaussian component on a sech profile 
\citep{vad06a}, 
which accounts for the burst of star formation.  
Extracting underlying structural parameters accordingly, \citet{mcc12b} showed that the two kinds of galaxies are indistinguishable in the Potential Plane (although the relationship between BCDs and dIs in clusters may be more complex:  see \citealt{mey14a}).
}

Except at their extremities, surface brightness profiles for dSphs are morphologically similar to those of dIs and BCDs.  {\color{black} In this paper, rather than utilizing a King law, Sersic function, or Gaussian \citep{kor16a}, they have been} modeled in the same way, {\color{black} partly because they may be related to LTDs, but especially} to minimize bias in the comparison of fitted parameters.  Thus, the surface brightness $I$ as a function of distance $r$ along the major axis was represented by
\begin{equation}
I(r)= \frac{2 I_{0}}{e^{r / r_{0}}+ e^{-{r / r_{0}}}}
\end{equation}  
where $I_{0}$ is the central surface brightness of the sech model and $r_{0}$ is the scale length.  To fit the profiles of Andromeda II and VII, a Gaussian component was added.  To fit the steeper core profiles of NGC~147, NGC~185, and NGC~205, an exponential component had to be added instead of a Gaussian.  {\color{black} To ensure that each fit was not skewed by anomalies in surface brightness in the core or periphery, inner and/or outer boundaries were set as necessary by eye.}

\begin{figure*}
\includegraphics[keepaspectratio=true,width=15cm]{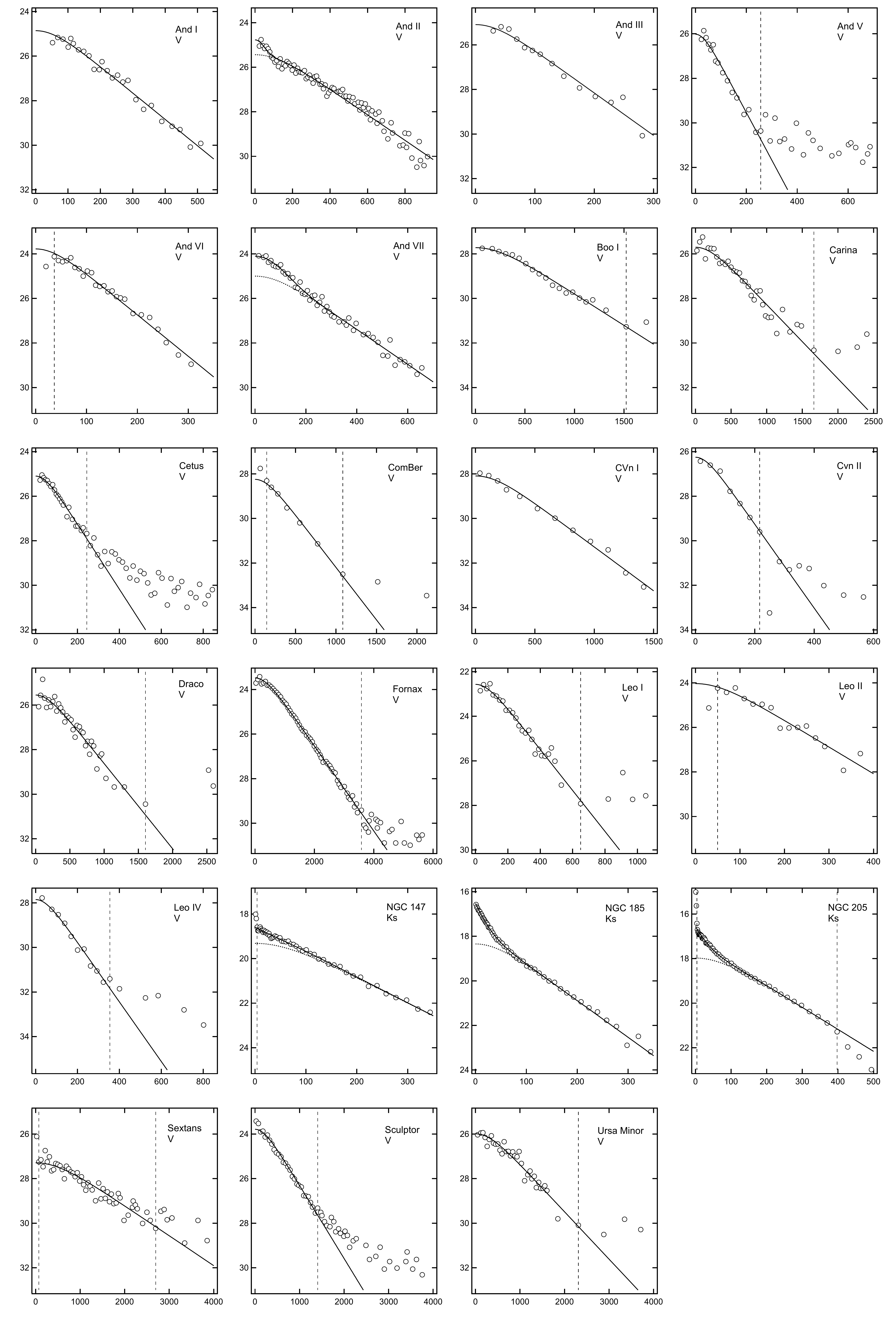}
\caption{\label{fig_sbp_grid}
Surface brightness profiles for dSphs.  The surface brightness in $V$ or $K_s$ in $\rm mag \, arcsec^{-2}$ (not corrected for extinction) is plotted against the radius in arcsec along the semi-major axis.  Solid black lines convey the overall fits, and in the case of profiles with two components, a thin dotted line traces the sech component.  
{\color{black} In panels where the range of radii fitted was restricted, inner and/or outer boundaries are marked by vertical dashed lines (set by eye).  An inner bound only was specified for And~VI, NGC~147, and Leo~II.  For the rest of the galaxies for which only one bound is specified, it is an outer bound.}
}
\end{figure*}

Fits to surface brightness profiles are shown in 
Figure~\ref{fig_sbp_grid}, 
and the parameters of the fits are given in Table~\ref{tab_obsdsph}. 
{\color{black} Because of the disparate ways in which data were presented in the literature, indicative uncertainties only are provided (in the table notes).  Surface brightnesses in $K_s$ for galaxies observed in $V$ typically have errors of 
$0.4 \, \rm mag$,  
$0.3 \, \rm mag$  
of which being due to background subtraction and 
$0.27 \, \rm mag$  
to the conversion from $V$ to $K_s$.  In comparison, photometric zero-point errors ($0.02 \, \rm mag$)  
and fitting errors 
($0.02 \, \rm mag$)  
are inconsequential.  For the three galaxies observed in $K_s$, the uncertainty in the central surface brightness is estimated to be 
$0.15 \, \rm mag$.   
Based upon experience with fitting, errors in scale lengths are typically 
5\%.}  

Generally, profiles are fitted well by the models.  At radii beyond the exponential zone, dSphs often show an enhancement in surface brightness.  {\color{black} For some, this may be due to underestimating the background level.  However, for those objects whose surface brightnesses steadily grow fainter, albeit at a slower rate, rather than leveling off (e.g., And~V, Cetus, CVn~II, and Sculptor), the enhancement may be a result of tides \citep{irw95a}.}  The only galaxy with a dubious fit is And~II, which has been shown to be undergoing a merger \citep{amo14a}.  In the centre of Sculptor there is a poorly-sampled spike in surface brightness analogous to what is seen in NGC 147, 185, and 205.  And~II is excluded from analyses of early-type dwarfs below, although it is marked in all figures.

To establish luminosities and masses of dSphs directly comparable to those of the LTDs used to define the Potential Plane, the apparent ``sech magnitude'' $m_\textit{sech}$ was computed by integrating the sech model out to infinity:
\begin{equation}
\label{appmag_sech}
m_\textit{sech} = -2.5 \log \left[ 11.51036 \, q \, r_{0}^2 \,I_{0}  \right]
\end{equation}  
Here, $q$ is the axis ratio, i.e., the ratio of the length of the semi-minor to semi-major axis of the isophotes.
As for LTDs, any core component was ignored.  {\color{black} Adopted axis ratios and sources are listed in Table~\ref{tab_obsdsph}.  Uncertainties in axis ratios are estimated to be 
10\%.  
Propagating all sources of error, uncertainties in the $K_s$ magnitudes are typically 
$0.43 \, \rm mag$ 
if they originate from observations in $V$ and 
$0.21 \, \rm mag$
otherwise (predominantly driven by the errors in the surface brightnesses).}

\subsubsection{Dwarf Elliptical Galaxies}

The distinction between the surface brightness profiles of dSphs and dEs is reminiscent of the difference between the profiles of dIs and BCDs.  For both kinds of galaxies, the surface brightness outside of the core declines more or less exponentially with radius (for the dEs studied here, Sersic indices range from 
0.8 to 1.7,  
with an average of 
1.1).   
It is hypothesized here that the framework of a dE is similar to that of a dSph, and that the foundational surface brightness profile resembles that described by a sech function.  Any excess in surface brightness towards the centre is regarded as an attribute which is not fundamental, but rather more like the excess seen in a BCD.  Even if this is not correct, it remains important that any comparison of dEs with dSphs be founded upon characterizations established in a like manner.  

The sech parameters for each dE were derived by matching both the slope and level of the surface brightness defined by a published Sersic model at a radius of 
2.5  
sech scale lengths (on average, 
1.7 times  
the half-light radii measured by \citet{jan14a}).  This radius is the minimum out to which a profile needs to be sampled to determine sech parameters robustly \citep{mcc12b}.

\citet{jan14a} carried out photometry of Virgo dEs in 2MASS $H$, and published Sersic indices, effective radii, axis ratios, and integrated magnitudes for Sersic models.  For galaxies with multi-component fits, the Sersic profile for the global/outer component was adopted to establish sech parameters.   \citet{for11a} employed photometry in SDSS $r$ for Virgo Cluster galaxies \citep{jan08a} and their own observations in $B$ and $R$ for the galaxy in the Leo I Group.   Sech parameters for the Virgo galaxies were derived from Sersic indices, effective semi-major axis lengths, axis ratios, and model magnitudes that were kindly conveyed by J. Janz (2015, personal communication).  For the Leo galaxy, \citet{for11a} fitted a Sersic model, but only the Sersic index and scale length were published.  To set the surface brightness scale in $r$, the integrated Sersic magnitude in $r$ was estimated from the Petrosian magnitude published in SDSS DR12 \citep{gra05a}.  

With sech parameters established from the Sersic profiles at 2.5 sech scale lengths, resulting sech profiles deviate on average from the \citet{jan14a} profiles by only 
$0.14 \, \rm mag$
at 5 sech scale lengths.  In the centres, the profiles differ on average by 
$1.0 \, \rm mag$,  
mostly because of the depression of the sech function (by 
$0.7 \, \rm mag$  
with respect to a pure exponential).  Despite the different core behaviours, integrated sech magnitudes deviate on average from the integrated Sersic magnitudes by only 
$0.14 \, \rm mag$,  
all but three being fainter.

For all but one of the dEs in the sample from \citet{jan14a}, an estimate of $H - K$ in the ARNICA system was obtainable from the GoldMine database \citep{gav03a}.  To estimate surface brightnesses and integrated sech magnitudes in $K_s$, the reddened colours were converted to the 2MASS system using the transformation equation provided by \citet{cut08a}.  For the remaining galaxy, the reddening-free $H-K_s$ colour in the 2MASS system was adopted to be 
$0.16 \pm 0.07$,  
which is the average for the other dEs.
For every galaxy studied by \citet{for11a}, the sech magnitude in $K_s$ was estimated by approximating the reddening-free $r - K_s$ colour to be 
$2.37 \pm 0.21$,  
which is the mean for the dEs from the \citet{jan14a} sample as evidenced by their SDSS Petrosian magnitudes in $r$.

Parameters of the sech models for dEs are summarized in Table~\ref{tab_obsde}.  {\color{black} Estimated uncertainties are given in the table notes.  By propagating errors, uncertainties in apparent sech magnitudes are judged to be 
$0.26 \, \rm mag$  
typically.}

\subsection{
Extinction Corrections
} 
\label{extinctions}

Magnitudes and colours of galaxies and the stars within them were corrected for extinction using the York Extinction Solver (YES)
after first removing any corrections included in published values.  In estimating the extinction of a target, YES properly accounts for the effective wavelength shifts plaguing observations through broad-band filters by accounting for the spectral energy distribution (SED) of both the probe of extinction and the target \citep{mcc04a}.  

For each ETD, the optical depth due to dust in the Milky Way was computed from $E(B-V)$ colour excesses mapped by 
\citet{sch98a}, 
which are based upon colours of elliptical galaxies corrected to a redshift of zero.  Adopted colour excesses and optical depths at $1 \, \rm \mu m$ are listed in Tables~\ref{tab_obsdsph} and \ref{tab_obsde}.  The extinctions and K-corrections in $V$, $r$, and $K_s$ were evaluated from the optical depth and the redshift using the spectral energy distribution of an elliptical galaxy \citep{mcc04a}.  Surface brightnesses were also corrected for redshift dimming.
The extinction in $I$ and reddening $E(V-I)$ of stars at the TRGB were evaluated using the spectral energy distribution of an 
M0 giant.   
The extinction in $V$ of RR Lyrae stars was judged using the spectral energy distribution of an 
F0 giant  
\citep{smi95a}.  SEDs for the stars were adopted from the library of 
\citet{pic98a}. 

\subsection{
Distances and Absolute Magnitudes
}
\label{distances}

Distances to dSphs were determined from the TRGB or RR Lyrae stars.  RR Lyrae distances were adopted for galaxies lacking a sufficient number of stars to locate the TRGB reliably (Draco, Ursa Minor, Cetus, Canes Venatici II, Leo IV, Canes Venatici I, Bootes I,  and Coma Berenices).  TRGB distances were adopted for the rest because they don't explicitly depend on \hbox{[Fe/H]}, the measurement of which is very difficult.

Just as in the analysis of LTDs by \citet{mcc12b}, the absolute magnitude in $I$ of the TRGB,  $M_\textit{I,TRGB}$, was computed from
\begin{align}
M_\textit{I,TRGB} &= (-4.053 \pm 0.028) \nonumber \\
& + (0.217 \pm 0.020) \lbrack (V-I)_\textit{TRGB} - 1.6 \rbrack
\end{align}  
where $(V-I)_\textit{TRGB}$ is the colour of TRGB stars corrected for extinction and redshift.  The slope is from \citet{riz07a}.  Following \citet{mcc14a}, the zero-point is set by the maser distance to NGC~4258, which was adopted to be 
$7.6 \, \rm Mpc$  
\citep{hum13a}.
The apparent magnitude in $I$ for the TRGB was extracted for each galaxy from the literature.  When the corresponding $V-I$ colour was not explicitly given, it was judged from a colour-magnitude diagram. 
 
The absolute magnitude in $V$ of the RR Lyrae stars,  $M_\textit{{V,RR}}$, was computed from
\begin{align}
M_\textit{V,RR} &= (0.511 \pm 0.044) \nonumber \\
& + (0.214 \pm 0.047) \lbrack \hbox{[Fe/H]} - (-1.5) \rbrack
\end{align}  
where $\hbox{[Fe/H]}$ is the iron abundance relative to the solar value. 
The slope is from \citet{gra04a}.  To ensure that distances aligned with the TRGB scale, the zero-point was determined by comparing TRGB and RR Lyrae distances to And~I, II, and III, the only three galaxies in the sample for which data could be guaranteed to be homogeneous (the red giant tip was measured by \citealt{mcco04a} and \citealt{mcco05a}, RR Lyrae mean magnitudes were measured by \citealt{pri04a} and \citealt{pri05a}, and iron abundances were measured by \citealt{var14a}).  The adopted value is 
$0.08 \, \rm mag$ brighter  
than the widely-used calibration of \citet{cac03a}.  

Adopted distance moduli for dSphs are listed in 
Table~\ref{tab_obsdsph}.  
Typically, the error in TRGB distances is 
$0.04 \, \rm mag$  
(excluding the zero-point uncertainty, which is 
$0.11 \, \rm mag$).  
Distances derived from RR Lyrae stars are somewhat more uncertain, 
$0.06 \, \rm mag$,  
because they require explicit knowledge of the metallicity.  

For each dE, the distance was assigned to be that of the host cluster or group.  The distance modulus for the Virgo Cluster was adopted to be 
$30.98 \pm 0.12$.   
This is the weighted mean of the Cepheid moduli derived for 
NGC~4535, 4536, 4548, 4321, and 4496A   
computed using the methodology of \citet{mcc14a}, converted to the maser scale described above.  The uncertainty here is the dispersion expected as a result of depth, given that the angular radius of the cluster is 
$6^\circ$.   
The distance modulus for the Leo Group was adopted to be  
$30.11 \pm 0.07$  
on the maser scale, 
which is the mean of the distance moduli for 
NGC 3351, 3368, 3377, 3379, and 3384  
published by \citet{mcc14a}.  

Absolute magnitudes of ETDs in $K_s$ were calculated from distance moduli and the apparent sech magnitudes {\color{black} computed from equation~\ref{appmag_sech}}.  They are given for dSphs in Table~\ref{tab_comdsph} and dEs in Table~\ref{tab_comde}.  {\color{black} Uncertainties were estimated by propagating the errors in the apparent sech magnitudes, extinctions, and distances.  Typical values for dSphs are 
$0.43 \, \rm mag$  
for those observed in $V$ and 
$0.22 \, \rm mag$  
for those observed in $K_s$.  They are 
$0.34 \, \rm mag$  
for dEs.}

\section{
Analysis   
}
\label{analysis}

\subsection{
Establishment of the Potential Plane
}

The Potential Plane for LTDs is the relationship between the potential, the HI line width, and the central surface brightness of the sech model.  
The index of the potential, i.e., referred to here as the {\it observed} potential $P^\textit{\,obs}$, is the ratio of the baryonic mass $\mathcal{M}_\textit{bar}$ to the scale length $r_{0}$ of the sech component:
\begin{equation} \label{potential}
P^\textit{\,obs} \equiv \lvert -\mathcal{M}_\textit{bar}  / r_{0} \rvert
\end{equation}  
The baryonic mass is given by the sum of the gaseous and stellar masses,
\begin{equation} \label{mass_baryons}
\mathcal{M}_\textit{bar} = \mathcal{M}_\textit{gas} + \mathcal{M}_\textit{str}
\end{equation}  
where the mass of stars is estimated from the stellar mass-to-light ratio $\Upsilon_\textit{str}$ and the luminosity 
$\mathcal{L}_\textit{sech}$ of the sech component:
\begin{equation} \label{mass_stars}
\mathcal{M}_\textit{str} = \Upsilon_\textit{str} \ \mathcal{L}_\textit{sech}
\end{equation}  

\citet{mcc12b} found that a prediction $P^\textit{\,pre}$ for $P^\textit{\,obs}$ that is accurate to within observational errors for dIs and BCDs alike with absolute magnitudes fainter than 
$-20$  
in $K_s$ is given by
\begin{align} \label{potplane}
\log P^\textit{\,pre} &= (5.720 \pm 0.065) \nonumber \\  
& + \, (1.134 \pm 0.080)(\log W_{20} - 1.8) \nonumber \\  
& + \, (-0.198 \pm 0.018)(\mu_0^{\textit{i=}0} - 20)  
\end{align}  
with
\begin{equation} \label{masstolight}
\Upsilon_\textit{str,Ks} = 0.88 \pm 0.20  
\end{equation}
given $M_{\textit{Ks},\odot} = 3.315$  
\citep{hol06a,fly06a}.
Here, $\mu_0^{\textit{i=}0}$ is the central surface brightness in $K_s$ of the sech model at zero tilt (judged from the axis ratio by assuming an oblate spheroidal geometry) and $W_{20}$ is the HI line width as observed, i.e., with no correction for tilt.
The units of $P^\textit{\,pre}$, $W_{20}$, and $\mu_0^{\textit{i=}0}$ are 
$\mathcal{M}_\odot \, pc^{-1}$, $\rm km \, s^{-1}$, and $\rm mag \, arcsec^{-2}$, respectively.  
Note that the zero-point is different from that in \citet{mcc12b}; it has been updated to account for an improved distance to NGC~4258 \citep{hum13a}.

The solution for $\Upsilon_\textit{str,Ks}$ was made possible by its impact on the balance between gaseous and stellar masses.  Comparable estimates have been obtained in $H$ from population syntheses (see \citealt{you14a} and references therein).  There was little improvement in the fit if line widths were allowed to vary with tilt, prompting \citet{mcc12b} to conclude that motions within the mass range sampled are primarily disordered and isotropic.   

\begin{figure}  
\includegraphics[keepaspectratio=true,width=8.5cm]{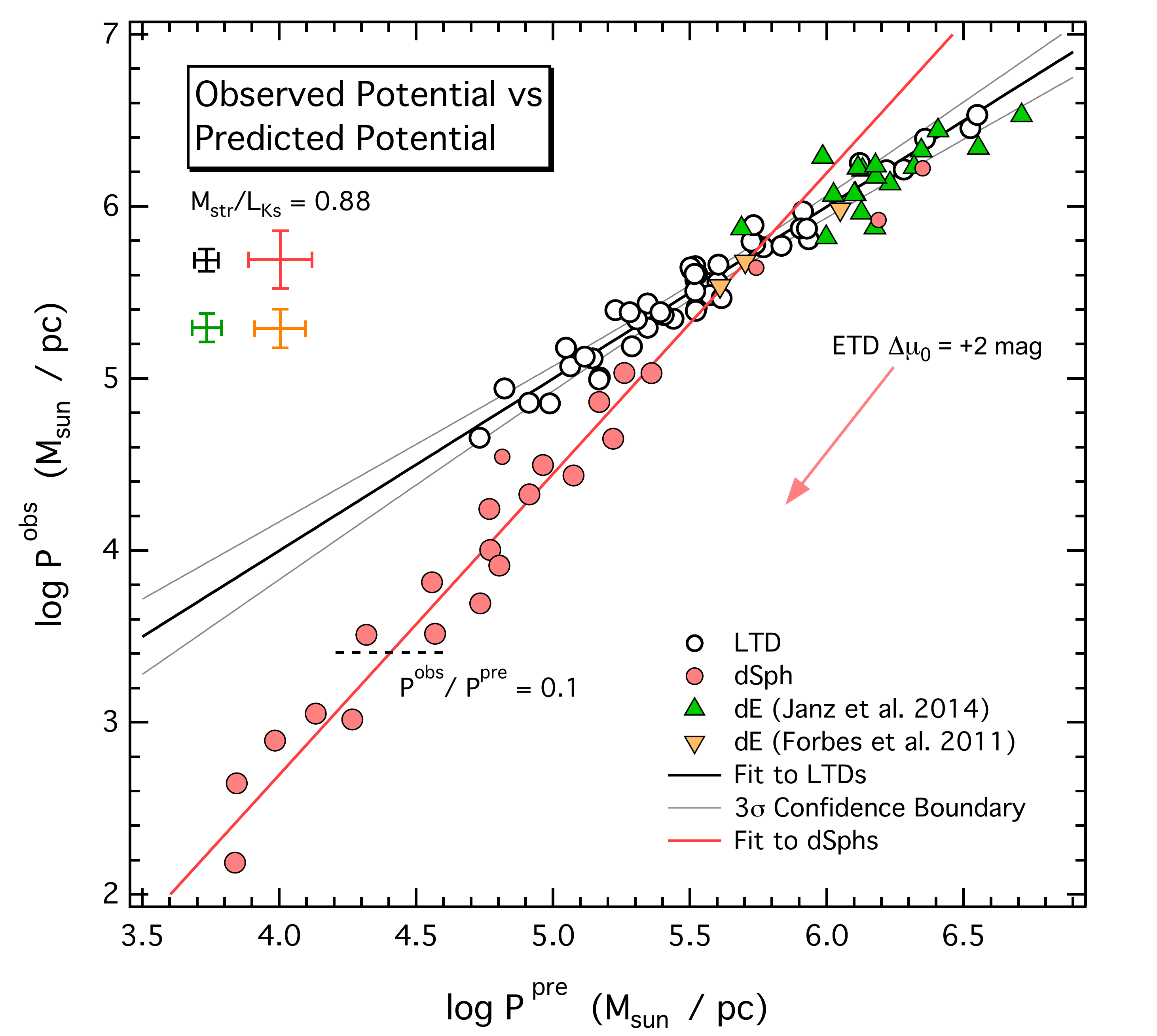}
\caption{\label{fig_potobs_vs_potpre}
Observed versus predicted potential for dwarfs.  The observed potential is established by the visible baryonic mass and the radial scale of a sech model of the light distribution in $K_s$.  The stellar component of the baryonic mass was estimated using a mass-to-light ratio in 
$K_s$ of 
0.88 in solar units.   
The predicted potential is established by internal motions and the central surface brightness of the sech model, the latter corrected to face-on.  Late-type dwarfs (LTDs), i.e., dIs and BCDs, are marked by open black circles.  They define the Potential Plane.  The corresponding fit is shown as a thick black line and $3\sigma$ confidence boundaries are marked by thin grey curves.  Solid pink circles mark dSphs.  Solid green triangles and solid yellow inverted triangles mark dEs.  For early-type dwarfs {\color{black} (ETDs), i.e., dSphs and dEs,} the gas mass was adopted to be zero.  The fit to the dSphs marked by large circles is shown as a thick pink line.  The intersecting horizontal dashed line elucidates where dSphs may retain only a tenth of their original baryonic content.  The galaxies excluded from the fit, which are marked with small circles, are those with absolute magnitudes brighter than 
$-16.5$  
in $K_s$  (NGC~147, NGC~185, and NGC~205) as well as And~II (a merger).     {\color{black}The pink arrow shows how the positioning of an ETD would change if its surface brightness were fainter by $2 \, \rm mag$.} Average observational uncertainties are indicated by the error crosses.
}
\end{figure}

The relationship between the observed and predicted potentials for LTDs is displayed in Figure~\ref{fig_potobs_vs_potpre}.
The 48 galaxies that \citet{mcc12b} employed in the fitting, which are marked by open black circles, delineate the Potential Plane from the side (the black line).  The dispersion 
($0.10 \, \rm dex$)  
is completely ascribable to observational errors \citep{mcc12b}.  {\color{black} Although 
five  
of the galaxies have velocity dispersions exceeding $30 \, \rm km \, s^{-1}$ by more than one sigma, they do not deviate from the trend.}

Equation~\ref{potplane} reveals that for LTDs,
\begin{equation}
P^\textit{\,pre} \propto \left[ W_{20} \right]^{1.1} \left[ I_{0}^{\textit{i=}0} \right] ^{0.5}  
\end{equation}  
where $I_{0}^{\textit{i=}0}$ is the central surface brightness in linear units.  For a virialized system, $P^\textit{\,pre}$ should vary as 
$\left[ W_{20} \right]^2$.   
\citet{mcc12b} suggested that the observed behaviour might be an artifact of varying mass-to-light ratios, compensation for which may be coming through the surface brightness term.  However, uncomfortably large fluctuations in the mass-to-light ratios are required to develop the virial expectation.  Regardless of its origin, the Potential Plane offers a baseline against which other kinds of dwarfs can be compared.

\subsection{
Placement of ETDs Relative to the Potential Plane
}
\label{placement}

The main objective of this paper is to establish the positioning of ETDs with respect to the potential plane.  ETDs lack gas, so $\mathcal{M}_\textit{bar}$ is determined by $\mathcal{L}_\textit{sech}$ alone.  Hence, $P^\textit{\,obs}$ is a proxy for 
$r_{0} \,I_{0}^{\textit{i=}0}$,  
so the relationship between $P^\textit{\,obs}$ and $P^\textit{\,pre}$ conveys, in effect, the relationship between $r_{0}$ and 
$\left[ W_{20} \right]^{1.1} \left[ I_{0}^{\textit{i=}0} \right]^{-0.5}$.  

To estimate $\log P^\textit{\,pre}$ for ETDs, central surface brightnesses were corrected to face-on by adopting an oblate spheroidal geometry, which is suggested by the distribution of axis ratios of dEs \citep{lis07a,san16a}.  Just as for LTDs, then,
\begin{equation}
\mu_0^{\textit{i=}0} = \mu_0 - 2.5 \log q
\end{equation}  
where {\color{black} $\mu_0$ is the central surface brightness of the sech model in magnitude units and} $q$ is the ratio of the semi-minor to semi-major axis of the isophotes. 

Since ETDs lack gas, it is necessary to utilize the line-of-sight velocity dispersion of stars to gauge $W_{20}$.  Comparisons with gas-based values for LTDs are valid, though, because dark matter is the predominant constituent of both dSphs and dIs \citep{kor16a}, its gravitational field establishing the motions of baryons of all forms {\color{black} (see \citealt{kir17a})}.
As for LTDs, motions in ETDs were assumed to be predominantly random and isotropic{\color{black}.  Even rotating ETDs (which tend to be the most luminous) are pressure supported in their inner regions \citep{lel14a}.  Thus,} values of $W_{20}$ determined from equation~\ref{w20_etd} were not adjusted for $q$.  The resulting values of $\log P^\textit{\,pre}$ for dSphs and dEs are provided in Tables~\ref{tab_comdsph} and \ref{tab_comde}, respectively.
{\color{black} Typical uncertainties arising from the errors in measurements are given in the table notes.  They were estimated by propagating the errors in the velocity dispersions, surface brightnesses, and axis ratios through equation~\ref{potplane}.}

Equations~\ref{potential}, 
\ref{mass_baryons}, and 
\ref{mass_stars} 
were utilized to estimate $P^\textit{\,obs}$ for ETDs for an appropriate choice of $\Upsilon_\textit{str,Ks}$ and with the approximation that $M_\textit{gas}$ is zero.  One way to evaluate $\Upsilon_\textit{str,Ks}$ and its range of variation is to use stellar population models.  However, an initial mass function must be assumed.  \citet{mar08a} 
derived mass-to-light ratios this way for faint satellites of the Milky Way.  For 11 dSphs with $\mathcal{L}_V / \mathcal{L}_{V,\odot}$ exceeding 
1000  
($-2.7 < \mathcal{M}_V <  -8.8$),  
the average value for $\Upsilon_\textit{str,Ks}$ in solar units is
$0.99 \pm 0.17$  
using the Salpeter IMF \citep{sal55a} and 
$0.50 \pm 0.07$  
using the Kroupa IMF \citep{kro93a} based upon the $V - K_s$ colour adopted in \S\ref{photometry_dsph}.  
Considering uncertainties, the range of variation for either choice of IMF is small.  Thus, it is approximated in this paper that $\Upsilon_\textit{str,Ks}$ for ETDs has a single value.   

There is cause to believe that the actual value {\color{black} of $\Upsilon_\textit{str,Ks}$} should be similar to that of LTDs.   The resolved population of nearby dIs (i.e., stars with $M_\textit{Ks} \le -7.5$) typically contributes only 5\% of the light in $K_s$, with the bulk of the flux coming from stars more than $4\, \rm Gyr$ old \citep{vad05a}.  Furthermore, the luminosity of an LTD inferred from the sech model must be representative of the old population, because there is no offset between BCDs and dIs in the Potential Plane \citep{mcc12b}.  Besides requiring no assumption about the IMF,  the value for $\Upsilon_\textit{str,Ks}$ derived for LTDs (equation~\ref{masstolight}) has the benefit of being tied directly to the observations establishing the Potential Plane.  Also, it is consistent with the mean value derived for dSphs by \citet{mar08a} using the Salpeter IMF.  Therefore, it is adopted for all ETDs.  

Computed stellar masses and quantities which follow from them are listed for dSphs in Table~\ref{tab_comdsph} and dEs in Table~\ref{tab_comde}.  {\color{black}  Typical uncertainties, which are listed in the table notes, were computed by propagating random errors in the measurements defining them.  Systematic errors common to both LTDs and ETDs, like those for the mass-to-light ratio and the zero-point of the distance scale, were ignored.  So, the error in $\mathcal{M}_\textit{str}$ was derived from the error in the absolute magnitude alone.  
The error in $P^\textit{\,obs}$ was derived from the errors in the surface brightness, axis ratio, angular scale length, and distance rather than the errors in the stellar mass and physical scale length in order to correctly account for the partial cancellation of the distance dependencies of the latter two quantities:
\begin{align}
\left[\delta \log P^\textit{\,obs}\right]^2 &= 
\left[\delta \log d\right]^2 + \left[\delta \log r_0\right]^2 \nonumber \\
& + \left[\delta \log q\right]^2 + \left[0.4 \, \delta \mu_0\right]^2
\end{align}
}

Figure~\ref{fig_potobs_vs_potpre} 
shows where the ETDs lie with respect to the Potential Plane {\color{black} defined by LTDs}.  Remarkably, dEs fall right on the plane.  The brightest dSphs (NGC 147, 185, and 205), all of whose absolute magnitudes are brighter than 
$-16.5$ in $K_s$,  
fall near the Plane, too, in the realm of the dEs. However, below $\log P^\textit{\,pre} = 5.5$, dSphs form a sequence that diverges below the Plane and its extrapolation.  Taking into account uncertainties in both the abscissae and ordinates, the relationship defined by the dSphs with absolute magnitudes fainter than 
$-16.5$  
(and excluding And~II) is given by
\begin{align} \label{potplane_dsph}
\log P^\textit{\,obs} (\hbox{dSph}) & = (5.320 \pm 0.122) \nonumber \\
& + (1.750 \pm 0.133) \left[ \log P^\textit{\,pre} - 5.5 \right] \nonumber \\
& \left( \log P^\textit{\,pre} \le 5.5 \right)
\end{align}  
The root mean square deviation is only 
$0.17 \, \rm dex$,  
comparable to observational errors.

Because dEs and the most massive dSphs lie so close to the Potential Plane, it is natural to posit that all dSphs started there.  Then, the positioning of dSphs that deviate is most easily explained if they are simply LTDs that have lost gas.  The trend observed in 
Figure~\ref{fig_potobs_vs_potpre} implies that proportionately more gas was lost by dSphs with shallower potentials, a result that could not be extracted from the two-parameter correlations studied by \citet{kor16a}. 

{\color{black}
Observed and predicted potentials for ETDs are coupled because the central surface brightness was employed to establish apparent magnitudes.  In Figure~\ref{fig_potobs_vs_potpre}, the trajectory that any of those galaxies would take if their surface brightness were fainter by $2 \, \rm mag$ is marked with an arrow.  The slope is close to that of the locus of dSphs with $\log P^\textit{\,pre} \leq 5.5$.  This means that errors in surface brightness are not responsible for the dispersion.  It also means that any aspects of evolution that solely affect the surface brightnesses of the galaxies are indiscernible. 
}

Unfortunately, LTDs in the sample do not extend to potentials as shallow as observed for dSphs.  It is conceivable that such galaxies don't exist if the point where dSphs begin to deviate represents an energy threshold below which LTDs begin to have difficulty retaining their gas.  Thus, inferences made about the smallest dSphs are at the mercy of extrapolation.  Nevertheless, the trend for LTDs is so tightly defined (as evidenced by the $3\sigma$ confidence boundaries displayed in Figure~\ref{fig_potobs_vs_potpre}) that any downturn within reason would be unlikely to eliminate the discrepancy with dSphs altogether.   In \S\ref{pcrit}, the location of the break-point is analyzed to gain insights into how dSphs lost their gas.

\begin{figure}  
\includegraphics[keepaspectratio=true,width=8.5cm]{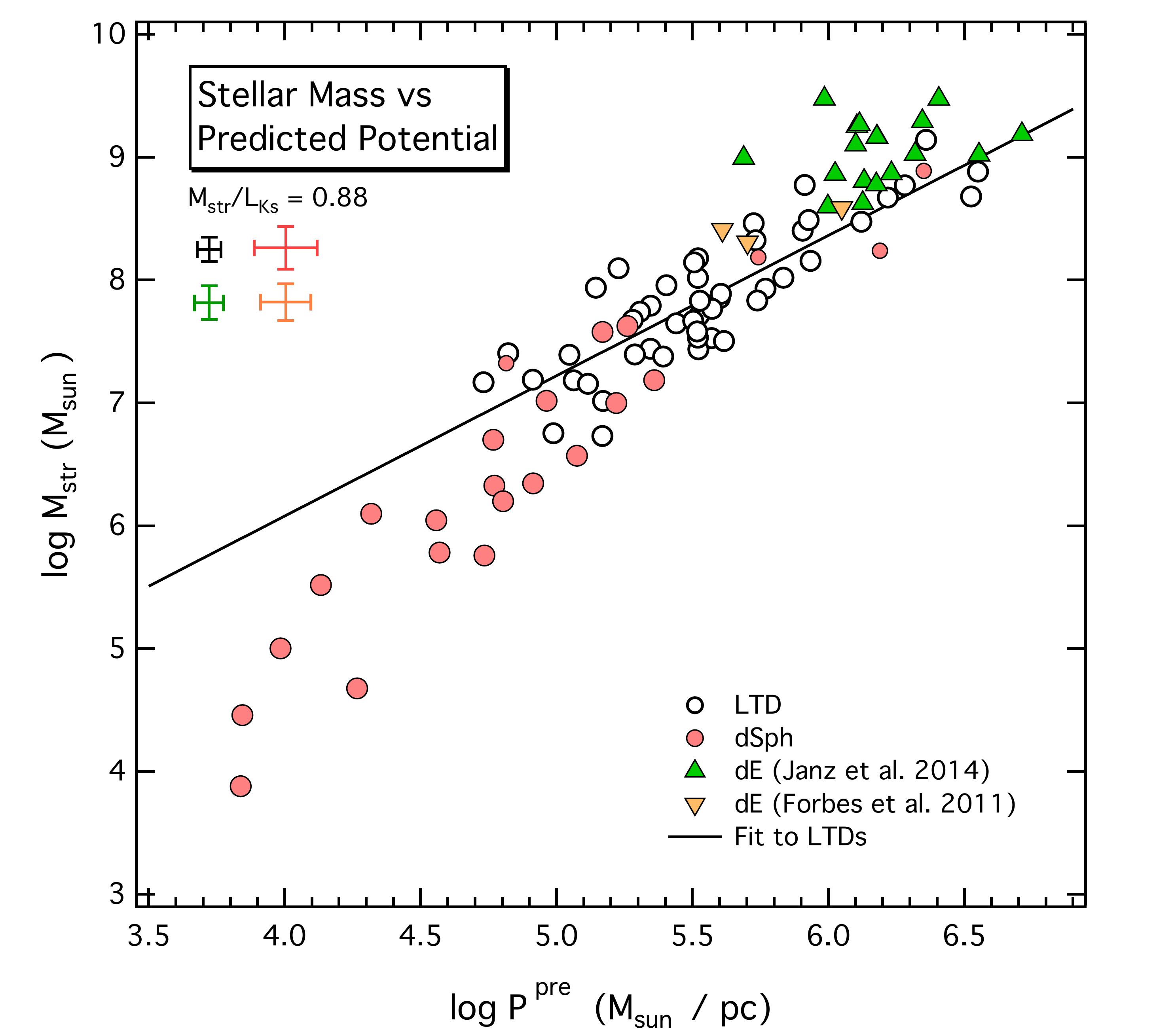}
\caption{\label{fig_mstr_vs_potpre}
Stellar mass versus predicted potential for dwarfs.  Symbols are as in Figure~\ref{fig_potobs_vs_potpre}. The black line is a least squares fit to the late-type dwarfs (LTDs).  Average observational uncertainties are indicated by the error crosses.}
\end{figure}

Figure~\ref{fig_mstr_vs_potpre} 
displays the correlation of stellar mass with the predicted potential. 
At any given value of the potential, dEs typically have more mass in stars than LTDs.  Because dEs lie on the Potential Plane, the excess is comparable to the mass of gas remaining in LTDs.  Thus, the stellar masses of dEs today should be close to their original baryonic totals.  On the other hand, the three most massive dSphs (NGC 147, 185, and 205) appear to be somewhat deficient in stars relative to dEs.  However, their proximity to the potential plane suggests that they were more efficient than most smaller dSphs in converting gas into stars, and in that sense they are like dEs.

\section{
Implications  
}
\label{implications}

The analysis above motivates the hypothesis that the Potential Plane of LTDs and its extrapolation define the starting point for ETDs, and that today's deviations of dSphs from the Plane are a consequence of gas loss.  In this section, implications are explored.

Given the prevailing evidence that dark matter controls dynamics \citep{wol10a,mcc12b,kor16a}, it is approximated here that the deviations of dSphs from the Potential Plane are close to vertical, i.e., that any loss of gas did not lead to a substantial perturbation of either the velocity dispersion or surface brightness of constituent stars.  If there were any response of the dark matter to the loss of gas, it likely would have been in the sense of becoming more dispersed.  The outcome would have been a lower velocity dispersion and/or a fainter surface brightness, reducing the predicted potential.   Consequently, the vertical offset from the Potential Plane that is observed today probably can be regarded as the {\color{black} {\it maximum}} possible deviation {\color{black} that could arise from gas loss}.  

\subsection{
Gas Masses for dSphs
}

If the only distinction between dSphs and LTDs is the gas content, then for a dSph the ratio of the observed potential to the predicted potential at a given value of the predicted potential can be used to evaluate the fraction $f_\textit{str}$ of baryonic matter that was in stellar form {\color{black} immediately before} gas loss, and in turn the gas fraction $f_\textit{gas}$. {\color{black} Accordingly,}
\begin{align} \label{potratio}
P^\textit{\,obs} / P^\textit{\,pre} &= \mathcal{M}_\textit{str} / (\mathcal{M}_\textit{gas} + \mathcal{M}_\textit{str}) \nonumber \\
& = f_\textit{str} \nonumber \\
& = 1 - f_\textit{gas}
\end{align}  
{\color{black} 
Thus, {\color{black} just prior to} gas loss, the baryonic mass was
\begin{equation} \label{mbar}
\mathcal{M}_\textit{bar} =  \mathcal{M}_\textit{str}  / \left[ P^\textit{\,obs} / P^\textit{\,pre} \right]
\end{equation}
and the mass of gas was
\begin{align}   \label{mgas}
\mathcal{M}_\textit{gas} &= \mathcal{M}_\textit{bar} - \mathcal{M}_\textit{str} \nonumber \\
& = \mathcal{M}_\textit{str} \left[ \frac{1 - P^\textit{\,obs} / P^\textit{\,pre}}{P^\textit{\,obs} / P^\textit{\,pre}} \right]
\end{align}  
}
The ratio of potentials (the gap between dSphs and LTDs in Figure~\ref{fig_potobs_vs_potpre}) can be expressed in terms of $P^\textit{\,pre}$ using equation~\ref{potplane_dsph}.  A black horizontal dashed line in Figure~\ref{fig_potobs_vs_potpre} marks where the ratio drops to 0.1.

\begin{figure}  
\includegraphics[keepaspectratio=true,width=8.5cm]{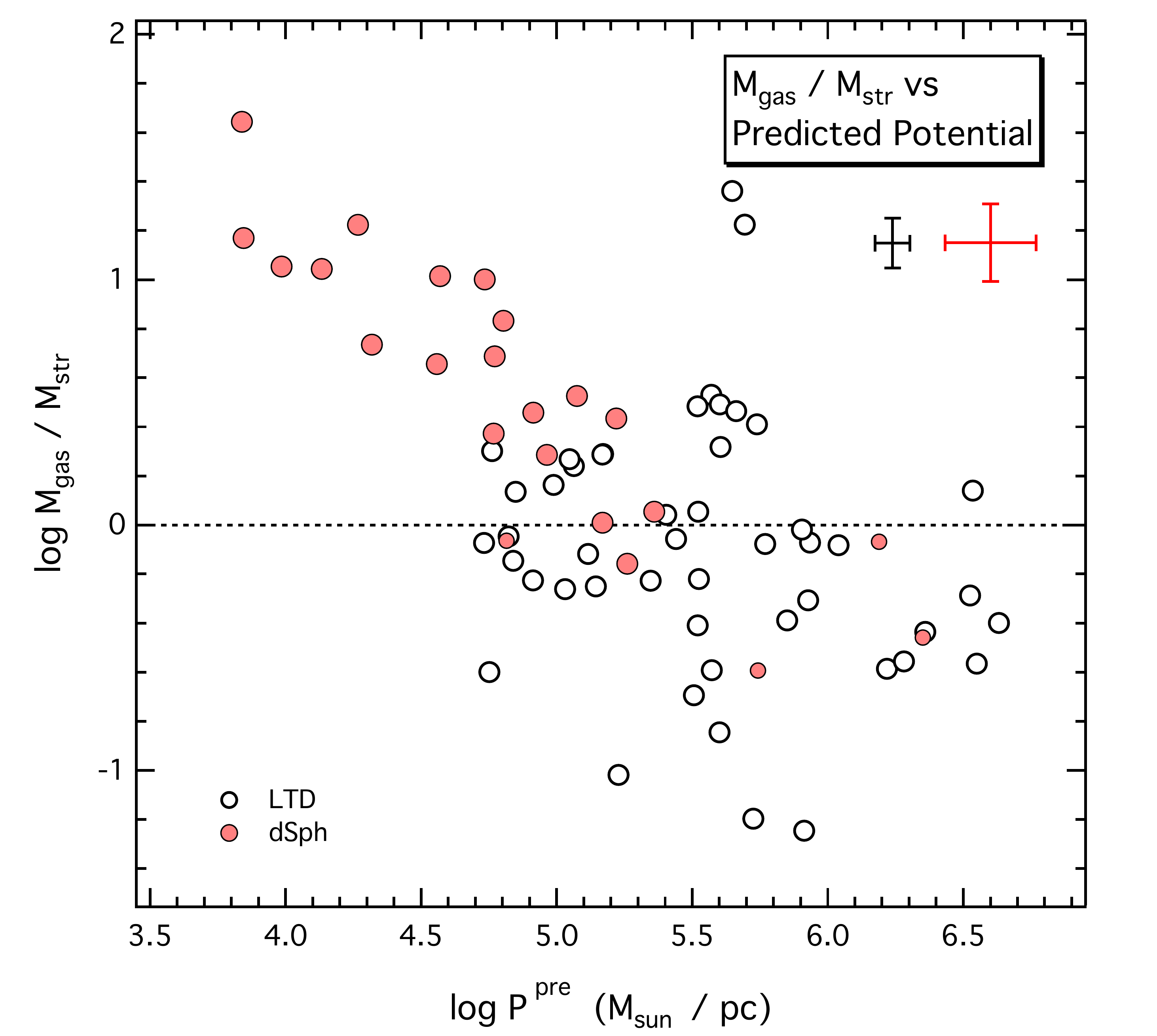}
\caption{\label{fig_mgas_to_mstr_vs_potpre} 
The ratio of the gas mass to the stellar mass as a function of the predicted potential.  Symbols are as in Figure~\ref{fig_potobs_vs_potpre}.  {\color{black} For late-type dwarfs (LTDs), the ratio of the gas mass to the stellar mass is that observed now, whereas for dwarf spheroidal galaxies (dSphs), it is that just prior to gas loss}.  Gas masses for LTDs are measured directly from $21 \, \rm cm$ fluxes.  Those for dSphs are inferred from offsets from the potential plane. 
{\color{black} The dotted horizontal black line marks where gas and stellar masses are equal.}  Average observational uncertainties are indicated by the error crosses.
}
\end{figure}

Figure~\ref{fig_mgas_to_mstr_vs_potpre} demonstrates that  $\mathcal{M}_\textit{gas} / \mathcal{M}_\textit{str}$ was higher in dSphs with shallower potentials, and suggests that those galaxies with
{\color{black}
$\log P^\textit{\,pre} < 5.5$
}
lost more than half of their baryons.
A system which loses half or more of its mass becomes gravitationally unbound \citep{hil80a}, so the mere existence of such objects supports the contention that dark matter is the dominant mass constituent (see also \citealt{kor16a}).

{\color{black}
Inferred gas masses computed from equation~\ref{mgas} are listed in Table~\ref{tab_comdsph}.  Representative uncertainties, given in the table notes, were computed from
\begin{align}
\left[ \delta \mathcal{M}_\textit{gas} / \mathcal{M}_\textit{bar}\right]^2 &= 
\left[(\delta d / d)^2 + (\delta r_0 / r_0)^2\right] \left[1 - 2 \, P^\textit{\,obs} / P^\textit{\,pre} \right]^2 \nonumber \\
& + \left[\delta q / q\right]^2 \left[ 2.5 c + P^\textit{\,obs} / P^\textit{\,pre} \right]^2 \nonumber \\
& +  \left[ \delta \mu_0 \ln 10 \right]^2 \left[ c + 0.4 \, P^\textit{\,obs} / P^\textit{\,pre} \right]^2 \nonumber \\
& + \left[ b \, \delta W_{20} / W_{20} \right]^2  + \left[ s \ln 10 \right]^2
\end{align}
where $b$ and $c$ are the coefficients in equation~\ref{potplane} defining the dependence of $\log P^\textit{\,pre}$ on $\log W_{20}$ and $\mu_0$, respectively, and where $s$ is the standard error of the mean for the fit ($0.01 \, \rm dex$).  The corresponding uncertainty in the baryonic mass is given by
\begin{align}
\left[ \delta \mathcal{M}_\textit{bar} / \mathcal{M}_\textit{bar} \right]^2 &= 
\left[\delta d / d \right]^2  + \left[\delta r_0 / r_0 \right]^2  \nonumber \\
&+ \left[ 2.5 c \, \delta q / q \right]^2 + \left[ (c \ln 10) \, \delta \mu_0 \right]^2  \nonumber \\
& + \left[ b \, \delta W_{20} / W_{20} \right]^2 + \left[ s \ln 10 \right]^2
\end{align}
The uncertainty in the gas fraction is given by
\begin{equation}
\delta (\mathcal{M}_\textit{gas} / \mathcal{M}_\textit{bar}) = \delta (P^\textit{\,obs} / P^\textit{\,pre})
\end{equation}
where
\begin{align}
\left[\delta \log (P^\textit{\,obs} / P^\textit{\,pre})\right]^2 &=
\left[\delta \log d\right]^2 + \left[\delta \log r_0\right]^2 \nonumber \\
& + \left[ (1 + 2.5 c)\delta \log q\right]^2 \nonumber \\
& + \left[(0.4 + c) \delta \mu_0\right]^2 \nonumber \\
& + \left[ b \, \delta \log W_{20} \right]^2 + s^2
\end{align}
}

\begin{figure}
\includegraphics[keepaspectratio=true,width=8.5cm]{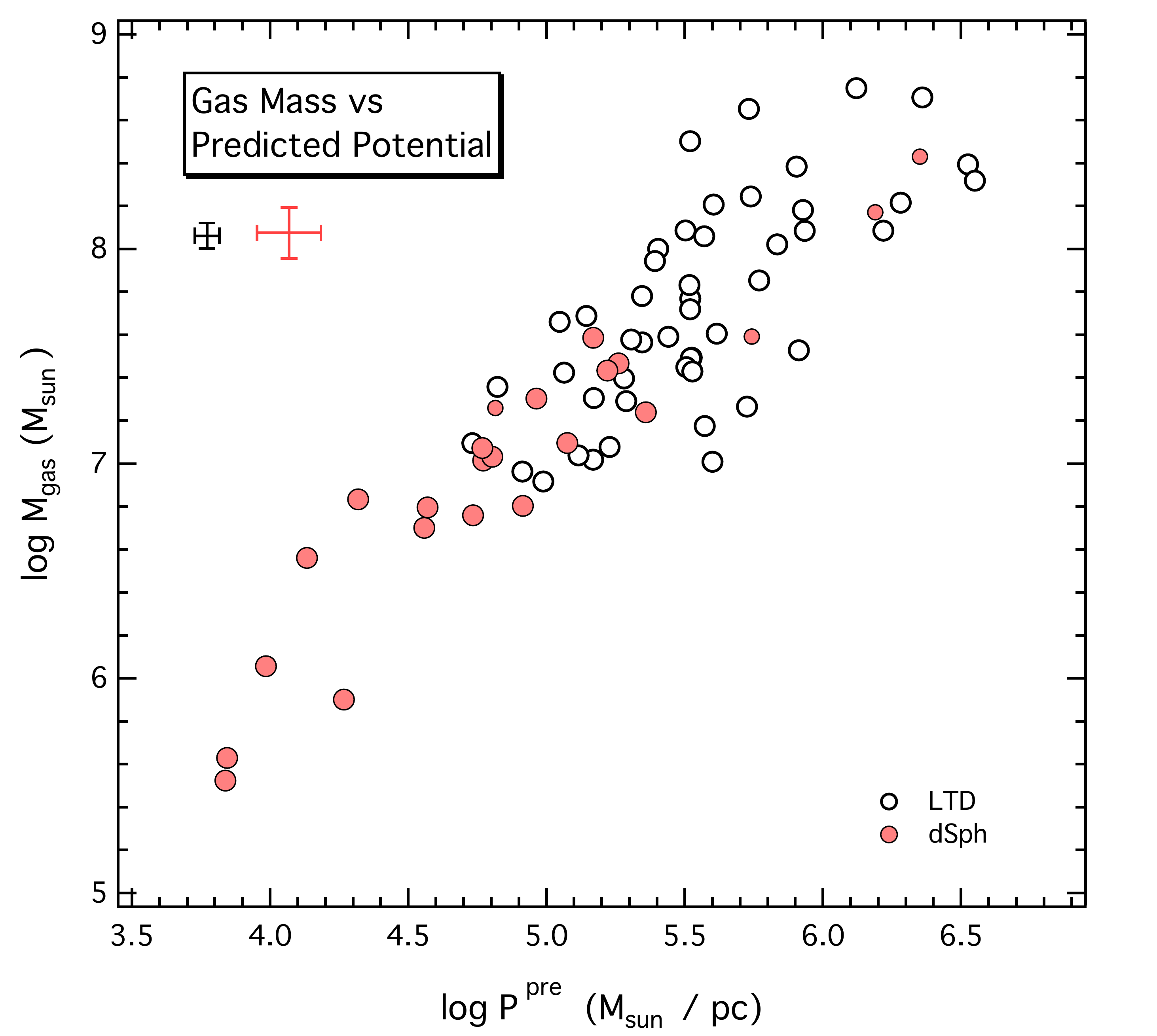}
\caption{\label{fig_mgas_vs_potpre}
The gas mass as a function of the predicted potential.   Symbols are as in Figure~\ref{fig_potobs_vs_potpre}.  Masses for late-type dwarfs (LTDs) are measured directly from $21 \, \rm cm$ fluxes, while those for dwarf spheroidal galaxies (dSphs) are inferred from offsets from the potential plane.
Average observational uncertainties are indicated by the error crosses.
}
\end{figure}

Figure~\ref{fig_mgas_vs_potpre} 
compares the inferred gas masses for dSphs with those determined from HI fluxes for LTDs.  
The gas masses for dSphs increase as $P^\textit{\,pre}$ increases, despite the declining gas fractions.  Especially significant is that the trend for dSphs meshes very well with that for LTDs, lending support for the hypothesis that the deviations of dSphs below the potential plane are a consequence of evacuation of gas from LTDs.  

\subsection{
Dynamical Masses
}

Dynamical mass-to-light ratios for the smallest dSphs can approach a thousand, giving the impression that baryons were incorporated into them with very low efficiency.  However, given the evidence that the galaxies lost much of their gas, it makes more sense to blame the extremes on the failure of luminosities to gauge the baryonic masses of the precursors. 

\citet{wol10a} showed that a ``dynamical'' mass $\mathcal{M}_{1/2}$ within the radius $r_{1/2}$ of the {\it volume} enclosing half of the light can be determined accurately from the measured line-of-sight velocity dispersion $\sigma_\textit{los}$ regardless of how mass is distributed.  Specifically, from the spherical Jeans equation,
\begin{equation}
{\mathcal M}_{1/2} / r_{1/2}  \cong 3 G^{-1} \sigma_\textit{los}^2
\end{equation}  
In this paper, a {\it dynamical mass} $\mathcal{M}_\textit{dyn}$ more indicative of the total is defined by
\begin{equation}
\mathcal{M}_\textit{dyn} \equiv 2 {\mathcal M}_{1/2}
\end{equation}  
How closely this dynamical mass comes to the total mass depends upon the density profile; it will underestimate the total mass if dark matter is more widely distributed than luminous matter.  
Nevertheless, a {\color{black} reliable} value of $\mathcal{M}_\textit{dyn} / \mathcal{M}_\textit{str}$ {\color{black} within $r_{1/2}$} should be given by $\mathcal{M}_\textit{dyn} / \mathcal{L}_\textit{sech}$ with an appropriate choice for $\mathcal{M}_\textit{str} / \mathcal{L}_\textit{Ks}$ (equation~\ref{masstolight}).  Then, $\mathcal{M}_\textit{dyn} / \mathcal{M}_\textit{bar}$ follows from a choice for $\mathcal{M}_\textit{gas} / \mathcal{M}_\textit{str}$ (for dSphs, from equation~\ref{mgas}).

The semi-major axis of the isophote encompassing half of the light from the stars, normally referred to as the effective radius $r_\textit{eff}$, is tightly correlated with the radius $r_{1/2}$.  To within 
2\% \citep{wol10a},  
\begin{equation}
\label{reff_to_rhalf}
r_\textit{eff} / r_{1/2} \cong 3/4
\end{equation}  
Also, for a sech profile with scale length $r_{0}$,
\begin{equation}
\label{reff_to_r0}
r_\textit{eff} / r_{0} = 1.80584
\end{equation}  
Then, the dynamical mass to associate with a sech profile is given by
\begin{equation}
\label{mdyn}
\mathcal{M}_\textit{dyn} \cong 6 G^{-1} \sigma_\textit{los}^2 r_{0} {\left[  \frac{r_\textit{eff}}{r_{0}} \right]} {\left[ \frac{r_{1/2}}{r_\textit{eff}} \right]}
\end{equation}  
Results for dSphs and dEs are given in Tables~\ref{tab_comdsph} and \ref{tab_comde}, respectively.  Values for dSphs range from 
$2 \times 10^6$  
to 
$2 \times 10^9 \, \rm \mathcal{M}_\odot$.   
{\color{black} Representative uncertainties listed in the table notes were computed from the errors in the angular scale lengths, distances, and velocity dispersions.  The uncertainty in the ratio of the dynamical mass to the stellar mass is given by
\begin{align}
\left[ \delta \log \mathcal{M}_\textit{dyn} / \mathcal{M}_\textit{str} \right]^2 &= 
\left[ \delta \log d \right]^2 + \left[ \delta \log r_0 \right]^2 + \left[ \delta \log q \right]^2  \nonumber \\
& +  \left[ 0.4 \, \delta \mu_0 \right]^2 + \left[ 2 \, \delta \log W_{20} \right]^2 \end{align}
}

\begin{figure}  
\includegraphics[keepaspectratio=true,width=8.5cm]{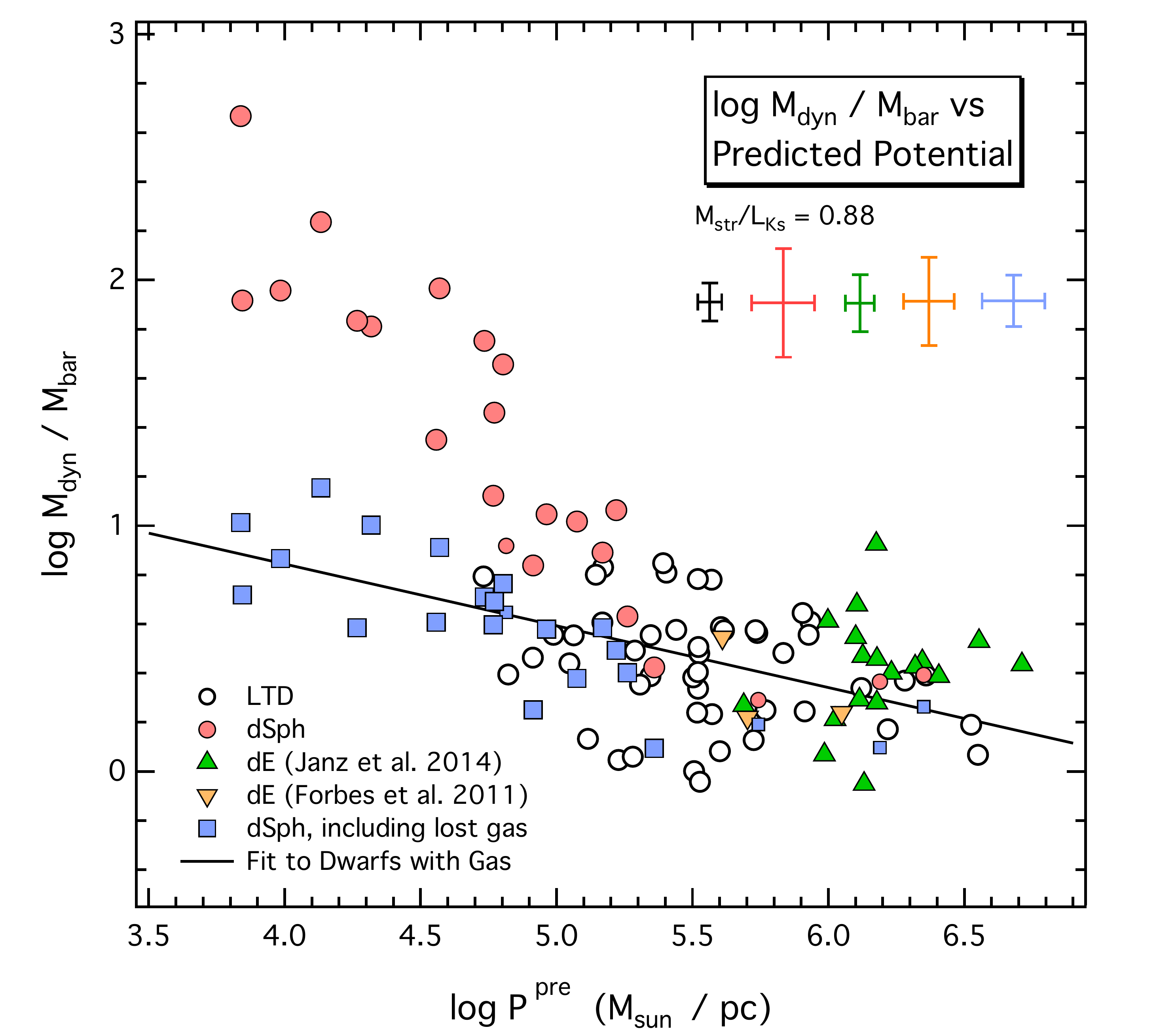}
\caption{\label{fig_mdyn_to_mbar_vs_potpre} 
The ratio of the dynamical mass to the baryonic mass as a function of the predicted potential.  Late-type dwarfs (LTDs) are shown as open black circles.  Solid pink circles, solid green triangles, and solid yellow inverted triangles mark where dwarf spheroidal galaxies (dSphs) and dwarf elliptical galaxies (dEs) lie when the baryonic mass is equated with the stellar mass.  Blue squares depict the positions of the dSphs after accounting for lost gas.  The black line is a combined fit to LTDs and dSphs with gas.    Ratios of dynamical to stellar masses can be converted to mass-to-light ratios in $K_s$ by multiplying by 0.88.  Average observational uncertainties are indicated by the error crosses.
}
\end{figure}

Figure~\ref{fig_mdyn_to_mbar_vs_potpre} shows, as a function of the predicted potential, how $\mathcal{M}_\textit{dyn} / \mathcal{M}_\textit{bar}$ for dSphs depends upon the choice for $\mathcal{M}_\textit{gas} / \mathcal{M}_\textit{str}$.  Results are compared with estimates for LTDs, for which $\mathcal{M}_\textit{gas} / \mathcal{M}_\textit{str}$ has been gauged from HI fluxes.
If it is approximated that dSphs did not lose any gas (pink circles), i.e., that stellar masses today are close to the baryonic masses of the precursors, then $\mathcal{M}_\textit{dyn} / \mathcal{M}_\textit{bar}$ rises rapidly as 
$\log P^\textit{\,pre}$ 
drops below 5.0,  
approaching 
500  
for the least massive systems.
Accounting for gas (blue squares) tempers the rise to match that observed for LTDs (open circles), because the less massive systems had proportionately more baryons in the past than the more massive ones.  Thus, the extremes exhibited by the least luminous dSphs should be regarded as an artifact of an incomplete census of baryons.

Combining the data for LTDs with the gas-corrected results for the dSphs with 
$\log P^\textit{\,pre}$ below 5.5, a least squares fit gives
\begin{align}
\label{mdyn_to_mbar}
\log \mathcal{M}_\textit{dyn} / \mathcal{M}_\textit{bar} &= (0.468 \pm 0.010) \nonumber \\
& + (-0.251 \pm 0.020) \left[ \log P^\textit{\,pre} - 5.5 \right]
\end{align}  
with an rms deviation of
$0.22 \, \rm dex$.  
The fit will be used in \S\ref{pcrit} to estimate escape velocities.

\subsection{
How Gas Was Lost by dSphs
}
\label{pcrit}

Figure~\ref{fig_potobs_vs_potpre} shows that there is a critical value of the predicted potential, $P^\textit{\,pre}_\textit{crit}$, below which ETDs break from the Potential Plane.  The corresponding escape velocity offers the means to discriminate why.   The escape velocity $v_\textit{esc}$ at $r_{1/2}$ is given by
\begin{align}
\label{vescape}
v_\textit{esc} &= \left[ 2 G \mathcal{M}_{1/2} / r_{1/2} \right]^{1/2} \\
&= \left[ G \mathcal{M}_\textit{dyn} / r_{1/2} \right]^{1/2}
\end{align}  
{\color{black} If the mass within radius $r$ grows linearly with $r$, as observed for rotating galaxies, then the escape velocity computed from equation~\ref{vescape} should be applicable over a wide range of radii.}
From the definition of the potential (equation~\ref{potential}), $P^\textit{\,pre}_\textit{crit}$ sets $\mathcal{M}_\textit{dyn} / r_{1/2}$ as follows:
\begin{equation}
{\frac{\mathcal{M}_\textit{dyn}}{r_{1/2}}} = 
P^\textit{\,pre}_\textit{crit}
{\left[ \frac{\mathcal{M}_\textit{dyn}}{\mathcal{M}_\textit{bar}} \right]}
{\left[ \frac{r_{0}}{r_\textit{eff}} \right]}
{\left[ \frac{r_\textit{eff}}{r_{1/2}} \right]}
\end{equation}  
The value of $\mathcal{M}_\textit{dyn} / \mathcal{M}_\textit{bar}$ can be evaluated from $P^\textit{\,pre}_\textit{crit}$ using equation~\ref{mdyn_to_mbar}, and $r_\textit{eff} / r_{1/2}$ and $r_{0} / r_\textit{eff}$ are defined by equations \ref{reff_to_rhalf} and \ref{reff_to_r0}, respectively.

In Figure~\ref{fig_potobs_vs_potpre}, the trend for dSphs intersects the Potential Plane at
$\log P^\textit{\,pre}_\textit{crit} = 5.74 \pm 0.19$   
($M_\textit{Ks} \cong -17.5$),  
where the units of $P^\textit{\,pre}_\textit{crit}$
are $\mathcal{M}_\odot \, \rm pc^{-1}$.  The corresponding value of
$\log \mathcal{M}_\textit{dyn} / \mathcal{M}_\textit{bar}$ is  
$0.41 \pm 0.06$.   
Thus, the escape velocity signified by $P^\textit{\,pre}_\textit{crit}$ is 
$50 \pm 8 \, \rm km \, s^{-1}$.   

\citet{kha53a} and \citet{pac53a} showed that a gas for which the ratio of specific heats is $\gamma$ expands into a vacuum at $2/(\gamma - 1)$ times the speed of sound in the gas.  For a monatomic gas, then, the terminal velocity should be 3 times the speed of sound (see also \citealp{che85a,mur05a,opp06a}).
Therefore, the escape velocity corresponds to that of a gas at a temperature of
{\color{black} $13,100 \pm 4,400 \, \rm K$}.   


The temperature is characteristic of a low-metallicity HII region, and it suggests that the ultraviolet radiation field stemming from a starburst is behind the evacuation of gas from dSphs with a potential shallower than $P^\textit{\,pre}_\textit{crit}$.  The gas would have been lost quickly.  For the galaxies studied here, the time required for {\color{black} uniformly expanding} gas with a temperature of 
$13,000 \, \rm K$  
to traverse 
$r_{1/2}$  
ranges from 
10 to 47 million years.  

\subsection{
dSph Starburst Masses
}
\label{starburst_mass}

To evacuate gas via photoionization, the mass of the instigating starburst would have had to have been great enough to ionize the bulk of the gas {\color{black} left} but not exceed the mass of stars seen today.  An idea of the minimum mass of the burst, $\mathcal{M}_\textit{burst}$, can be judged from ionization equilibrium, because the luminosity of photons with energies sufficient to ionize hydrogen is tied to the number of stars that have spectral type O.  In equilibrium, the luminosity by number of photons energetic enough to ionize neutral hydrogen, $Q(\hbox{H}^0)$, is related to the recombination rate by
\begin{equation} \label{qh0}
Q(\hbox{H}^0) \ge \alpha_\textit{B} \, n_e X \mathcal{M}_\textit{gas} / m_p
\end{equation}  
where $\alpha_B$ is the Case B recombination coefficient, $n_e$ is the number density of electrons in the ionized gas, $X$ is the hydrogen mass fraction, $\mathcal{M}_\textit{gas}$ is the mass of gas in the galaxy (equation~\ref{mgas}),  and $m_p$ is the mass of the proton.  The inequality arises from the fact that there is a finite amount of gas to ionize.  Case B is applicable to a galaxy in which $Q(\hbox{H}^0)$ is  just sufficient to ionize all of the available gas.  

An estimate for $n_e$ can be judged from the scale length of the light distribution.  Define $n_H$ to be the local number density of hydrogen atoms and ions.  If a fraction $f$ of the gas were confined to a volume with a radius equal to $r_{1/2}$, then the mean density $\langle n_H \rangle_{1/2}$ over that volume would be
\begin{equation} \label{hdensity}
{\langle n_H \rangle}_{1/2} = (f X \mathcal{M}_\textit{gas} / m_p) / (4 \pi r_{1/2}^3 / 3 )
\end{equation}  
Everywhere, ionization would be nearly complete, so $n_e \cong n_H$.  However, because the interstellar medium is clumpy, 
$n_e > \langle n_H \rangle_{1/2}$.  
Fortunately, as can be seen from equation~\ref{qh0}, adoption of 
$\langle n_H \rangle_{1/2}$ as a lower limit to $n_e$ still guarantees a lower limit to $Q(\hbox{H}^0)$ and, in turn, a lower limit for the mass of the starburst.

For each dSph, the minimum value of $Q(\hbox{H}^0)$ required to ionize all pre-existing gas was computed from equations \ref{qh0} and \ref{hdensity} using 
$X = 0.738$  
(appropriate for metallicity 
$Z = 0.002$),  
$\alpha_\textit{B} = 2.04 \times 10^{-13} \, \rm cm^3 \, s^{-1}$  
(appropriate for a temperature of 
$13,100 \ \rm K$:  
\citealt{peq91a}), and 
$f = 1/2$.   
Estimates for ${\langle n_H \rangle}_{1/2}$ range from 
0.05  
to
$2.8 \, \rm \hbox{cm}^{-3}$, and   
results for $\log Q(\hbox{H}^0)$ span
49.5 to 51.8  
($\, \rm s^{-1}$).
Starburst99 v7.0.1  
was employed to scale $\log Q(\hbox{H}^0)$ to $\mathcal{M}_\textit{burst}$
\citep{lei99a, vaz05a, lei10a, lei14a}.
For a zero-age starburst of mass 
$1 \times 10^6 \, \rm \mathcal{M}_\odot$  
with 
$Z = 0.002$  
and with stellar masses distributed according to a Salpeter IMF, the predicted value of $\log Q(\hbox{H}^0)$ is 
52.444 (2,040 O stars).  

\begin{figure}  
\includegraphics[keepaspectratio=true,width=8.5cm]{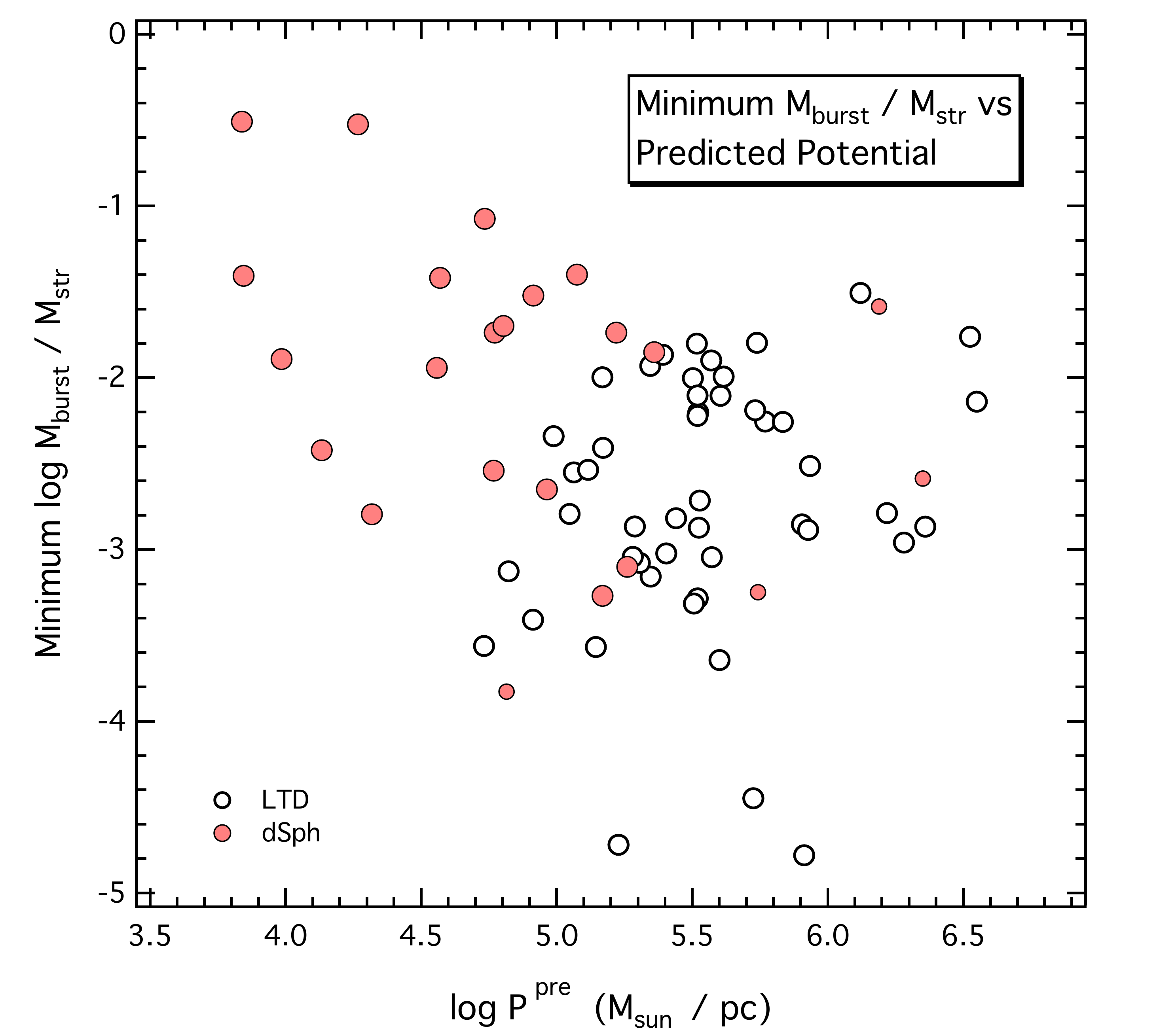}
\caption{\label{fig_mburst_to_mstr_vs_potpre} 
As a proportion of the present-day stellar mass, the minimum starburst mass required to photoionize the gas of a dwarf is plotted against the predicted potential.  Symbols are as in Figure~\ref{fig_potobs_vs_potpre}.  Starburst masses are well below stellar masses, so it is plausible that photoionization could have {\color{black} instigated} the evacuation of gas from dwarf spheroidal galaxies.
}
\end{figure}

Computed lower limits for $\mathcal{M}_\textit{burst}/ \mathcal{M}_\textit{str}$ are plotted against the predicted potential in 
Figure~\ref{fig_mburst_to_mstr_vs_potpre}.
Typically, the estimates for the starburst masses are only about 
1\%   
of the stellar masses of dSphs today.  Thus, it is plausible that photoionization {\color{black} initiated} the loss of gas from dSphs.  The results for 
Canes Venatici II and Coma Berenices   
are 
30\%  
of their stellar masses, suggesting that the star formation events required to evacuate their gas could have been responsible for the formation of a large fraction of their stars.

\subsection{
The Minimum Mass of a Galaxy
}
\label{minimum_mass}

Celestial bodies that are defined to be galaxies contain stars.  {\color{black} They contain stars because at some time in the past star formation ignited in a body of gas that had accumulated in a dark matter halo.  It is fair to ask, then, ``Are there galaxies that failed?".  More pointedly, ``Could there {\color{black} have been} a threshold for the baryonic mass below which stars {\color{black} could not} form?''.

Figure~\ref{fig_mgas_to_mstr_vs_potpre} demonstrates that baryons in dSphs were not converted to stars with 100\% efficiency, and Figure~\ref{fig_mgas_vs_mstr} shows that
the amount of gas remaining just prior to gas loss correlated strongly with the stellar mass.
\begin{figure}
\includegraphics[keepaspectratio=true,width=8.5cm]{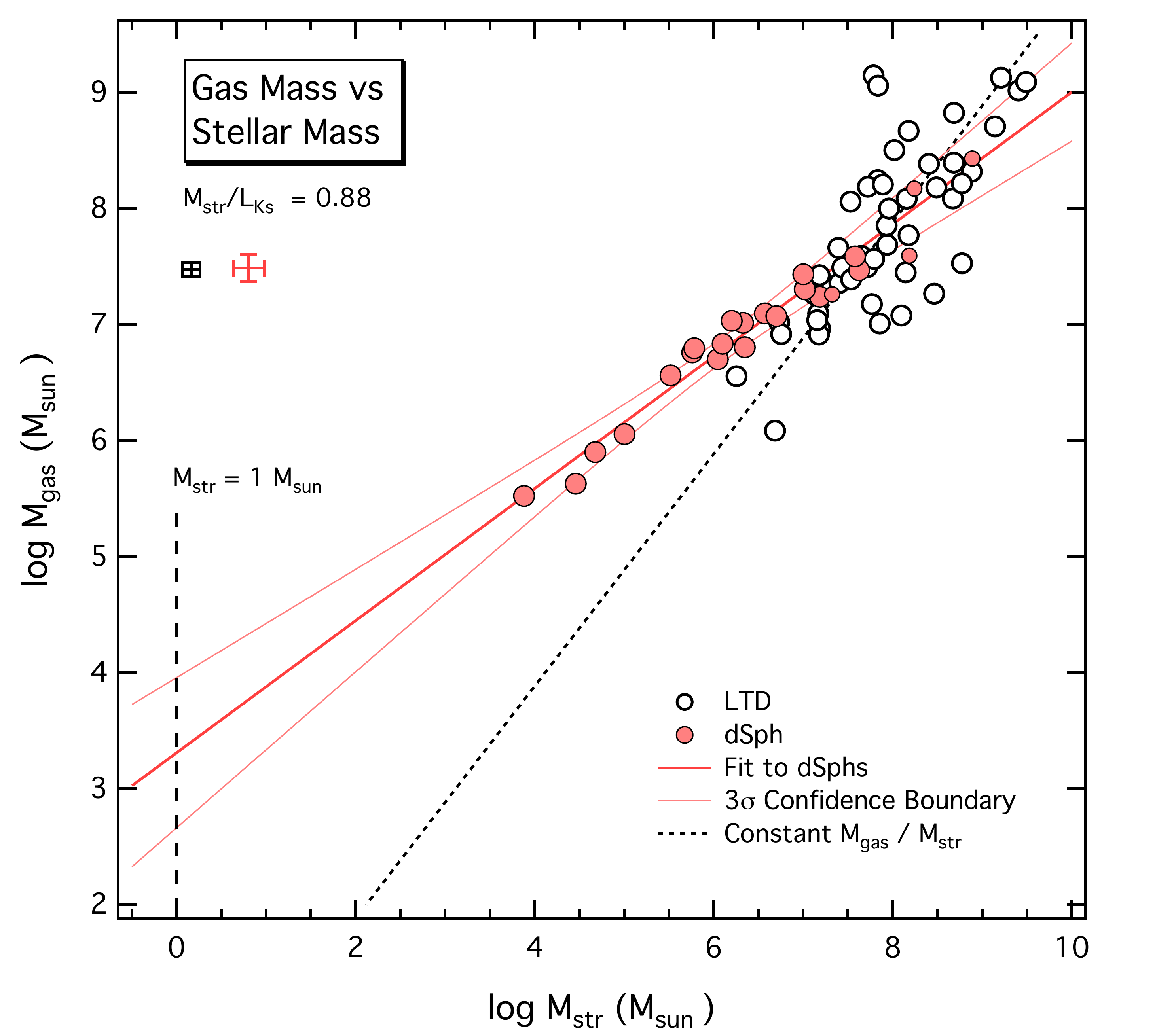}
\caption{\label{fig_mgas_vs_mstr}
The gas mass versus the stellar mass.  Symbols are as in Figure~\ref{fig_potobs_vs_potpre}.  Gas masses for late-type dwarfs (LTDs) are measured directly from $21 \, \rm cm$ fluxes, while those for dwarf spheroidal galaxies (dSphs) are inferred from offsets from the potential plane.  A least squares fit to the dSphs is shown as a thick pink line, and $3\sigma$ confidence boundaries are marked by thin pink curves.  {\color{black} The dotted black line depicts a least squares fit to the LTDs with $\mathcal{M}_\textit{gas} / \mathcal{M}_\textit{str}$ constrained to be constant.}
The vertical dashed line marks a stellar mass of $1 \, \rm \mathcal{M}_\odot$.
Average observational uncertainties are indicated by the error crosses.  {\color{black} The positioning of the dSphs with respect to the dotted black line shows that the efficiency of star formation declines with the baryonic mass.}
}
\end{figure}
Both Figures~\ref{fig_mgas_to_mstr_vs_potpre} and \ref{fig_mgas_vs_mstr} reveal that} the efficiency of conversion {\color{black} diminished with shallower potentials, a result that extends downward  by
three  
orders of magnitude in stellar mass the trend established from reconstructed star formation histories for galaxies  \citep{beh13a}.}  This opens up the possibility that there was {\color{black}indeed} a minimum {\color{black} baryonic} mass below which stars did not form.  Even if the efficiency were invariant (but less than 100\%), there would have been a threshold in the baryonic mass below which the stellar mass could be considered negligible.   With knowledge of how the gas masses inferred for the progenitors of dSphs correlate with stellar masses, it is possible in principle to evaluate what the minimum baryonic mass of a galaxy might be.  Gross extrapolations are required, but it is interesting to see what present data confer.

A linear least squares fit to the dSphs {\color{black} in Figure~\ref{fig_mgas_vs_mstr}} gives
{\color{black}
\begin{align}   \label{mgas2}
\log \mathcal{M}_\textit{gas} &= (3.312 \pm 0.190) \nonumber \\
& + (0.569 \pm 0.031) \log \mathcal{M}_\textit{str}
\end{align}  
where masses are in solar units.  The rms deviation of points about the fit} is
$0.12 \, \rm dex$.  
The minimum baryonic mass of a galaxy might be considered to be the baryonic mass at which the stellar mass is 
$1 \, \rm \mathcal{M}_\odot$.   
In Figure~\ref{fig_mgas_vs_mstr}, the fit and $3 \sigma$ confidence boundaries are shown extrapolated to that value.  Of course, underlying the extrapolation is the assumption that the relationship between $P^\textit{\,obs}$ and $P^\textit{\,pre}$ observed for LTDs can be extrapolated not only into the realm of the faintest dSphs studied in this paper, which was the premise behind evaluating their gas masses in the first place, but even beyond to the stellar mass minimum.  The corresponding gas mass, which can be regarded as the minimum baryonic mass of a galaxy, lies between 
500  
and 
$10,000 \, \rm \mathcal{M}_\odot$,  
{\color{black} the preferred value being 
$2,000 \, \rm \mathcal{M}_\odot$.}  

{\color{black}
Determining the halo mass associated with the minimum baryonic mass is more problematic, because surveys and simulations used to relate stellar masses to halo masses through density matching aren't deep enough or have insufficient resolution to include galaxies like those that are the focus of this paper (see \citealt{beh13a}).  Dynamical masses estimated from equation~\ref{mdyn} gauge total masses within $r_{1/2}$, but it is difficult to extrapolate these results to determine halo masses because neither the mass distributions nor their edges are known.  

To constrain at least the dynamical mass of the halo housing the minimum baryonic mass, dynamical masses for dwarfs are plotted against stellar masses in Figure~\ref{fig_mdyn_vs_mstr}.
\begin{figure}
\includegraphics[keepaspectratio=true,width=8.5cm]{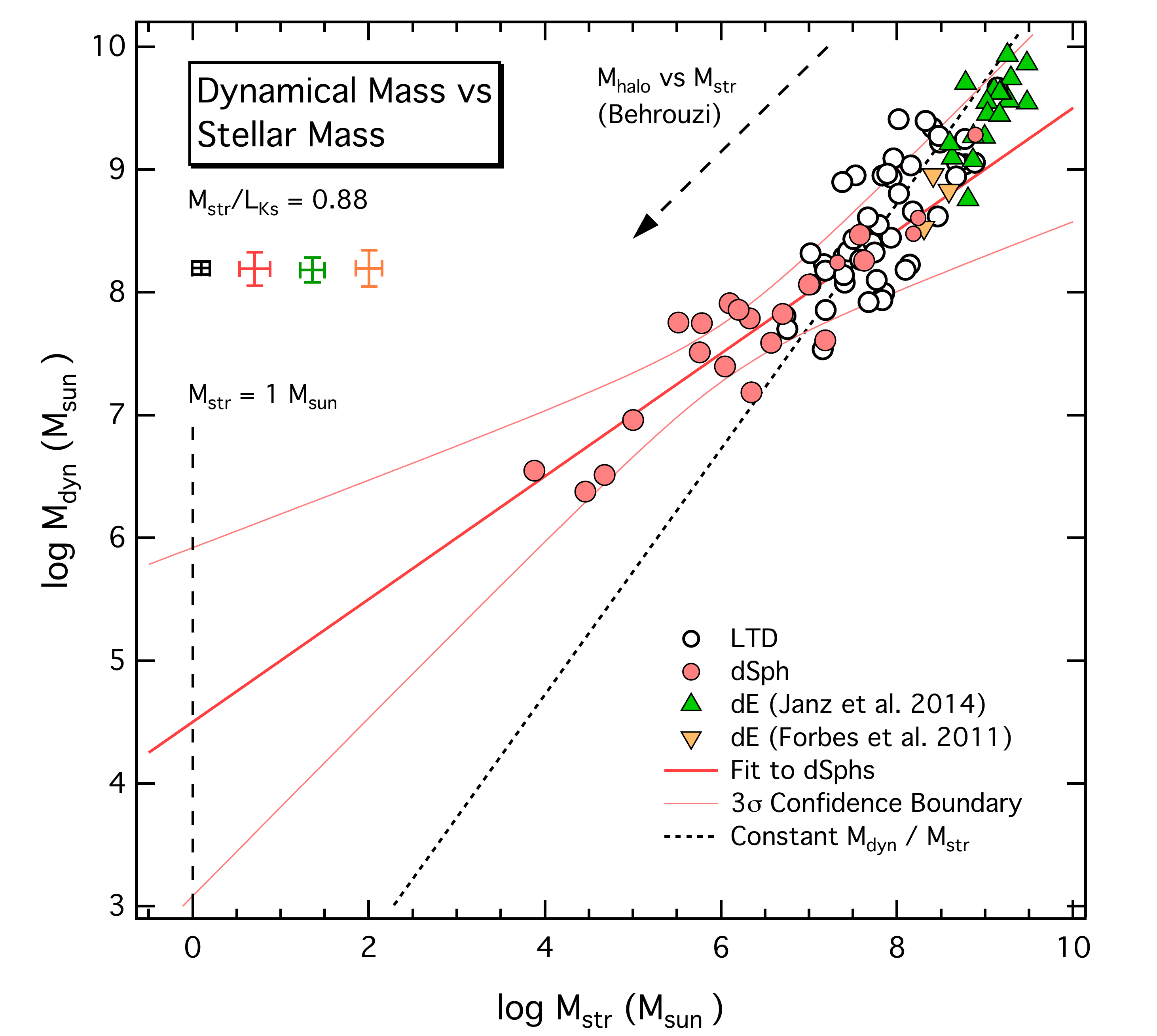}
\caption{\label{fig_mdyn_vs_mstr}
{\color{black}
The dynamical mass versus the stellar mass.  Symbols are as in Figure~\ref{fig_potobs_vs_potpre}.   A least squares fit to the dwarf spheroidal galaxies (dSphs) is shown as a thick pink line, and $3\sigma$ confidence boundaries are marked by thin pink curves.  The dotted black line depicts a least squares fit to the late-type dwarfs (LTDs) with $\mathcal{M}_\textit{dyn} / \mathcal{M}_\textit{str}$ constrained to be constant.  The vertical dashed line marks a stellar mass of $1 \, \rm \mathcal{M}_\odot$.    {\color{black} The dashed arrow shows the extrapolated trend for halo masses that has been derived from density matching \citep{beh13a}; the slope is comparable to that observed for dSph dynamical masses.}
Average observational uncertainties are indicated by the error crosses.
The positioning of the dSphs with respect to the dotted black line shows that galaxies with less massive haloes form proportionately fewer stars.
}
}
\end{figure}
A linear least squares fit to the dSphs gives
\begin{align}
\log \mathcal{M}_\textit{dyn} &= (4.502 \pm 0.417) \nonumber \\
& + (0.500 \pm 0.068) \log \mathcal{M}_\textit{str}
\end{align}  
where masses are in solar units.
The rms deviation of points about the line is
$0.27 \, \rm dex$.  
Extrapolating the $3 \sigma$ confidence boundaries,
the dynamical mass of a galaxy with a stellar mass of only $1 \, \rm \mathcal{M}_\odot$ lies between
$10^3$  
and 
$10^6 \, \rm \mathcal{M}_\odot$.   

Figure~\ref{fig_mdyn_vs_mstr} also displays with a dotted black line the trend of $\mathcal{M}_\textit{dyn}$ with $\mathcal{M}_\textit{str}$ that would be expected if $\mathcal{M}_\textit{dyn} / \mathcal{M}_\textit{str}$ were a constant.  Stellar masses fall farther and farther to the left of this line as dynamical masses decrease, showing that haloes with lower {\color{black} dynamical} masses produce stars less efficiently. {\color{black} In fact, there is reason to believe dSphs with lower {\it halo} masses produce stars less efficiently, because the correlation of halo {\color{black} masses} with stellar masses gleaned for much larger galaxies \citep{beh13a}, the extrapolation of which is displayed by the dashed arrow in Figure~\ref{fig_mdyn_vs_mstr}, has a slope similar to that for the trend of $\mathcal{M}_\textit{dyn}$ with $\mathcal{M}_\textit{str}$.}
 
Given the extrapolations required, results {\color{black} for minimum masses} are  at best tentative and at worst meaningless.  Perhaps with more confidence, equation~\ref{mgas2} provides a means to estimate the cumulative effects of gas loss from an ensemble of dSphs described by a luminosity function.

\section{
Discussion
}
\label{discussion}

{\color{black} 
Supernovae have been flagged often as being responsible for evacuating gas from ETDs, either individually or through the cumulative effects of many.  However, models by \citet{fer00a} showed that complete loss of gas (blowaway) only occurs for galaxies with a total mass at or below $5 \times 10^6 \, \rm \mathcal{M}_\odot$. As was recognized by {\color{black} those authors}, the hitch is that most dSphs, including all of those studied in this paper, have higher dynamical masses.  Recent models of the Ursa Minor dSph confirm that supernovae do not completely remove gas \citep{cap17a}.  Although supernovae must have played a role in driving some gas out of Ursa Minor, the mass of gas {\color{black} predicted to remain after} 3 Gyr exceeded the observational upper limit by two orders of magnitude.  Notably, in the first 600~Myr, supernovae-driven winds pushed more than half of the original gas out to large radii, where it would have been less strongly bound. \citet{cap17a} concluded that an additional mechanism was necessary to completely remove the gas.

\citet{dek03a} examined the role of feedback in determining scaling relations for low surface brightness galaxies and dwarfs (i.e., galaxies with {\color{black} stellar} masses down to $6 \times 10^5 \, \rm \mathcal{M}_\odot$).  By assuming that the energy input from supernovae was proportional to the mass of stars and then equating it to the binding energy of the gas, they were able to reproduce well the observed scaling relations.  They argued that the binding energy distinguishes ETDs from LTDs, with the former {\color{black} housed in} haloes with {\color{black} a mass, expressed as a circular velocity,} between 10 and $30 \, \rm km \, s^{-1}$ 
($4.3 < \log P^\textit{\,pre} < 5.6$)
{\color{black} and the latter residing in more massive haloes with a velocity above $30 \, \rm km \, s^{-1}$}.   
This is because the intergalactic UV radiation field established by the first stars and galaxies would have photo-evaporated gas in bodies with {\color{black}  a velocity} less than about $20 \, \rm km \, s^{-1}$ 
($\log P^\textit{\,pre} < 5.1$)  
and halted infall of gas into systems with {\color{black} a velocity} less than $30 \, \rm km \, s^{-1}$.  Consequently, ETDs would have had to form most of their stars prior to reionization, whereas LTDs could have retained gas to the present.
The problem with this scenario is that it cannot readily explain how there can be ETDs and LTDs today with overlapping values of $P^\textit{\,pre}$.  Photo-evaporation initiated internally by a starburst could have been the impetus for evacuation of gas from the precursors of larger ETDs.

\citet{wyi13a} developed a model for the rate density of star formation at high redshift incorporating a prescription for the suppression of star formation by supernovae feedback.  They found that individual starbursts in low-mass galaxies came to an end after only $2 \times 10^7$ years, a time short enough to suggest that subsequent star formation was quenched by massive stars. How is open to question.  Supernovae might have been involved, but photoionization {\color{black} probably played a role, too}.

\citet{spa97a} demonstrated the importance of a multi-phase interstellar medium to the evolution of dwarfs.  They claim that heating by supernovae created a hot phase with a temperature of $\sim 10^7 \, \rm K$ and a filling factor that reached close to unity by a redshift of $\sim 1$ (see also \citealt{cla02a}).  {\color{black} The gas was so hot that it would have been} able to escape.  {\color{black} However, the work presented here reveals that gas in dSphs could have begun to escape before a hot phase developed.  {\color{black} Nevertheless,} the loss of gas could have been assisted or even accelerated by supernovae, either directly by the pressure of ejecta or indirectly through additional heating.}

Tides and/or ram pressure stripping have been proposed as an alternative to outflows driven in one way or another by star formation.  Simulations by \citet{may01a} of the evolution of a small disk galaxy in an NFW halo \citep{nav96a} orbiting a massive host revealed that
tidal stirring could lead to structural and kinematic properties comparable to {\color{black} those of} dwarf spheroidals after just a few orbits (see also \citealt{arr14a}).  
{\color{black}
Ram pressure had to be invoked to completely evacuate gas, though \citep{may06a}, which worked for galaxies with a circular velocity below $30 \, \rm km \, s^{-1}$ 
($\log P^\textit{\,pre} < 5.6$). 
Interestingly, it is the ETDs in that velocity range that are observed to deviate from the Potential Plane.
To evacuate a galaxy with a circular velocity exceeding $30 \, \rm km \, s^{-1}$, 
}
a heating source appears to have been necessary to counteract inflow stimulated by tides and to maintain the extension of the gas.   \citet{may06a} propose that the heating could have been provided by the cosmic UV background.  
{\color{black} However, the issue may be moot.  ETDs this massive do not deviate significantly from the Potential Plane, suggesting that most of their gas was converted into stars.
}

Besides naturally leading to a relationship between morphology and environment, {\color{black} the combination of tidal stirring and ram pressure} admits the possibility of star formation over several gigayears (one or more orbits), which indeed has been observed for dwarfs in the Local Volume \citep{wei11a}.  In fact, \citet{wei11a} goes so far as to say that the transformation from gas-rich to gas-poor could not have been caused by stellar feedback alone because internal processes cannot explain the connection to environment.  

Figure~\ref{fig_mgas_to_mstr_vs_potpre} showed that for dSphs there was a strong correlation between the star formation efficiency and the baryonic mass.  It seems unlikely that the combination of tides and ram pressure could have led to such a relationship because orbital parameters vary from galaxy to galaxy.  {\color{black} Also, it is hard to see how a dSph could be a tidally stirred LTD given the structural similarities of the two kinds of galaxies today.} For {\color{black} these reasons}, it is preferable to {\color{black} surmise} that the absence of gas in dSphs stems from a star formation history different from that of LTDs.

The three most massive dSphs, all satellites of M31, as well as dEs, have {\color{black} circular} velocities close to or above $30 \, \rm km \, s^{-1}$.  All have potentials deeper than $P^\textit{\,pre}_\textit{crit}$ and congregate near the potential plane.  Their positioning is consistent with the theoretical evidence that their gas would have been difficult to strip by any means.  As suggested in \S\ref{placement} (see Figure~\ref{fig_mstr_vs_potpre}), such galaxies appear to have converted a much higher fraction of gas into stars than objects with a potential shallower than $P^\textit{\,pre}_\textit{crit}$.
}



{\color{black}
If gas loss from dSphs stemmed from photoionization by a starburst, it remains to be explained why there is such a tight relationship between $\mathcal{M}_\textit{str}/ \mathcal{M}_\textit{bar}$ and the potential at the time of gas loss (Figures~\ref{fig_potobs_vs_potpre} and \ref{fig_mgas_to_mstr_vs_potpre}).  It is as if the progenitor of a dSph  {\color{black} ``knew''}what fraction of its baryons it had to convert into stars before losing its gas.  In fact, relating gas loss to photionization offers some insights.}
For gas to be ejected, most of it had to be ionized, so the mass of the instigating starburst had to be connected at some level to the total mass of gas expelled.  Since most baryons were gaseous, then
$\mathcal{M}_\textit{str}/ \mathcal{M}_\textit{bar} \propto Q(\hbox{H}^0) / \mathcal{M}_\textit{gas}$.  
Based upon equation~\ref{qh0}, the trend in $\mathcal{M}_\textit{str}/ \mathcal{M}_\textit{bar}$ could be explained if $n_e$ were higher in dwarfs with deeper potentials.  {\color{black} However,} no correlation of ${\langle n_H \rangle}_{1/2}$ with $P^\textit{\,pre}$ is apparent for dSphs.  {\color{black} Nevertheless,} the trend still could have developed if gas in more massive dwarfs tended to be clumpier {\color{black} (i.e., if the filling factor were lower), {\color{black} since $n_e / {\langle n_H \rangle}_{1/2}$ would be raised}.}



{\color{black}
Those LTDs with $P^\textit{\,pre}$ below $P^\textit{\,pre}_\textit{crit}$ risk losing their gas quickly if they form enough stars at once to photoionize most of their gas.  It is because there is no long-lived intermediate stage that LTDs and ETDs with potentials shallower than $P^\textit{\,pre}_\textit{crit}$ are so cleanly separated.
So long as LTDs form stars gradually, they should be able to remain LTDs until their gas supply is exhausted.  Indeed, \citet{chi15a} observed that the bulk of the low-ionized gas in dwarf irregular galaxies is not moving fast enough to escape. In fact, \citet{tas08a} was able to reproduce scaling relations for LTDs without invoking gas loss at all, but rather through a combination of increasingly inefficient star formation towards lower masses and efficient mixing of metals into haloes.
}

\citet{kor16a} found that baryons are dynamically unimportant in galaxies in which the outer circular velocity in the dark matter halo is below 
$42 \pm 4 \, \rm km \, s^{-1}$.   
The escape velocity at $P^\textit{\,pre}_\textit{crit}$ 
corresponds to a circular velocity of 
$35 \pm 6 \, \rm km \, s^{-1}$.  
Although one may question how well the estimated escape velocity  is gauging the circular velocity in the outer halo, the agreement {\color{black} may arise because dSphs with smaller circular velocities have expelled most of the baryons that they once had}.

{\color{black}
The minimum mass of a galaxy has a bearing on the missing satellite problem.
A cosmological simulation of the Local Group by \cite{saw16a} showed that reionization would have prevented stars from forming in haloes with masses below about $10^{7.5} \, \rm \mathcal{M}_\odot$ and would have led to the loss of gas by photo-evaporation.  This is because molecular hydrogen would have been dissociated and, in the absence of metals, cooling would have been curtailed \citep{dek03a}.  Galaxies that survived to the present were interpreted to be those with the most concentrated distributions of baryons.  

In \S\ref{minimum_mass}, the dynamical mass corresponding to the estimated minimum baryonic mass of a galaxy was estimated to be at most 
$\sim 10^6 \, \rm \mathcal{M}_\odot$.   
The value rises by a factor of 
six  
if conditions were such that the mass of stars formed in the first star formation event was $10^3 \, \rm \mathcal{M}_\odot$ rather than $1 \, \rm \mathcal{M}_\odot$ (within a factor of 
seven  
of the stellar mass in the least luminous dSphs studied in this paper).  Also, the limit would be higher if halo masses are larger than inferred from $\mathcal{M}_\textit{dyn}$, {\color{black} as \citet{beh13a} indicate (see Figure~\ref{fig_mdyn_vs_mstr})}.  Thus, the observations suggest that the theoretically-predicted minimum {\color{black} halo mass} is not unreasonably high.}

\section{
Conclusions
}
\label{conclusions}

The positioning of dSphs and dEs (early-type dwarfs) with respect to the Potential Plane defined by dIs and BCDs (late-type dwarfs) has been examined.  With the mass-to-light ratio in $K_s$ fixed at the fitted value for late-type dwarfs, dEs and massive dSphs are clustered very near the Plane, {\color{black} suggesting that they evolved from late-type dwarfs that managed to transform the bulk of their gas into stars.}  There is a critical value of the potential at which dSphs with shallower potentials fall away {\color{black} from the Plane}.  The deviations are attributed to an incomplete census of baryons, which can be interpreted to be a consequence of gas loss.  As such, the magnitude of the offset of a dSph from the plane gauges the amount of gas that the galaxy once had.  The observations show that dSphs of lower mass must have {\color{black} converted baryons into stars with lower efficiency and} lost proportionately more gas.  Ratios of dynamical to baryonic masses fall in line with observations of late-type dwarfs once the missing gas is accommodated.  

The escape velocity defined by the critical potential is 
$50 \pm 8 \, \rm km \, s^{-1}$,  
which implies that the temperature of escaping gas would have been
$13,000 \pm 4,000 \, \rm K$.  
The temperature is typical of metal-poor HII regions, suggesting that {\color{black} near-complete} photoionization {\color{black} of neutral gas} following a starburst {\color{black}played a role in driving} gas loss.  This is plausible because, for every dSph sampled,
the minimum mass of a starburst required to ionize the entirety of the gas is smaller than the observed mass of stars today.   {\color{black} If unimpeded,} evacuation of gas following such a starburst would have occurred over a few tens of millions of years {\color{black} (less if aided by stellar winds or supernovae).   This can explain} why it is that dwarfs in a state of transition are rare and consequently why it is that dwarfs with a potential shallower than the critical value form two distinct sequences.    The escape velocity implies a circular velocity of 
$35 \pm 6 \, \rm km \, s^{-1}$,  
close to the halo value below which baryons are believed to be dynamically irrelevant.

Because smaller dSphs were less efficient at converting gas into stars, there should be a minimum baryonic mass for {\color{black} galaxies, i.e., below which no stars ever formed}.  Grossly extrapolating the trend of the inferred gas mass with stellar mass down to a stellar mass of $1 \, \rm \mathcal{M}_\odot$, the minimum baryonic mass comes out to be between
500 
and 
$10,000 \, \rm \mathcal{M}_\odot$.  
{\color{black} The corresponding dynamical mass is less than $10^6 \, \rm \mathcal{M}_\odot$.}   {\color{black} Even though there may have been smaller bodies of baryonic matter in the early Universe, it is possible that only their dark matter haloes remain today if the gas was photo-evaporated during reionization.}

\section*{
Acknowledgements
}
MLM conducted part of this research while a Visiting (York) Scholar at Massey College, and is grateful to the College for the use of its facilities.   The Natural Sciences and Engineering Research Council of Canada is thanked for its continued financing.  Thanks are conveyed to B. Schuman for his advice about titling.  MLM is particularly grateful to B. Nathoo, F. Shariff, H. Tanzer, and J. Vecchiarelli for their dedication to preserving his health long enough to complete this paper.  This research has made use of the databases GOLDMine and HyperLEDA, as well as the software Igor Pro and Starburst99.



\bibliographystyle{mnras}
\bibliography{mccall_references}

\begin{thebibliography}{}
\makeatletter
\relax
\def\mn@urlcharsother{\let\do\@makeother \do\$\do\&\do\#\do\^\do\_\do\%\do\~}
\def\mn@doi{\begingroup\mn@urlcharsother \@ifnextchar [ {\mn@doi@}
  {\mn@doi@[]}}
\def\mn@doi@[#1]#2{\def\@tempa{#1}\ifx\@tempa\@empty \href
  {http://dx.doi.org/#2} {doi:#2}\else \href {http://dx.doi.org/#2} {#1}\fi
  \endgroup}
\def\mn@eprint#1#2{\mn@eprint@#1:#2::\@nil}
\def\mn@eprint@arXiv#1{\href {http://arxiv.org/abs/#1} {{\tt arXiv:#1}}}
\def\mn@eprint@dblp#1{\href {http://dblp.uni-trier.de/rec/bibtex/#1.xml}
  {dblp:#1}}
\def\mn@eprint@#1:#2:#3:#4\@nil{\def\@tempa {#1}\def\@tempb {#2}\def\@tempc
  {#3}\ifx \@tempc \@empty \let \@tempc \@tempb \let \@tempb \@tempa \fi \ifx
  \@tempb \@empty \def\@tempb {arXiv}\fi \@ifundefined
  {mn@eprint@\@tempb}{\@tempb:\@tempc}{\expandafter \expandafter \csname
  mn@eprint@\@tempb\endcsname \expandafter{\@tempc}}}

\bibitem[\protect\citeauthoryear{{Amorisco}, {Evans}  \& {van de
  Ven}}{{Amorisco} et~al.}{2014}]{amo14a}
{Amorisco} N.~C.,  {Evans} N.~W.,   {van de Ven} G.,  2014, \mn@doi [\nat]
  {10.1038/nature12995}, \href
  {http://adsabs.harvard.edu/abs/2014Natur.507..335A} {507, 335}

\bibitem[\protect\citeauthoryear{{Arraki}, {Klypin}, {More}  \&
  {Trujillo-Gomez}}{{Arraki} et~al.}{2014}]{arr14a}
{Arraki} K.~S.,  {Klypin} A.,  {More} S.,   {Trujillo-Gomez} S.,  2014, \mn@doi
  [\mnras] {10.1093/mnras/stt2279}, \href
  {http://adsabs.harvard.edu/abs/2014MNRAS.438.1466A} {438, 1466}

\bibitem[\protect\citeauthoryear{{Bailin} \& {Ford}}{{Bailin} \&
  {Ford}}{2007}]{bai07a}
{Bailin} J.,  {Ford} A.,  2007, \mn@doi [\mnras]
  {10.1111/j.1745-3933.2006.00271.x}, \href
  {http://adsabs.harvard.edu/abs/2007MNRAS.375L..41B} {375, L41}

\bibitem[\protect\citeauthoryear{{Behroozi}, {Wechsler}  \&
  {Conroy}}{{Behroozi} et~al.}{2013}]{beh13a}
{Behroozi} P.~S.,  {Wechsler} R.~H.,   {Conroy} C.,  2013, \mn@doi [\apjl]
  {10.1088/2041-8205/762/2/L31}, \href
  {http://adsabs.harvard.edu/abs/2013ApJ...762L..31B} {762, L31}

\bibitem[\protect\citeauthoryear{{Bellazzini}, {Ferraro}, {Origlia}, {Pancino},
  {Monaco}  \& {Oliva}}{{Bellazzini} et~al.}{2002}]{bel02a}
{Bellazzini} M.,  {Ferraro} F.~R.,  {Origlia} L.,  {Pancino} E.,  {Monaco} L.,
   {Oliva} E.,  2002, \mn@doi [\aj] {10.1086/344794}, \href
  {http://adsabs.harvard.edu/abs/2002AJ....124.3222B} {124, 3222}

\bibitem[\protect\citeauthoryear{{Bellazzini}, {Gennari}, {Ferraro}  \&
  {Sollima}}{{Bellazzini} et~al.}{2004}]{bel04a}
{Bellazzini} M.,  {Gennari} N.,  {Ferraro} F.~R.,   {Sollima} A.,  2004,
  \mn@doi [\mnras] {10.1111/j.1365-2966.2004.08226.x}, \href
  {http://adsabs.harvard.edu/abs/2004MNRAS.354..708B} {354, 708}

\bibitem[\protect\citeauthoryear{{Bellazzini}, {Gennari}  \&
  {Ferraro}}{{Bellazzini} et~al.}{2005}]{bel05a}
{Bellazzini} M.,  {Gennari} N.,   {Ferraro} F.~R.,  2005, \mn@doi [\mnras]
  {10.1111/j.1365-2966.2005.09027.x}, \href
  {http://adsabs.harvard.edu/abs/2005MNRAS.360..185B} {360, 185}

\bibitem[\protect\citeauthoryear{{Bernard} et~al.,}{{Bernard}
  et~al.}{2009}]{ber09a}
{Bernard} E.~J.,  et~al., 2009, \mn@doi [\apj] {10.1088/0004-637X/699/2/1742},
  \href {http://adsabs.harvard.edu/abs/2009ApJ...699.1742B} {699, 1742}

\bibitem[\protect\citeauthoryear{{Bonanos}, {Stanek}, {Szentgyorgyi},
  {Sasselov}  \& {Bakos}}{{Bonanos} et~al.}{2004}]{bon04a}
{Bonanos} A.~Z.,  {Stanek} K.~Z.,  {Szentgyorgyi} A.~H.,  {Sasselov} D.~D.,
  {Bakos} G.~{\'A}.,  2004, \mn@doi [\aj] {10.1086/381073}, \href
  {http://adsabs.harvard.edu/abs/2004AJ....127..861B} {127, 861}

\bibitem[\protect\citeauthoryear{{Butler} \&
  {Mart{\'{\i}}nez-Delgado}}{{Butler} \&
  {Mart{\'{\i}}nez-Delgado}}{2005}]{but05a}
{Butler} D.~J.,  {Mart{\'{\i}}nez-Delgado} D.,  2005, \mn@doi [\aj]
  {10.1086/429524}, \href {http://adsabs.harvard.edu/abs/2005AJ....129.2217B}
  {129, 2217}

\bibitem[\protect\citeauthoryear{{Cacciari} \& {Clementini}}{{Cacciari} \&
  {Clementini}}{2003}]{cac03a}
{Cacciari} C.,  {Clementini} G.,  2003, in {Alloin} D.,  {Gieren} W.,  eds,
  Lecture Notes in Physics, Berlin Springer Verlag Vol. 635, Stellar Candles
  for the Extragalactic Distance Scale. pp 105--122 (\mn@eprint {}
  {astro-ph/0301550}), \mn@doi{10.1007/978-3-540-39882-0_6}

\bibitem[\protect\citeauthoryear{{Cair{\'o}s}, {Caon}  \&
  {Weilbacher}}{{Cair{\'o}s} et~al.}{2015}]{cai15a}
{Cair{\'o}s} L.~M.,  {Caon} N.,   {Weilbacher} P.~M.,  2015, \mn@doi [\aap]
  {10.1051/0004-6361/201322518}, \href
  {http://adsabs.harvard.edu/abs/2015A%26A...577A..21C} {577, A21}

\bibitem[\protect\citeauthoryear{{Caproni}, {Amaral Lanfranchi}, {Campos Baio},
  {Kowal}  \& {Falceta-Gon{\c c}alves}}{{Caproni} et~al.}{2017}]{cap17a}
{Caproni} A.,  {Amaral Lanfranchi} G.,  {Campos Baio} G.~H.,  {Kowal} G.,
  {Falceta-Gon{\c c}alves} D.,  2017, \mn@doi [\apj]
  {10.3847/1538-4357/aa6002}, \href
  {http://adsabs.harvard.edu/abs/2017ApJ...838...99C} {838, 99}

\bibitem[\protect\citeauthoryear{{Carrera}, {Aparicio},
  {Mart{\'{\i}}nez-Delgado}  \& {Alonso-Garc{\'{\i}}a}}{{Carrera}
  et~al.}{2002}]{car02a}
{Carrera} R.,  {Aparicio} A.,  {Mart{\'{\i}}nez-Delgado} D.,
  {Alonso-Garc{\'{\i}}a} J.,  2002, \mn@doi [\aj] {10.1086/340702}, \href
  {http://adsabs.harvard.edu/abs/2002AJ....123.3199C} {123, 3199}

\bibitem[\protect\citeauthoryear{{Chevalier} \& {Clegg}}{{Chevalier} \&
  {Clegg}}{1985}]{che85a}
{Chevalier} R.~A.,  {Clegg} A.~W.,  1985, \mn@doi [\nat] {10.1038/317044a0},
  \href {http://adsabs.harvard.edu/abs/1985Natur.317...44C} {317, 44}

\bibitem[\protect\citeauthoryear{{Chisholm}, {Tremonti}, {Leitherer}, {Chen},
  {Wofford}  \& {Lundgren}}{{Chisholm} et~al.}{2015}]{chi15a}
{Chisholm} J.,  {Tremonti} C.~A.,  {Leitherer} C.,  {Chen} Y.,  {Wofford} A.,
  {Lundgren} B.,  2015, \mn@doi [\apj] {10.1088/0004-637X/811/2/149}, \href
  {http://adsabs.harvard.edu/abs/2015ApJ...811..149C} {811, 149}

\bibitem[\protect\citeauthoryear{{Clarke} \& {Oey}}{{Clarke} \&
  {Oey}}{2002}]{cla02a}
{Clarke} C.,  {Oey} M.~S.,  2002, \mn@doi [\mnras]
  {10.1046/j.1365-8711.2002.05976.x}, \href
  {http://adsabs.harvard.edu/abs/2002MNRAS.337.1299C} {337, 1299}

\bibitem[\protect\citeauthoryear{{Collins} et~al.,}{{Collins}
  et~al.}{2013}]{col13a}
{Collins} M.~L.~M.,  et~al., 2013, \mn@doi [\apj]
  {10.1088/0004-637X/768/2/172}, \href
  {http://adsabs.harvard.edu/abs/2013ApJ...768..172C} {768, 172}

\bibitem[\protect\citeauthoryear{{Conn} et~al.,}{{Conn} et~al.}{2012}]{con12a}
{Conn} A.~R.,  et~al., 2012, \mn@doi [\apj] {10.1088/0004-637X/758/1/11}, \href
  {http://adsabs.harvard.edu/abs/2012ApJ...758...11C} {758, 11}

\bibitem[\protect\citeauthoryear{{Cutri} et~al.,}{{Cutri}
  et~al.}{2008}]{cut08a}
{Cutri} R.~M.,  et~al., 2008, Technical report, {Explanatory Supplement to the
  2MASS All Sky Data Release and Extended Mission Products}

\bibitem[\protect\citeauthoryear{{Dall'Ora} et~al.,}{{Dall'Ora}
  et~al.}{2006}]{dal06a}
{Dall'Ora} M.,  et~al., 2006, \mn@doi [\apjl] {10.1086/510665}, \href
  {http://adsabs.harvard.edu/abs/2006ApJ...653L.109D} {653, L109}

\bibitem[\protect\citeauthoryear{{Dekel} \& {Silk}}{{Dekel} \&
  {Silk}}{1986}]{dek86a}
{Dekel} A.,  {Silk} J.,  1986, \mn@doi [\apj] {10.1086/164050}, \href
  {http://adsabs.harvard.edu/abs/1986ApJ...303...39D} {303, 39}

\bibitem[\protect\citeauthoryear{{Dekel} \& {Woo}}{{Dekel} \&
  {Woo}}{2003}]{dek03a}
{Dekel} A.,  {Woo} J.,  2003, \mn@doi [\mnras]
  {10.1046/j.1365-8711.2003.06923.x}, \href
  {http://adsabs.harvard.edu/abs/2003MNRAS.344.1131D} {344, 1131}

\bibitem[\protect\citeauthoryear{{Djorgovski} \& {Davis}}{{Djorgovski} \&
  {Davis}}{1987}]{djo87a}
{Djorgovski} S.,  {Davis} M.,  1987, \mn@doi [\apj] {10.1086/164948}, \href
  {http://adsabs.harvard.edu/abs/1987ApJ...313...59D} {313, 59}

\bibitem[\protect\citeauthoryear{{Dressler}, {Lynden-Bell}, {Burstein},
  {Davies}, {Faber}, {Terlevich}  \& {Wegner}}{{Dressler}
  et~al.}{1987}]{dre87a}
{Dressler} A.,  {Lynden-Bell} D.,  {Burstein} D.,  {Davies} R.~L.,  {Faber}
  S.~M.,  {Terlevich} R.,   {Wegner} G.,  1987, \mn@doi [\apj]
  {10.1086/164947}, \href {http://adsabs.harvard.edu/abs/1987ApJ...313...42D}
  {313, 42}

\bibitem[\protect\citeauthoryear{{Dubath}, {Meylan}  \& {Mayor}}{{Dubath}
  et~al.}{1997}]{dub97a}
{Dubath} P.,  {Meylan} G.,   {Mayor} M.,  1997, \aap, \href
  {http://adsabs.harvard.edu/abs/1997A%26A...324..505D} {324, 505}

\bibitem[\protect\citeauthoryear{{Einasto}, {Saar}, {Kaasik}  \&
  {Chernin}}{{Einasto} et~al.}{1974}]{ein74a}
{Einasto} J.,  {Saar} E.,  {Kaasik} A.,   {Chernin} A.~D.,  1974, \mn@doi
  [\nat] {10.1038/252111a0}, \href
  {http://adsabs.harvard.edu/abs/1974Natur.252..111E} {252, 111}

\bibitem[\protect\citeauthoryear{{Ferrara} \& {Tolstoy}}{{Ferrara} \&
  {Tolstoy}}{2000}]{fer00a}
{Ferrara} A.,  {Tolstoy} E.,  2000, \mn@doi [\mnras]
  {10.1046/j.1365-8711.2000.03209.x}, \href
  {http://adsabs.harvard.edu/abs/2000MNRAS.313..291F} {313, 291}

\bibitem[\protect\citeauthoryear{{Flynn}, {Holmberg}, {Portinari}, {Fuchs}  \&
  {Jahrei{\ss}}}{{Flynn} et~al.}{2006}]{fly06a}
{Flynn} C.,  {Holmberg} J.,  {Portinari} L.,  {Fuchs} B.,   {Jahrei{\ss}} H.,
  2006, \mn@doi [\mnras] {10.1111/j.1365-2966.2006.10911.x}, \href
  {http://adsabs.harvard.edu/abs/2006MNRAS.372.1149F} {372, 1149}

\bibitem[\protect\citeauthoryear{{Forbes}, {Spitler}, {Graham}, {Foster}, {Hau}
   \& {Benson}}{{Forbes} et~al.}{2011}]{for11a}
{Forbes} D.~A.,  {Spitler} L.~R.,  {Graham} A.~W.,  {Foster} C.,  {Hau}
  G.~K.~T.,   {Benson} A.,  2011, \mn@doi [\mnras]
  {10.1111/j.1365-2966.2011.18335.x}, \href
  {http://adsabs.harvard.edu/abs/2011MNRAS.413.2665F} {413, 2665}

\bibitem[\protect\citeauthoryear{{Gavazzi}, {Boselli}, {Donati}, {Franzetti}
  \& {Scodeggio}}{{Gavazzi} et~al.}{2003}]{gav03a}
{Gavazzi} G.,  {Boselli} A.,  {Donati} A.,  {Franzetti} P.,   {Scodeggio} M.,
  2003, \mn@doi [\aap] {10.1051/0004-6361:20030026}, \href
  {http://adsabs.harvard.edu/abs/2003A%26A...400..451G} {400, 451}

\bibitem[\protect\citeauthoryear{{Geha}, {Guhathakurta}, {Rich}  \&
  {Cooper}}{{Geha} et~al.}{2006a}]{geh06a}
{Geha} M.,  {Guhathakurta} P.,  {Rich} R.~M.,   {Cooper} M.~C.,  2006a, \mn@doi
  [\aj] {10.1086/498686}, \href
  {http://adsabs.harvard.edu/abs/2006AJ....131..332G} {131, 332}

\bibitem[\protect\citeauthoryear{{Geha}, {Blanton}, {Masjedi}  \&
  {West}}{{Geha} et~al.}{2006b}]{geh06b}
{Geha} M.,  {Blanton} M.~R.,  {Masjedi} M.,   {West} A.~A.,  2006b, \mn@doi
  [\apj] {10.1086/508604}, \href
  {http://adsabs.harvard.edu/abs/2006ApJ...653..240G} {653, 240}

\bibitem[\protect\citeauthoryear{{Geha}, {van der Marel}, {Guhathakurta},
  {Gilbert}, {Kalirai}  \& {Kirby}}{{Geha} et~al.}{2010}]{geh10a}
{Geha} M.,  {van der Marel} R.~P.,  {Guhathakurta} P.,  {Gilbert} K.~M.,
  {Kalirai} J.,   {Kirby} E.~N.,  2010, \mn@doi [\apj]
  {10.1088/0004-637X/711/1/361}, \href
  {http://adsabs.harvard.edu/abs/2010ApJ...711..361G} {711, 361}

\bibitem[\protect\citeauthoryear{{Graham}, {Driver}, {Petrosian}, {Conselice},
  {Bershady}, {Crawford}  \& {Goto}}{{Graham} et~al.}{2005}]{gra05a}
{Graham} A.~W.,  {Driver} S.~P.,  {Petrosian} V.,  {Conselice} C.~J.,
  {Bershady} M.~A.,  {Crawford} S.~M.,   {Goto} T.,  2005, \mn@doi [\aj]
  {10.1086/444475}, \href {http://adsabs.harvard.edu/abs/2005AJ....130.1535G}
  {130, 1535}

\bibitem[\protect\citeauthoryear{{Gratton}, {Bragaglia}, {Clementini},
  {Carretta}, {Di Fabrizio}, {Maio}  \& {Taribello}}{{Gratton}
  et~al.}{2004}]{gra04a}
{Gratton} R.~G.,  {Bragaglia} A.,  {Clementini} G.,  {Carretta} E.,  {Di
  Fabrizio} L.,  {Maio} M.,   {Taribello} E.,  2004, \mn@doi [\aap]
  {10.1051/0004-6361:20035840}, \href
  {http://adsabs.harvard.edu/abs/2004A%26A...421..937G} {421, 937}

\bibitem[\protect\citeauthoryear{{Grebel}}{{Grebel}}{2001}]{gre01a}
{Grebel} E.~K.,  2001, \mn@doi [Astrophysics and Space Science Supplement]
  {10.1023/A:1012742903265}, \href
  {http://adsabs.harvard.edu/abs/2001ApSSS.277..231G} {277, 231}

\bibitem[\protect\citeauthoryear{{Greco} et~al.,}{{Greco}
  et~al.}{2008}]{gre08a}
{Greco} C.,  et~al., 2008, \mn@doi [\apjl] {10.1086/533585}, \href
  {http://adsabs.harvard.edu/abs/2008ApJ...675L..73G} {675, L73}

\bibitem[\protect\citeauthoryear{{Han} et~al.,}{{Han} et~al.}{1997}]{han97a}
{Han} M.,  et~al., 1997, \mn@doi [\aj] {10.1086/118316}, \href
  {http://adsabs.harvard.edu/abs/1997AJ....113.1001H} {113, 1001}

\bibitem[\protect\citeauthoryear{{Hills}}{{Hills}}{1980}]{hil80a}
{Hills} J.~G.,  1980, \mn@doi [\apj] {10.1086/157703}, \href
  {http://adsabs.harvard.edu/abs/1980ApJ...235..986H} {235, 986}

\bibitem[\protect\citeauthoryear{{Ho}, {Geha}, {Tollerud}, {Zinn},
  {Guhathakurta}  \& {Vargas}}{{Ho} et~al.}{2015}]{ho15a}
{Ho} N.,  {Geha} M.,  {Tollerud} E.~J.,  {Zinn} R.,  {Guhathakurta} P.,
  {Vargas} L.~C.,  2015, \mn@doi [\apj] {10.1088/0004-637X/798/2/77}, \href
  {http://adsabs.harvard.edu/abs/2015ApJ...798...77H} {798, 77}

\bibitem[\protect\citeauthoryear{{Holmberg}, {Flynn}  \&
  {Portinari}}{{Holmberg} et~al.}{2006}]{hol06a}
{Holmberg} J.,  {Flynn} C.,   {Portinari} L.,  2006, \mn@doi [\mnras]
  {10.1111/j.1365-2966.2005.09832.x}, \href
  {http://adsabs.harvard.edu/abs/2006MNRAS.367..449H} {367, 449}

\bibitem[\protect\citeauthoryear{{Huchtmeier}, {Karachentsev}  \&
  {Karachentseva}}{{Huchtmeier} et~al.}{2003}]{huc03a}
{Huchtmeier} W.~K.,  {Karachentsev} I.~D.,   {Karachentseva} V.~E.,  2003,
  \mn@doi [\aap] {10.1051/0004-6361:20030138}, \href
  {http://adsabs.harvard.edu/abs/2003A%26A...401..483H} {401, 483}

\bibitem[\protect\citeauthoryear{{Humphreys}, {Reid}, {Moran}, {Greenhill}  \&
  {Argon}}{{Humphreys} et~al.}{2013}]{hum13a}
{Humphreys} E.~M.~L.,  {Reid} M.~J.,  {Moran} J.~M.,  {Greenhill} L.~J.,
  {Argon} A.~L.,  2013, \mn@doi [\apj] {10.1088/0004-637X/775/1/13}, \href
  {http://adsabs.harvard.edu/abs/2013ApJ...775...13H} {775, 13}

\bibitem[\protect\citeauthoryear{{Irwin} \& {Hatzidimitriou}}{{Irwin} \&
  {Hatzidimitriou}}{1995}]{irw95a}
{Irwin} M.,  {Hatzidimitriou} D.,  1995, \mnras, \href
  {http://adsabs.harvard.edu/abs/1995MNRAS.277.1354I} {277, 1354}

\bibitem[\protect\citeauthoryear{{Janz} \& {Lisker}}{{Janz} \&
  {Lisker}}{2008}]{jan08a}
{Janz} J.,  {Lisker} T.,  2008, \mn@doi [\apjl] {10.1086/595720}, \href
  {http://adsabs.harvard.edu/abs/2008ApJ...689L..25J} {689, L25}

\bibitem[\protect\citeauthoryear{{Janz} et~al.,}{{Janz} et~al.}{2014}]{jan14a}
{Janz} J.,  et~al., 2014, \mn@doi [\apj] {10.1088/0004-637X/786/2/105}, \href
  {http://adsabs.harvard.edu/abs/2014ApJ...786..105J} {786, 105}

\bibitem[\protect\citeauthoryear{{Jarrett}, {Chester}, {Cutri}, {Schneider}  \&
  {Huchra}}{{Jarrett} et~al.}{2003}]{jar03a}
{Jarrett} T.~H.,  {Chester} T.,  {Cutri} R.,  {Schneider} S.~E.,   {Huchra}
  J.~P.,  2003, \mn@doi [\aj] {10.1086/345794}, \href
  {http://adsabs.harvard.edu/abs/2003AJ....125..525J} {125, 525}

\bibitem[\protect\citeauthoryear{{Khare}}{{Khare}}{1953}]{kha53a}
{Khare} R.~C.,  1953, \zap, \href
  {http://adsabs.harvard.edu/abs/1953ZA.....33..251K} {33, 251}

\bibitem[\protect\citeauthoryear{{Kinemuchi}, {Harris}, {Smith}, {Silbermann},
  {Snyder}, {La Cluyz{\'e}}  \& {Clark}}{{Kinemuchi} et~al.}{2008}]{kin08a}
{Kinemuchi} K.,  {Harris} H.~C.,  {Smith} H.~A.,  {Silbermann} N.~A.,  {Snyder}
  L.~A.,  {La Cluyz{\'e}} A.~P.,   {Clark} C.~L.,  2008, \mn@doi [\aj]
  {10.1088/0004-6256/136/5/1921}, \href
  {http://adsabs.harvard.edu/abs/2008AJ....136.1921K} {136, 1921}

\bibitem[\protect\citeauthoryear{{King}}{{King}}{1962}]{kin62a}
{King} I.,  1962, \mn@doi [\aj] {10.1086/108756}, \href
  {http://adsabs.harvard.edu/abs/1962AJ.....67..471K} {67, 471}

\bibitem[\protect\citeauthoryear{{Kirby}, {Cohen}, {Guhathakurta}, {Cheng},
  {Bullock}  \& {Gallazzi}}{{Kirby} et~al.}{2013}]{kir13a}
{Kirby} E.~N.,  {Cohen} J.~G.,  {Guhathakurta} P.,  {Cheng} L.,  {Bullock}
  J.~S.,   {Gallazzi} A.,  2013, \mn@doi [\apj] {10.1088/0004-637X/779/2/102},
  \href {http://adsabs.harvard.edu/abs/2013ApJ...779..102K} {779, 102}

\bibitem[\protect\citeauthoryear{{Kirby}, {Bullock}, {Boylan-Kolchin},
  {Kaplinghat}  \& {Cohen}}{{Kirby} et~al.}{2014}]{kir14a}
{Kirby} E.~N.,  {Bullock} J.~S.,  {Boylan-Kolchin} M.,  {Kaplinghat} M.,
  {Cohen} J.~G.,  2014, \mn@doi [\mnras] {10.1093/mnras/stu025}, \href
  {http://adsabs.harvard.edu/abs/2014MNRAS.439.1015K} {439, 1015}

\bibitem[\protect\citeauthoryear{{Kirby}, {Rizzi}, {Held}, {Cohen}, {Cole},
  {Manning}, {Skillman}  \& {Weisz}}{{Kirby} et~al.}{2017}]{kir17a}
{Kirby} E.~N.,  {Rizzi} L.,  {Held} E.~V.,  {Cohen} J.~G.,  {Cole} A.~A.,
  {Manning} E.~M.,  {Skillman} E.~D.,   {Weisz} D.~R.,  2017, \mn@doi [\apj]
  {10.3847/1538-4357/834/1/9}, \href
  {http://adsabs.harvard.edu/abs/2017ApJ...834....9K} {834, 9}

\bibitem[\protect\citeauthoryear{{Koch}, {Kleyna}, {Wilkinson}, {Grebel},
  {Gilmore}, {Evans}, {Wyse}  \& {Harbeck}}{{Koch} et~al.}{2007}]{koc07a}
{Koch} A.,  {Kleyna} J.~T.,  {Wilkinson} M.~I.,  {Grebel} E.~K.,  {Gilmore}
  G.~F.,  {Evans} N.~W.,  {Wyse} R.~F.~G.,   {Harbeck} D.~R.,  2007, \mn@doi
  [\aj] {10.1086/519380}, \href
  {http://adsabs.harvard.edu/abs/2007AJ....134..566K} {134, 566}

\bibitem[\protect\citeauthoryear{{Koch}, {Grebel}, {Gilmore}, {Wyse}, {Kleyna},
  {Harbeck}, {Wilkinson}  \& {Wyn Evans}}{{Koch} et~al.}{2008}]{koc08a}
{Koch} A.,  {Grebel} E.~K.,  {Gilmore} G.~F.,  {Wyse} R.~F.~G.,  {Kleyna}
  J.~T.,  {Harbeck} D.~R.,  {Wilkinson} M.~I.,   {Wyn Evans} N.,  2008, \mn@doi
  [\aj] {10.1088/0004-6256/135/4/1580}, \href
  {http://adsabs.harvard.edu/abs/2008AJ....135.1580K} {135, 1580}

\bibitem[\protect\citeauthoryear{{Koposov} et~al.,}{{Koposov}
  et~al.}{2011}]{kop11a}
{Koposov} S.~E.,  et~al., 2011, \mn@doi [\apj] {10.1088/0004-637X/736/2/146},
  \href {http://adsabs.harvard.edu/abs/2011ApJ...736..146K} {736, 146}

\bibitem[\protect\citeauthoryear{{Kormendy}}{{Kormendy}}{1985}]{kor85a}
{Kormendy} J.,  1985, \mn@doi [\apj] {10.1086/163350}, \href
  {http://adsabs.harvard.edu/abs/1985ApJ...295...73K} {295, 73}

\bibitem[\protect\citeauthoryear{{Kormendy} \& {Freeman}}{{Kormendy} \&
  {Freeman}}{2016}]{kor16a}
{Kormendy} J.,  {Freeman} K.~C.,  2016, \mn@doi [\apj]
  {10.3847/0004-637X/817/2/84}, \href
  {http://adsabs.harvard.edu/abs/2016ApJ...817...84K} {817, 84}

\bibitem[\protect\citeauthoryear{{Kroupa}, {Tout}  \& {Gilmore}}{{Kroupa}
  et~al.}{1993}]{kro93a}
{Kroupa} P.,  {Tout} C.~A.,   {Gilmore} G.,  1993, \mnras, \href
  {http://adsabs.harvard.edu/abs/1993MNRAS.262..545K} {262, 545}

\bibitem[\protect\citeauthoryear{{Kuehn} et~al.,}{{Kuehn}
  et~al.}{2008}]{kue08a}
{Kuehn} C.,  et~al., 2008, \mn@doi [\apjl] {10.1086/529137}, \href
  {http://adsabs.harvard.edu/abs/2008ApJ...674L..81K} {674, L81}

\bibitem[\protect\citeauthoryear{{Lai}, {Lee}, {Bolte}, {Lucatello}, {Beers},
  {Johnson}, {Sivarani}  \& {Rockosi}}{{Lai} et~al.}{2011}]{lai11a}
{Lai} D.~K.,  {Lee} Y.~S.,  {Bolte} M.,  {Lucatello} S.,  {Beers} T.~C.,
  {Johnson} J.~A.,  {Sivarani} T.,   {Rockosi} C.~M.,  2011, \mn@doi [\apj]
  {10.1088/0004-637X/738/1/51}, \href
  {http://adsabs.harvard.edu/abs/2011ApJ...738...51L} {738, 51}

\bibitem[\protect\citeauthoryear{{Lanzoni} et~al.,}{{Lanzoni}
  et~al.}{2013}]{lan13a}
{Lanzoni} B.,  et~al., 2013, \mn@doi [\apj] {10.1088/0004-637X/769/2/107},
  \href {http://adsabs.harvard.edu/abs/2013ApJ...769..107L} {769, 107}

\bibitem[\protect\citeauthoryear{{Lee} et~al.,}{{Lee} et~al.}{2003}]{lee03a}
{Lee} M.~G.,  et~al., 2003, \mn@doi [\aj] {10.1086/379171}, \href
  {http://adsabs.harvard.edu/abs/2003AJ....126.2840L} {126, 2840}

\bibitem[\protect\citeauthoryear{{Lee}, {Kennicutt}, {Funes}, {Sakai}  \&
  {Akiyama}}{{Lee} et~al.}{2007}]{lee07a}
{Lee} J.~C.,  {Kennicutt} R.~C.,  {Funes} Jos{\'e}~G. S.~J.,  {Sakai} S.,
  {Akiyama} S.,  2007, \mn@doi [\apjl] {10.1086/526341}, \href
  {http://adsabs.harvard.edu/abs/2007ApJ...671L.113L} {671, L113}

\bibitem[\protect\citeauthoryear{{Leitherer} et~al.,}{{Leitherer}
  et~al.}{1999}]{lei99a}
{Leitherer} C.,  et~al., 1999, \mn@doi [\apjs] {10.1086/313233}, \href
  {http://adsabs.harvard.edu/abs/1999ApJS..123....3L} {123, 3}

\bibitem[\protect\citeauthoryear{{Leitherer}, {Ortiz Ot{\'a}lvaro}, {Bresolin},
  {Kudritzki}, {Lo Faro}, {Pauldrach}, {Pettini}  \& {Rix}}{{Leitherer}
  et~al.}{2010}]{lei10a}
{Leitherer} C.,  {Ortiz Ot{\'a}lvaro} P.~A.,  {Bresolin} F.,  {Kudritzki}
  R.-P.,  {Lo Faro} B.,  {Pauldrach} A.~W.~A.,  {Pettini} M.,   {Rix} S.~A.,
  2010, \mn@doi [\apjs] {10.1088/0067-0049/189/2/309}, \href
  {http://adsabs.harvard.edu/abs/2010ApJS..189..309L} {189, 309}

\bibitem[\protect\citeauthoryear{{Leitherer}, {Ekstr{\"o}m}, {Meynet},
  {Schaerer}, {Agienko}  \& {Levesque}}{{Leitherer} et~al.}{2014}]{lei14a}
{Leitherer} C.,  {Ekstr{\"o}m} S.,  {Meynet} G.,  {Schaerer} D.,  {Agienko}
  K.~B.,   {Levesque} E.~M.,  2014, \mn@doi [\apjs]
  {10.1088/0067-0049/212/1/14}, \href
  {http://adsabs.harvard.edu/abs/2014ApJS..212...14L} {212, 14}

\bibitem[\protect\citeauthoryear{{Lelli}, {Fraternali}  \& {Verheijen}}{{Lelli}
  et~al.}{2014}]{lel14a}
{Lelli} F.,  {Fraternali} F.,   {Verheijen} M.,  2014, \mn@doi [\aap]
  {10.1051/0004-6361/201322658}, \href
  {http://adsabs.harvard.edu/abs/2014A%26A...563A..27L} {563, A27}

\bibitem[\protect\citeauthoryear{{Lewis}, {Ibata}, {Chapman}, {McConnachie},
  {Irwin}, {Tolstoy}  \& {Tanvir}}{{Lewis} et~al.}{2007}]{lew07a}
{Lewis} G.~F.,  {Ibata} R.~A.,  {Chapman} S.~C.,  {McConnachie} A.,  {Irwin}
  M.~J.,  {Tolstoy} E.,   {Tanvir} N.~R.,  2007, \mn@doi [\mnras]
  {10.1111/j.1365-2966.2007.11395.x}, \href
  {http://adsabs.harvard.edu/abs/2007MNRAS.375.1364L} {375, 1364}

\bibitem[\protect\citeauthoryear{{Lieder}, {Lisker}, {Hilker}, {Misgeld}  \&
  {Durrell}}{{Lieder} et~al.}{2012}]{lie12a}
{Lieder} S.,  {Lisker} T.,  {Hilker} M.,  {Misgeld} I.,   {Durrell} P.,  2012,
  \mn@doi [\aap] {10.1051/0004-6361/201117163}, \href
  {http://cdsads.u-strasbg.fr/abs/2012A%26A...538A..69L} {538, A69}

\bibitem[\protect\citeauthoryear{{Lin} \& {Faber}}{{Lin} \&
  {Faber}}{1983}]{lin83a}
{Lin} D.~N.~C.,  {Faber} S.~M.,  1983, \mn@doi [\apjl] {10.1086/183971}, \href
  {http://adsabs.harvard.edu/abs/1983ApJ...266L..21L} {266, L21}

\bibitem[\protect\citeauthoryear{{Lisker}, {Grebel}, {Binggeli}  \&
  {Glatt}}{{Lisker} et~al.}{2007}]{lis07a}
{Lisker} T.,  {Grebel} E.~K.,  {Binggeli} B.,   {Glatt} K.,  2007, \mn@doi
  [\apj] {10.1086/513090}, \href
  {http://adsabs.harvard.edu/abs/2007ApJ...660.1186L} {660, 1186}

\bibitem[\protect\citeauthoryear{{L{\"u}tzgendorf}, {Gebhardt}, {Baumgardt},
  {Noyola}, {Neumayer}, {Kissler-Patig}  \& {de Zeeuw}}{{L{\"u}tzgendorf}
  et~al.}{2015}]{lut15a}
{L{\"u}tzgendorf} N.,  {Gebhardt} K.,  {Baumgardt} H.,  {Noyola} E.,
  {Neumayer} N.,  {Kissler-Patig} M.,   {de Zeeuw} T.,  2015, \mn@doi [\aap]
  {10.1051/0004-6361/201425524}, \href
  {http://adsabs.harvard.edu/abs/2015A%26A...581A...1L} {581, A1}

\bibitem[\protect\citeauthoryear{{Lux}, {Read}  \& {Lake}}{{Lux}
  et~al.}{2010}]{lux10a}
{Lux} H.,  {Read} J.~I.,   {Lake} G.,  2010, \mn@doi [\mnras]
  {10.1111/j.1365-2966.2010.16877.x}, \href
  {http://adsabs.harvard.edu/abs/2010MNRAS.406.2312L} {406, 2312}

\bibitem[\protect\citeauthoryear{{Martin}, {de Jong}  \& {Rix}}{{Martin}
  et~al.}{2008}]{mar08a}
{Martin} N.~F.,  {de Jong} J.~T.~A.,   {Rix} H.-W.,  2008, \mn@doi [\apj]
  {10.1086/590336}, \href {http://adsabs.harvard.edu/abs/2008ApJ...684.1075M}
  {684, 1075}

\bibitem[\protect\citeauthoryear{{Mayer}, {Governato}, {Colpi}, {Moore},
  {Quinn}, {Wadsley}, {Stadel}  \& {Lake}}{{Mayer} et~al.}{2001}]{may01a}
{Mayer} L.,  {Governato} F.,  {Colpi} M.,  {Moore} B.,  {Quinn} T.,  {Wadsley}
  J.,  {Stadel} J.,   {Lake} G.,  2001, \mn@doi [\apj] {10.1086/322356}, \href
  {http://adsabs.harvard.edu/abs/2001ApJ...559..754M} {559, 754}

\bibitem[\protect\citeauthoryear{{Mayer}, {Mastropietro}, {Wadsley}, {Stadel}
  \& {Moore}}{{Mayer} et~al.}{2006}]{may06a}
{Mayer} L.,  {Mastropietro} C.,  {Wadsley} J.,  {Stadel} J.,   {Moore} B.,
  2006, \mn@doi [\mnras] {10.1111/j.1365-2966.2006.10403.x}, \href
  {http://adsabs.harvard.edu/abs/2006MNRAS.369.1021M} {369, 1021}

\bibitem[\protect\citeauthoryear{{McCall}}{{McCall}}{2004}]{mcc04a}
{McCall} M.~L.,  2004, \mn@doi [\aj] {10.1086/424933}, \href
  {http://adsabs.harvard.edu/abs/2004AJ....128.2144M} {128, 2144}

\bibitem[\protect\citeauthoryear{{McCall}}{{McCall}}{2012}]{mcc12a}
{McCall} M.~L.,  2012, in American Astronomical Society Meeting Abstracts
  \#219. p. 244.14

\bibitem[\protect\citeauthoryear{{McCall}}{{McCall}}{2014}]{mcc14a}
{McCall} M.~L.,  2014, \mn@doi [\mnras] {10.1093/mnras/stu199}, \href
  {http://adsabs.harvard.edu/abs/2014MNRAS.440..405M} {440, 405}

\bibitem[\protect\citeauthoryear{{McCall}, {Vaduvescu}, {Pozo Nunez}, {Barr
  Dominguez}, {Fingerhut}, {Unda-Sanzana}, {Li}  \& {Albrecht}}{{McCall}
  et~al.}{2012}]{mcc12b}
{McCall} M.~L.,  {Vaduvescu} O.,  {Pozo Nunez} F.,  {Barr Dominguez} A.,
  {Fingerhut} R.,  {Unda-Sanzana} E.,  {Li} B.,   {Albrecht} M.,  2012, \mn@doi
  [\aap] {10.1051/0004-6361/201117669}, \href
  {http://adsabs.harvard.edu/abs/2012A%26A...540A..49M} {540, A49}

\bibitem[\protect\citeauthoryear{{McConnachie}}{{McConnachie}}{2012}]{mcco12a}
{McConnachie} A.~W.,  2012, \mn@doi [\aj] {10.1088/0004-6256/144/1/4}, \href
  {http://adsabs.harvard.edu/abs/2012AJ....144....4M} {144, 4}

\bibitem[\protect\citeauthoryear{{McConnachie}}{{McConnachie}}{2016}]{mcco16a}
{McConnachie} A.~W.,  2016, {The Observed Properties of Dwarf Galaxies in and
  around the Local Group},
  \url{https://www.astrosci.ca/users/alan/Nearby_Dwarfs_Database.html}

\bibitem[\protect\citeauthoryear{{McConnachie} \& {Irwin}}{{McConnachie} \&
  {Irwin}}{2006}]{mcco06a}
{McConnachie} A.~W.,  {Irwin} M.~J.,  2006, \mn@doi [\mnras]
  {10.1111/j.1365-2966.2005.09806.x}, \href
  {http://adsabs.harvard.edu/abs/2006MNRAS.365.1263M} {365, 1263}

\bibitem[\protect\citeauthoryear{{McConnachie}, {Irwin}, {Ferguson}, {Ibata},
  {Lewis}  \& {Tanvir}}{{McConnachie} et~al.}{2004}]{mcco04a}
{McConnachie} A.~W.,  {Irwin} M.~J.,  {Ferguson} A.~M.~N.,  {Ibata} R.~A.,
  {Lewis} G.~F.,   {Tanvir} N.,  2004, \mn@doi [\mnras]
  {10.1111/j.1365-2966.2004.07637.x}, \href
  {http://adsabs.harvard.edu/abs/2004MNRAS.350..243M} {350, 243}

\bibitem[\protect\citeauthoryear{{McConnachie}, {Irwin}, {Ferguson}, {Ibata},
  {Lewis}  \& {Tanvir}}{{McConnachie} et~al.}{2005}]{mcco05a}
{McConnachie} A.~W.,  {Irwin} M.~J.,  {Ferguson} A.~M.~N.,  {Ibata} R.~A.,
  {Lewis} G.~F.,   {Tanvir} N.,  2005, \mn@doi [\mnras]
  {10.1111/j.1365-2966.2004.08514.x}, \href
  {http://adsabs.harvard.edu/abs/2005MNRAS.356..979M} {356, 979}

\bibitem[\protect\citeauthoryear{{Meyer}, {Lisker}, {Janz}  \&
  {Papaderos}}{{Meyer} et~al.}{2014}]{mey14a}
{Meyer} H.~T.,  {Lisker} T.,  {Janz} J.,   {Papaderos} P.,  2014, \mn@doi
  [\aap] {10.1051/0004-6361/201220700}, \href
  {http://adsabs.harvard.edu/abs/2014A%26A...562A..49M} {562, A49}

\bibitem[\protect\citeauthoryear{{Minor}, {Martinez}, {Bullock}, {Kaplinghat}
  \& {Trainor}}{{Minor} et~al.}{2010}]{min10a}
{Minor} Q.~E.,  {Martinez} G.,  {Bullock} J.,  {Kaplinghat} M.,   {Trainor} R.,
   2010, \mn@doi [\apj] {10.1088/0004-637X/721/2/1142}, \href
  {http://adsabs.harvard.edu/abs/2010ApJ...721.1142M} {721, 1142}

\bibitem[\protect\citeauthoryear{{Misgeld}, {Mieske}  \& {Hilker}}{{Misgeld}
  et~al.}{2008}]{mis08a}
{Misgeld} I.,  {Mieske} S.,   {Hilker} M.,  2008, \mn@doi [\aap]
  {10.1051/0004-6361:200810014}, \href
  {http://adsabs.harvard.edu/abs/2008A%26A...486..697M} {486, 697}

\bibitem[\protect\citeauthoryear{{Misgeld}, {Hilker}  \& {Mieske}}{{Misgeld}
  et~al.}{2009}]{mis09a}
{Misgeld} I.,  {Hilker} M.,   {Mieske} S.,  2009, \mn@doi [\aap]
  {10.1051/0004-6361/200811451}, \href
  {http://adsabs.harvard.edu/abs/2009A%26A...496..683M} {496, 683}

\bibitem[\protect\citeauthoryear{{Moretti} et~al.,}{{Moretti}
  et~al.}{2009}]{mor09a}
{Moretti} M.~I.,  et~al., 2009, \mn@doi [\apjl] {10.1088/0004-637X/699/2/L125},
  \href {http://adsabs.harvard.edu/abs/2009ApJ...699L.125M} {699, L125}

\bibitem[\protect\citeauthoryear{{Mu{\~n}oz}, {Geha}  \& {Willman}}{{Mu{\~n}oz}
  et~al.}{2010}]{mun10a}
{Mu{\~n}oz} R.~R.,  {Geha} M.,   {Willman} B.,  2010, \mn@doi [\aj]
  {10.1088/0004-6256/140/1/138}, \href
  {http://adsabs.harvard.edu/abs/2010AJ....140..138M} {140, 138}

\bibitem[\protect\citeauthoryear{{Murray}, {Quataert}  \& {Thompson}}{{Murray}
  et~al.}{2005}]{mur05a}
{Murray} N.,  {Quataert} E.,   {Thompson} T.~A.,  2005, \mn@doi [\apj]
  {10.1086/426067}, \href {http://adsabs.harvard.edu/abs/2005ApJ...618..569M}
  {618, 569}

\bibitem[\protect\citeauthoryear{{Musella} et~al.,}{{Musella}
  et~al.}{2009}]{mus09a}
{Musella} I.,  et~al., 2009, \mn@doi [\apjl] {10.1088/0004-637X/695/1/L83},
  \href {http://adsabs.harvard.edu/abs/2009ApJ...695L..83M} {695, L83}

\bibitem[\protect\citeauthoryear{{Navarro}, {Frenk}  \& {White}}{{Navarro}
  et~al.}{1996}]{nav96a}
{Navarro} J.~F.,  {Frenk} C.~S.,   {White} S.~D.~M.,  1996, \mn@doi [\apj]
  {10.1086/177173}, \href {http://adsabs.harvard.edu/abs/1996ApJ...462..563N}
  {462, 563}

\bibitem[\protect\citeauthoryear{{Nemec}, {Wehlau}  \& {Mendes de
  Oliveira}}{{Nemec} et~al.}{1988}]{nem88a}
{Nemec} J.~M.,  {Wehlau} A.,   {Mendes de Oliveira} C.,  1988, \mn@doi [\aj]
  {10.1086/114830}, \href {http://adsabs.harvard.edu/abs/1988AJ.....96..528N}
  {96, 528}

\bibitem[\protect\citeauthoryear{{Nowotny}, {Kerschbaum}, {Olofsson}  \&
  {Schwarz}}{{Nowotny} et~al.}{2003}]{now03a}
{Nowotny} W.,  {Kerschbaum} F.,  {Olofsson} H.,   {Schwarz} H.~E.,  2003,
  \mn@doi [\aap] {10.1051/0004-6361:20030282}, \href
  {http://adsabs.harvard.edu/abs/2003A%26A...403...93N} {403, 93}

\bibitem[\protect\citeauthoryear{{Okamoto}, {Arimoto}, {Yamada}  \&
  {Onodera}}{{Okamoto} et~al.}{2012}]{oka12a}
{Okamoto} S.,  {Arimoto} N.,  {Yamada} Y.,   {Onodera} M.,  2012, \mn@doi
  [\apj] {10.1088/0004-637X/744/2/96}, \href
  {http://adsabs.harvard.edu/abs/2012ApJ...744...96O} {744, 96}

\bibitem[\protect\citeauthoryear{{Oppenheimer} \& {Dav{\'e}}}{{Oppenheimer} \&
  {Dav{\'e}}}{2006}]{opp06a}
{Oppenheimer} B.~D.,  {Dav{\'e}} R.,  2006, \mn@doi [\mnras]
  {10.1111/j.1365-2966.2006.10989.x}, \href
  {http://adsabs.harvard.edu/abs/2006MNRAS.373.1265O} {373, 1265}

\bibitem[\protect\citeauthoryear{{Pack}}{{Pack}}{1953}]{pac53a}
{Pack} D.~C.,  1953, \mnras, \href
  {http://adsabs.harvard.edu/abs/1953MNRAS.113...43P} {113, 43}

\bibitem[\protect\citeauthoryear{{Paturel}, {Petit}, {Prugniel}, {Theureau},
  {Rousseau}, {Brouty}, {Dubois}  \& {Cambr{\'e}sy}}{{Paturel}
  et~al.}{2003}]{pat03a}
{Paturel} G.,  {Petit} C.,  {Prugniel} P.,  {Theureau} G.,  {Rousseau} J.,
  {Brouty} M.,  {Dubois} P.,   {Cambr{\'e}sy} L.,  2003, \mn@doi [\aap]
  {10.1051/0004-6361:20031411}, \href
  {http://adsabs.harvard.edu/abs/2003A%26A...412...45P} {412, 45}

\bibitem[\protect\citeauthoryear{{Pequignot}, {Petitjean}  \&
  {Boisson}}{{Pequignot} et~al.}{1991}]{peq91a}
{Pequignot} D.,  {Petitjean} P.,   {Boisson} C.,  1991, \aap, \href
  {http://adsabs.harvard.edu/abs/1991A%26A...251..680P} {251, 680}

\bibitem[\protect\citeauthoryear{{Pickles}}{{Pickles}}{1998}]{pic98a}
{Pickles} A.~J.,  1998, \mn@doi [\pasp] {10.1086/316197}, \href
  {http://adsabs.harvard.edu/abs/1998PASP..110..863P} {110, 863}

\bibitem[\protect\citeauthoryear{{Pritzl}, {Armandroff}, {Jacoby}  \& {Da
  Costa}}{{Pritzl} et~al.}{2004}]{pri04a}
{Pritzl} B.~J.,  {Armandroff} T.~E.,  {Jacoby} G.~H.,   {Da Costa} G.~S.,
  2004, \mn@doi [\aj] {10.1086/380613}, \href
  {http://adsabs.harvard.edu/abs/2004AJ....127..318P} {127, 318}

\bibitem[\protect\citeauthoryear{{Pritzl}, {Armandroff}, {Jacoby}  \& {Da
  Costa}}{{Pritzl} et~al.}{2005}]{pri05a}
{Pritzl} B.~J.,  {Armandroff} T.~E.,  {Jacoby} G.~H.,   {Da Costa} G.~S.,
  2005, \mn@doi [\aj] {10.1086/428372}, \href
  {http://adsabs.harvard.edu/abs/2005AJ....129.2232P} {129, 2232}

\bibitem[\protect\citeauthoryear{{Richer} \& {McCall}}{{Richer} \&
  {McCall}}{1995}]{ric95a}
{Richer} M.~G.,  {McCall} M.~L.,  1995, \mn@doi [\apj] {10.1086/175727}, \href
  {http://adsabs.harvard.edu/abs/1995ApJ...445..642R} {445, 642}

\bibitem[\protect\citeauthoryear{{Rizzi}, {Held}, {Saviane}, {Tully}  \&
  {Gullieuszik}}{{Rizzi} et~al.}{2007a}]{riz07b}
{Rizzi} L.,  {Held} E.~V.,  {Saviane} I.,  {Tully} R.~B.,   {Gullieuszik} M.,
  2007a, \mn@doi [\mnras] {10.1111/j.1365-2966.2007.12196.x}, \href
  {http://adsabs.harvard.edu/abs/2007MNRAS.380.1255R} {380, 1255}

\bibitem[\protect\citeauthoryear{{Rizzi}, {Tully}, {Makarov}, {Makarova},
  {Dolphin}, {Sakai}  \& {Shaya}}{{Rizzi} et~al.}{2007b}]{riz07a}
{Rizzi} L.,  {Tully} R.~B.,  {Makarov} D.,  {Makarova} L.,  {Dolphin} A.~E.,
  {Sakai} S.,   {Shaya} E.~J.,  2007b, \mn@doi [\apj] {10.1086/516566}, \href
  {http://adsabs.harvard.edu/abs/2007ApJ...661..815R} {661, 815}

\bibitem[\protect\citeauthoryear{{Robles-Valdez},
  {Rodr{\'{\i}}guez-Gonz{\'a}lez}, {Hern{\'a}ndez-Mart{\'{\i}}nez}  \&
  {Esquivel}}{{Robles-Valdez} et~al.}{2016}]{rob16a}
{Robles-Valdez} F.,  {Rodr{\'{\i}}guez-Gonz{\'a}lez} A.,
  {Hern{\'a}ndez-Mart{\'{\i}}nez} L.,   {Esquivel} A.,  2016, preprint, \href
  {http://adsabs.harvard.edu/abs/2016arXiv161202787R} {} (\mn@eprint {arXiv}
  {1612.02787})

\bibitem[\protect\citeauthoryear{{Ry{\'s}}, {Falc{\'o}n-Barroso}  \& {van de
  Ven}}{{Ry{\'s}} et~al.}{2013}]{rys13a}
{Ry{\'s}} A.,  {Falc{\'o}n-Barroso} J.,   {van de Ven} G.,  2013, \mn@doi
  [\mnras] {10.1093/mnras/sts245}, \href
  {http://adsabs.harvard.edu/abs/2013MNRAS.428.2980R} {428, 2980}

\bibitem[\protect\citeauthoryear{{Salpeter}}{{Salpeter}}{1955}]{sal55a}
{Salpeter} E.~E.,  1955, \mn@doi [\apj] {10.1086/145971}, \href
  {http://adsabs.harvard.edu/abs/1955ApJ...121..161S} {121, 161}

\bibitem[\protect\citeauthoryear{{S{\'a}nchez-Janssen}
  et~al.,}{{S{\'a}nchez-Janssen} et~al.}{2016}]{san16a}
{S{\'a}nchez-Janssen} R.,  et~al., 2016, \mn@doi [\apj]
  {10.3847/0004-637X/820/1/69}, \href
  {http://adsabs.harvard.edu/abs/2016ApJ...820...69S} {820, 69}

\bibitem[\protect\citeauthoryear{{Sandage} \& {Binggeli}}{{Sandage} \&
  {Binggeli}}{1984}]{san84a}
{Sandage} A.,  {Binggeli} B.,  1984, \mn@doi [\aj] {10.1086/113588}, \href
  {http://adsabs.harvard.edu/abs/1984AJ.....89..919S} {89, 919}

\bibitem[\protect\citeauthoryear{{Sawala} et~al.,}{{Sawala}
  et~al.}{2016}]{saw16a}
{Sawala} T.,  et~al., 2016, \mn@doi [\mnras] {10.1093/mnras/stv2597}, \href
  {http://adsabs.harvard.edu/abs/2016MNRAS.456...85S} {456, 85}

\bibitem[\protect\citeauthoryear{{Schlegel}, {Finkbeiner}  \&
  {Davis}}{{Schlegel} et~al.}{1998}]{sch98a}
{Schlegel} D.~J.,  {Finkbeiner} D.~P.,   {Davis} M.,  1998, \mn@doi [\apj]
  {10.1086/305772}, \href {http://adsabs.harvard.edu/abs/1998ApJ...500..525S}
  {500, 525}

\bibitem[\protect\citeauthoryear{{S{\'e}rsic}}{{S{\'e}rsic}}{1963}]{ser63a}
{S{\'e}rsic} J.~L.,  1963, Boletin de la Asociacion Argentina de Astronomia La
  Plata Argentina, \href {http://adsabs.harvard.edu/abs/1963BAAA....6...41S}
  {6, 41}

\bibitem[\protect\citeauthoryear{{Simon} \& {Geha}}{{Simon} \&
  {Geha}}{2007}]{sim07a}
{Simon} J.~D.,  {Geha} M.,  2007, \mn@doi [\apj] {10.1086/521816}, \href
  {http://adsabs.harvard.edu/abs/2007ApJ...670..313S} {670, 313}

\bibitem[\protect\citeauthoryear{{Smecker-Hane}, {Stetson}, {Hesser}  \&
  {Lehnert}}{{Smecker-Hane} et~al.}{1994}]{sme94a}
{Smecker-Hane} T.~A.,  {Stetson} P.~B.,  {Hesser} J.~E.,   {Lehnert} M.~D.,
  1994, \mn@doi [\aj] {10.1086/117087}, \href
  {http://adsabs.harvard.edu/abs/1994AJ....108..507S} {108, 507}

\bibitem[\protect\citeauthoryear{{Smith}}{{Smith}}{1995}]{smi95a}
{Smith} H.~A.,  1995, Cambridge Astrophysics Series, \href
  {http://adsabs.harvard.edu/abs/1995CAS....27.....S} {27}

\bibitem[\protect\citeauthoryear{{Sohn}, {Besla}, {van der Marel},
  {Boylan-Kolchin}, {Majewski}  \& {Bullock}}{{Sohn} et~al.}{2013}]{soh13a}
{Sohn} S.~T.,  {Besla} G.,  {van der Marel} R.~P.,  {Boylan-Kolchin} M.,
  {Majewski} S.~R.,   {Bullock} J.~S.,  2013, \mn@doi [\apj]
  {10.1088/0004-637X/768/2/139}, \href
  {http://adsabs.harvard.edu/abs/2013ApJ...768..139S} {768, 139}

\bibitem[\protect\citeauthoryear{{Spaans} \& {Norman}}{{Spaans} \&
  {Norman}}{1997}]{spa97a}
{Spaans} M.,  {Norman} C.~A.,  1997, \mn@doi [\apj] {10.1086/304231}, \href
  {http://adsabs.harvard.edu/abs/1997ApJ...483...87S} {483, 87}

\bibitem[\protect\citeauthoryear{{Spekkens}, {Urbancic}, {Mason}, {Willman}  \&
  {Aguirre}}{{Spekkens} et~al.}{2014}]{spe14a}
{Spekkens} K.,  {Urbancic} N.,  {Mason} B.~S.,  {Willman} B.,   {Aguirre}
  J.~E.,  2014, \mn@doi [\apjl] {10.1088/2041-8205/795/1/L5}, \href
  {http://adsabs.harvard.edu/abs/2014ApJ...795L...5S} {795, L5}

\bibitem[\protect\citeauthoryear{{Stark}, {McGaugh}  \& {Swaters}}{{Stark}
  et~al.}{2009}]{sta09a}
{Stark} D.~V.,  {McGaugh} S.~S.,   {Swaters} R.~A.,  2009, \mn@doi [\aj]
  {10.1088/0004-6256/138/2/392}, \href
  {http://adsabs.harvard.edu/abs/2009AJ....138..392S} {138, 392}

\bibitem[\protect\citeauthoryear{{Tassis}, {Kravtsov}  \& {Gnedin}}{{Tassis}
  et~al.}{2008}]{tas08a}
{Tassis} K.,  {Kravtsov} A.~V.,   {Gnedin} N.~Y.,  2008, \mn@doi [\apj]
  {10.1086/523880}, \href {http://adsabs.harvard.edu/abs/2008ApJ...672..888T}
  {672, 888}

\bibitem[\protect\citeauthoryear{{Tollerud}, {Bullock}, {Graves}  \&
  {Wolf}}{{Tollerud} et~al.}{2011}]{tol11a}
{Tollerud} E.~J.,  {Bullock} J.~S.,  {Graves} G.~J.,   {Wolf} J.,  2011,
  \mn@doi [\apj] {10.1088/0004-637X/726/2/108}, \href
  {http://adsabs.harvard.edu/abs/2011ApJ...726..108T} {726, 108}

\bibitem[\protect\citeauthoryear{{Tollerud} et~al.,}{{Tollerud}
  et~al.}{2012}]{tol12a}
{Tollerud} E.~J.,  et~al., 2012, \mn@doi [\apj] {10.1088/0004-637X/752/1/45},
  \href {http://adsabs.harvard.edu/abs/2012ApJ...752...45T} {752, 45}

\bibitem[\protect\citeauthoryear{{Toloba}, {Boselli}, {Cenarro}, {Peletier},
  {Gorgas}, {Gil de Paz}  \& {Mu{\~n}oz-Mateos}}{{Toloba}
  et~al.}{2011}]{tol11b}
{Toloba} E.,  {Boselli} A.,  {Cenarro} A.~J.,  {Peletier} R.~F.,  {Gorgas} J.,
  {Gil de Paz} A.,   {Mu{\~n}oz-Mateos} J.~C.,  2011, \mn@doi [\aap]
  {10.1051/0004-6361/201015344}, \href
  {http://adsabs.harvard.edu/abs/2011A%26A...526A.114T} {526, A114}

\bibitem[\protect\citeauthoryear{{Toloba} et~al.,}{{Toloba}
  et~al.}{2014}]{tol14a}
{Toloba} E.,  et~al., 2014, \mn@doi [\apjs] {10.1088/0067-0049/215/2/17}, \href
  {http://adsabs.harvard.edu/abs/2014ApJS..215...17T} {215, 17}

\bibitem[\protect\citeauthoryear{{Toloba} et~al.,}{{Toloba}
  et~al.}{2018}]{tol18a}
{Toloba} E.,  et~al., 2018, \mn@doi [\apjl] {10.3847/2041-8213/aab603}, \href
  {http://adsabs.harvard.edu/abs/2018ApJ...856L..31T} {856, L31}

\bibitem[\protect\citeauthoryear{{Trentham} \& {Hodgkin}}{{Trentham} \&
  {Hodgkin}}{2002}]{tre02a}
{Trentham} N.,  {Hodgkin} S.,  2002, \mn@doi [\mnras]
  {10.1046/j.1365-8711.2002.05440.x}, \href
  {http://adsabs.harvard.edu/abs/2002MNRAS.333..423T} {333, 423}

\bibitem[\protect\citeauthoryear{{Vaduvescu} \& {McCall}}{{Vaduvescu} \&
  {McCall}}{2008}]{vad08a}
{Vaduvescu} O.,  {McCall} M.~L.,  2008, \mn@doi [\aap]
  {10.1051/0004-6361:200809706}, \href
  {http://adsabs.harvard.edu/abs/2008A%26A...487..147V} {487, 147}

\bibitem[\protect\citeauthoryear{{Vaduvescu}, {McCall}, {Richer}  \&
  {Fingerhut}}{{Vaduvescu} et~al.}{2005}]{vad05a}
{Vaduvescu} O.,  {McCall} M.~L.,  {Richer} M.~G.,   {Fingerhut} R.~L.,  2005,
  \mn@doi [\aj] {10.1086/444498}, \href
  {http://adsabs.harvard.edu/abs/2005AJ....130.1593V} {130, 1593}

\bibitem[\protect\citeauthoryear{{Vaduvescu}, {Richer}  \&
  {McCall}}{{Vaduvescu} et~al.}{2006}]{vad06a}
{Vaduvescu} O.,  {Richer} M.~G.,   {McCall} M.~L.,  2006, \mn@doi [\aj]
  {10.1086/498723}, \href {http://adsabs.harvard.edu/abs/2006AJ....131.1318V}
  {131, 1318}

\bibitem[\protect\citeauthoryear{{Vargas}, {Geha}  \& {Tollerud}}{{Vargas}
  et~al.}{2014}]{var14a}
{Vargas} L.~C.,  {Geha} M.~C.,   {Tollerud} E.~J.,  2014, \mn@doi [\apj]
  {10.1088/0004-637X/790/1/73}, \href
  {http://adsabs.harvard.edu/abs/2014ApJ...790...73V} {790, 73}

\bibitem[\protect\citeauthoryear{{V{\'a}zquez} \& {Leitherer}}{{V{\'a}zquez} \&
  {Leitherer}}{2005}]{vaz05a}
{V{\'a}zquez} G.~A.,  {Leitherer} C.,  2005, \mn@doi [\apj] {10.1086/427866},
  \href {http://adsabs.harvard.edu/abs/2005ApJ...621..695V} {621, 695}

\bibitem[\protect\citeauthoryear{{Verheijen} \& {Sancisi}}{{Verheijen} \&
  {Sancisi}}{2001}]{ver01a}
{Verheijen} M.~A.~W.,  {Sancisi} R.,  2001, \mn@doi [\aap]
  {10.1051/0004-6361:20010090}, \href
  {http://adsabs.harvard.edu/abs/2001A%26A...370..765V} {370, 765}

\bibitem[\protect\citeauthoryear{{Walker}, {Mateo}, {Olszewski}, {Sen}  \&
  {Woodroofe}}{{Walker} et~al.}{2009a}]{wal09a}
{Walker} M.~G.,  {Mateo} M.,  {Olszewski} E.~W.,  {Sen} B.,   {Woodroofe} M.,
  2009a, \mn@doi [\aj] {10.1088/0004-6256/137/2/3109}, \href
  {http://adsabs.harvard.edu/abs/2009AJ....137.3109W} {137, 3109}

\bibitem[\protect\citeauthoryear{{Walker}, {Mateo}, {Olszewski},
  {Pe{\~n}arrubia}, {Wyn Evans}  \& {Gilmore}}{{Walker} et~al.}{2009b}]{wal09b}
{Walker} M.~G.,  {Mateo} M.,  {Olszewski} E.~W.,  {Pe{\~n}arrubia} J.,  {Wyn
  Evans} N.,   {Gilmore} G.,  2009b, \mn@doi [\apj]
  {10.1088/0004-637X/704/2/1274}, \href
  {http://adsabs.harvard.edu/abs/2009ApJ...704.1274W} {704, 1274}

\bibitem[\protect\citeauthoryear{{Walker}, {Mateo}, {Olszewski},
  {Pe{\~n}arrubia}, {Wyn Evans}  \& {Gilmore}}{{Walker} et~al.}{2010}]{wal10a}
{Walker} M.~G.,  {Mateo} M.,  {Olszewski} E.~W.,  {Pe{\~n}arrubia} J.,  {Wyn
  Evans} N.,   {Gilmore} G.,  2010, \mn@doi [\apj]
  {10.1088/0004-637X/710/1/886}, \href
  {http://adsabs.harvard.edu/abs/2010ApJ...710..886W} {710, 886}

\bibitem[\protect\citeauthoryear{{Weisz} et~al.,}{{Weisz}
  et~al.}{2011}]{wei11a}
{Weisz} D.~R.,  et~al., 2011, \mn@doi [\apj] {10.1088/0004-637X/739/1/5}, \href
  {http://adsabs.harvard.edu/abs/2011ApJ...739....5W} {739, 5}

\bibitem[\protect\citeauthoryear{{Wolf}, {Martinez}, {Bullock}, {Kaplinghat},
  {Geha}, {Mu{\~n}oz}, {Simon}  \& {Avedo}}{{Wolf} et~al.}{2010}]{wol10a}
{Wolf} J.,  {Martinez} G.~D.,  {Bullock} J.~S.,  {Kaplinghat} M.,  {Geha} M.,
  {Mu{\~n}oz} R.~R.,  {Simon} J.~D.,   {Avedo} F.~F.,  2010, \mn@doi [\mnras]
  {10.1111/j.1365-2966.2010.16753.x}, \href
  {http://adsabs.harvard.edu/abs/2010MNRAS.406.1220W} {406, 1220}

\bibitem[\protect\citeauthoryear{{Woo}, {Courteau}  \& {Dekel}}{{Woo}
  et~al.}{2008}]{woo08a}
{Woo} J.,  {Courteau} S.,   {Dekel} A.,  2008, \mn@doi [\mnras]
  {10.1111/j.1365-2966.2008.13770.x}, \href
  {http://adsabs.harvard.edu/abs/2008MNRAS.390.1453W} {390, 1453}

\bibitem[\protect\citeauthoryear{{Wyithe} \& {Loeb}}{{Wyithe} \&
  {Loeb}}{2013}]{wyi13a}
{Wyithe} J.~S.~B.,  {Loeb} A.,  2013, \mn@doi [\mnras] {10.1093/mnras/sts242},
  \href {http://adsabs.harvard.edu/abs/2013MNRAS.428.2741W} {428, 2741}

\bibitem[\protect\citeauthoryear{{Young}, {Jerjen}, {L{\'o}pez-S{\'a}nchez}  \&
  {Koribalski}}{{Young} et~al.}{2014}]{you14a}
{Young} T.,  {Jerjen} H.,  {L{\'o}pez-S{\'a}nchez} {\'A}.~R.,   {Koribalski}
  B.~S.,  2014, \mn@doi [\mnras] {10.1093/mnras/stu1646}, \href
  {http://adsabs.harvard.edu/abs/2014MNRAS.444.3052Y} {444, 3052}

\bibitem[\protect\citeauthoryear{{Zaritsky}, {Gonzalez}  \&
  {Zabludoff}}{{Zaritsky} et~al.}{2006a}]{zar06a}
{Zaritsky} D.,  {Gonzalez} A.~H.,   {Zabludoff} A.~I.,  2006a, \mn@doi [\apj]
  {10.1086/498672}, \href {http://adsabs.harvard.edu/abs/2006ApJ...638..725Z}
  {638, 725}

\bibitem[\protect\citeauthoryear{{Zaritsky}, {Gonzalez}  \&
  {Zabludoff}}{{Zaritsky} et~al.}{2006b}]{zar06b}
{Zaritsky} D.,  {Gonzalez} A.~H.,   {Zabludoff} A.~I.,  2006b, \mn@doi [\apjl]
  {10.1086/504352}, \href {http://adsabs.harvard.edu/abs/2006ApJ...642L..37Z}
  {642, L37}

\bibitem[\protect\citeauthoryear{{Zaritsky}, {Zabludoff}  \&
  {Gonzalez}}{{Zaritsky} et~al.}{2011}]{zar11a}
{Zaritsky} D.,  {Zabludoff} A.~I.,   {Gonzalez} A.~H.,  2011, \mn@doi [\apj]
  {10.1088/0004-637X/727/2/116}, \href
  {http://adsabs.harvard.edu/abs/2011ApJ...727..116Z} {727, 116}

\bibitem[\protect\citeauthoryear{{van Dokkum} et~al.,}{{van Dokkum}
  et~al.}{2016}]{dok16a}
{van Dokkum} P.,  et~al., 2016, \mn@doi [\apjl] {10.3847/2041-8205/828/1/L6},
  \href {http://adsabs.harvard.edu/abs/2016ApJ...828L...6V} {828, L6}

\makeatother
\end{thebibliography}




\clearpage
\onecolumn
\setcounter{table}{0}

\begin{landscape}
\begin{center}
\begin{ThreePartTable}
\small

\begin{TableNotes}

\item[]
(1) Name of galaxy;
(2) $B-V$ colour excess, from \citet{sch98a} via NED;
(3) Optical depth at $1 \, \rm \mu m$, computed from $E(B-V)$ using an elliptical
SED at $z = 0$;
(4) Heliocentric velocity, from \citet{mcco12a};
(5) Distance modulus and uncertainty;
(6) Method used to determine distance modulus (TRGB:  tip of the red giant branch; RR:  RR Lyrae stars);
(7) {\color{black} Source of the data used by the authors to determine the distance modulus;}
(8) [Fe/H] and uncertainty;
(9) Source of [Fe/H];
(10) Line-of-sight velocity dispersion and uncertainty, in $\rm km \, s^{-1}$;
(11) Source of the velocity dispersion {\color{black} and uncertainty};
(12): Extinction-corrected central surface brightness in $K_s$ from sech fit, in $\rm mag \, arcsec^{-2}$ {\color{black}($\pm 0.15$ for NGC 147, 185, and 205, and $\pm 0.40$ otherwise)};
(13): Scale length from sech fit, in arc seconds ($\pm 5\%$);
(14): Ratio of the length of the semi-minor to semi-major axis ($\pm 10\%$);
(15) {\color{black} Source of the axis ratio and the data that was fitted by the authors to derive the central surface brightness and scale length}.
\smallskip
\item[]
Source Translations: \\
amo14 \citep{amo14a};
bel02 \citep{bel02a};
bel04 \citep{bel04a};
bel05 \citep{bel05a};
ber09 \citep{ber09a};
bon04 \citep{bon04a};
but05 \citep{but05a};
car02 \citep{car02a};
col13 \citep{col13a};
con12 \citep{con12a};
dal06 \citep{dal06a};
geh06 \citep{geh06a};
geh10 \citep{geh10a};
gre08 \citep{gre08a};
han97 \citep{han97a};
ho15 \citep{ho15a};
irw95 \citep{irw95a};
jar03 \citep{jar03a};
kin08 \citep{kin08a};
kir13 \citep{kir13a};
kir14 \citep{kir14a};
koc07 \citep{koc07a};
koc08 \citep{koc08a};
kop11 \citep{kop11a};
kue08 \citep{kue08a};
lai11 \citep{lai11a};
lee03 \citep{lee03a};
lew07 \citep{lew07a};
mcco04 \citep{mcco04a};
mcco05 \citep{mcco05a};
mcco06 \citep{mcco06a};
mcco16 \citep{mcco16a};
mor09 \citep{mor09a};
mun10 \citep{mun10a};
mus09 \citep{mus09a};
nem88 \citep{nem88a};
now03 \citep{now03a};
oka12 \citep{oka12a};
oka12 \citep{oka12a};
riz07a \citep{riz07a};
riz07b \citep{riz07b};
sim07 \citep{sim07a};
sme94 \citep{sme94a};
tol12 \citep{tol12a};
var14 \citep{var14a};
wal09a \citep{wal09a};
wal09b \citep{wal09b};
wal10 \citep{wal10a}.

\end{TableNotes}

\setlength{\jot}{-0.0pt}

\begin{longtable}{lcccccccccccccc}

\caption{
Observed Properties of Dwarf Spheroidal Galaxies
\label{tab_obsdsph}
}
\\

\hline
\noalign{\smallskip}

Galaxy &
$E$ &
$\tau_{1}$ &
$v_\odot$ &
$\mathit{DM}$ &
Method &
Foundation &
[Fe/H] &
Source &
$\sigma_\textit{los}$ &
Source &
$\mu_{0}$ &
$r_{0}$ &
$q$ &
Foundation
\\


(1) &
(2) &
(3) &
(4) &
(5) &
(6) &
(7) &
(8) &
(9) &
(10) &
(11) &
(12) &
(13) &
(14) &
(15)
\\

\noalign{\smallskip}
\hline
\noalign{\smallskip}

\endfirsthead

\caption{cont'd.} \\

\hline
\noalign{\smallskip}

Galaxy &
$E$ &
$\tau_{1}$ &
$v_\odot$ &
$\mathit{DM}$ &
Method &
Foundation &
[Fe/H] &
Source &
$\sigma_\textit{los}$ &
Source &
$\mu_{0}$ &
$r_{0}$ &
$q$ &
Foundation
\\


(1) &
(2) &
(3) &
(4) &
(5) &
(6) &
(7) &
(8) &
(9) &
(10) &
(11) &
(12) &
(13) &
(14) &
(15)
\\

\noalign{\smallskip}
\hline
\endhead

\noalign{\smallskip}
\hline

\endfoot
\hline
\noalign{\smallskip}
\insertTableNotes

\endlastfoot


And~I	&	0.054	&	0.061	&	$		-375.8	$	&	$	24.366	\pm	0.031	$	&	TRGB	&	mcco04,mcco05	&	$	-1.11	\pm	0.12	$	&	var14	&	$		10.2	\pm	1.9	$	&	tol12	&	22.20	&	\phantom{~}	92	&	0.78	&	mcco06	\\
And~II	&	0.062	&	0.070	&	$		-193.6	$	&	$	24.086	\pm	0.021	$	&	TRGB	&	mcco04,mcco05	&	$	-1.37	\pm	0.12	$	&	var14	&	$	\phantom{1}	9.3	\pm	0.6	$	&	amo14	&	22.75	&		189	&	0.80	&	mcco06	\\
And~III	&	0.057	&	0.065	&	$		-345.6	$	&	$	24.424	\pm	0.041	$	&	TRGB	&	mcco05	&	$	-1.81	\pm	0.12	$	&	var14	&	$	\phantom{~}	9.3	\pm	1.4	$	&	tol12	&	22.42	&	\phantom{~}	57	&	0.48	&	mcco06	\\
And~V	&	0.124	&	0.141	&	$		-403.0	$	&	$	24.483	\pm	0.041	$	&	TRGB	&	mcco05	&	$	-1.71	\pm	0.12	$	&	var14	&	$		10.5	\pm	1.1	$	&	tol12	&	23.14	&	\phantom{~}	51	&	0.82	&	mcco06	\\
And~VI	&	0.064	&	0.073	&	$		-354.0	$	&	$	24.490	\pm	0.041	$	&	TRGB	&	mcco05	&	$	-1.50	\pm	0.10	$	&	col13,mcco16	&	$		12.4	\pm	1.4	$	&	col13	&	21.09	&	\phantom{~}	58	&	0.59	&	mcco06	\\
And~VII	&	0.196	&	0.223	&	$		-309.4	$	&	$	24.442	\pm	0.050	$	&	TRGB	&	mcco05	&	$	-1.24	\pm	0.12	$	&	var14	&	$		13.0	\pm	1.0	$	&	tol12	&	21.89	&		138	&	0.87	&	mcco06	\\
Bootes~I	&	0.018	&	0.020	&	$	\phantom{-1}	99.0	$	&	$	19.193	\pm	0.051	$	&	RR	&	dal06	&	$	-2.51	\pm	0.03	$	&	lai11	&	$	\phantom{1}	4.6	\pm	0.7	$	&	kop11	&	25.17	&		384	&	0.78	&	oka12	\\
Carina	&	0.063	&	0.072	&	$	\phantom{-}	222.9	$	&	$	20.154	\pm	0.055	$	&	TRGB	&	sme94	&	$	-1.69	\pm	0.17	$	&	koc08	&	$	\phantom{1}	6.6	\pm	1.2	$	&	wal09a	&	23.01	&		326	&	0.67	&	irw95	\\
Cetus	&	0.029	&	0.033	&	$	\phantom{}	-87.0	$	&	$	24.508	\pm	0.029	$	&	RR	&	ber09	&	$	-1.90	\pm	0.10	$	&	lew07	&	$	\phantom{1}	8.3	\pm	1.0	$	&	kir14	&	22.52	&	\phantom{~}	74	&	0.67	&	mcco06	\\
Coma~Ber	&	0.017	&	0.019	&	$	\phantom{-1}	98.1	$	&	$	18.235	\pm	0.054	$	&	RR	&	mus09	&	$	-2.25	\pm	0.04	$	&	kir13	&	$	\phantom{1}	4.6	\pm	0.8	$	&	sim07	&	25.70	&		231	&	0.64	&	mun10	\\
CVn~I	&	0.014	&	0.016	&	$	\phantom{-1}	30.9	$	&	$	21.702	\pm	0.028	$	&	RR	&	kue08	&	$	-1.91	\pm	0.01	$	&	kir13	&	$	\phantom{1}	7.6	\pm	0.4	$	&	sim07	&	25.55	&		276	&	0.70	&	oka12	\\
CVn~II	&	0.010	&	0.011	&	$		-128.9	$	&	$	21.066	\pm	0.068	$	&	RR	&	gre08,oka12	&	$	-2.12	\pm	0.05	$	&	kir13	&	$	\phantom{1}	4.6	\pm	1.0	$	&	sim07	&	23.73	&	\phantom{~}	58	&	0.77	&	oka12	\\
Draco	&	0.027	&	0.031	&	$		-291.0	$	&	$	19.643	\pm	0.046	$	&	RR	&	kin08,bon04	&	$	-1.98	\pm	0.01	$	&	kir13	&	$	\phantom{1}	9.1	\pm	1.2	$	&	wal09b,wal10	&	22.98	&		284	&	0.71	&	irw95	\\
Fornax	&	0.020	&	0.023	&	$	\phantom{-1}	55.3	$	&	$	20.764	\pm	0.021	$	&	TRGB	&	riz07b	&	$	-1.04	\pm	0.01	$	&	kir13	&	$		11.7	\pm	0.9	$	&	wal09a	&	20.92	&		570	&	0.70	&	irw95	\\
Leo~I	&	0.036	&	0.041	&	$	\phantom{-}	282.5	$	&	$	21.985	\pm	0.040	$	&	TRGB	&	bel04	&	$	-1.45	\pm	0.01	$	&	kir13	&	$	\phantom{1}	9.2	\pm	1.4	$	&	wal09b,wal10	&	19.96	&		118	&	0.79	&	irw95	\\
Leo~II	&	0.017	&	0.019	&	$	\phantom{-1}	78.0	$	&	$	21.889	\pm	0.031	$	&	TRGB	&	bel05	&	$	-1.63	\pm	0.01	$	&	kir13	&	$	\phantom{1}	6.6	\pm	0.7	$	&	koc07	&	21.49	&	\phantom{~}	90	&	0.87	&	irw95	\\
Leo~IV	&	0.025	&	0.028	&	$	\phantom{-}	132.3	$	&	$	21.092	\pm	0.056	$	&	RR	&	mor09	&	$	-2.45	\pm	0.07	$	&	kir13	&	$	\phantom{1}	3.3	\pm	1.7	$	&	sim07	&	25.28	&	\phantom{~}	81	&	0.96	&	oka12	\\
NGC~147	&	0.173	&	0.197	&	$		-193.1	$	&	$	24.396	\pm	0.028	$	&	TRGB	&	han97,now03,con12	&	$	-0.49	\pm	0.12	$	&	var14	&	$		16.0	\pm	1.0	$	&	geh10	&	19.25	&	\phantom{~}	95	&	0.69	&	jar03	\\
NGC~185	&	0.182	&	0.207	&	$		-203.8	$	&	$	24.072	\pm	0.061	$	&	TRGB	&	now03,riz07a,con12	&	$	-0.92	\pm	0.12	$	&	var14	&	$		24.0	\pm	1.0	$	&	geh10	&	18.28	&	\phantom{~}	66	&	0.89	&	jar03	\\
NGC~205	&	0.062	&	0.070	&	$		-246.0	$	&	$	24.709	\pm	0.100	$	&	TRGB	&	but05	&	$	-0.87	\pm	0.05	$	&	ho15	&	$		35.0	\pm	5.0	$	&	geh06	&	17.95	&		110	&	0.59	&	jar03	\\
Sculptor	&	0.018	&	0.020	&	$	\phantom{-}	111.4	$	&	$	19.647	\pm	0.031	$	&	TRGB	&	riz07a	&	$	-1.68	\pm	0.01	$	&	kir13	&	$	\phantom{1}	9.2	\pm	1.1	$	&	wal09a	&	21.23	&		331	&	0.68	&	irw95	\\
Sextans	&	0.050	&	0.057	&	$	\phantom{-}	224.2	$	&	$	19.979	\pm	0.041	$	&	TRGB	&	lee03	&	$	-1.94	\pm	0.01	$	&	kir13	&	$	\phantom{1}	7.9	\pm	1.3	$	&	wal09a	&	24.63	&		809	&	0.65	&	irw95	\\
Ursa~Min	&	0.032	&	0.036	&	$		-246.9	$	&	$	19.369	\pm	0.085	$	&	RR	&	bel02,car02,nem88	&	$	-2.13	\pm	0.01	$	&	kir13	&	$	\phantom{1}	9.5	\pm	1.2	$	&	wal09b,wal10	&	23.40	&		510	&	0.44	&	irw95	\\

\noalign{\smallskip}

\end{longtable}

\end{ThreePartTable}

\end{center}
\end{landscape}

\clearpage
\onecolumn

\begin{center}
\begin{ThreePartTable}

\begin{TableNotes}

\item[]
(1) Name of galaxy;
(2) Nuclear type (N = nucleated; nN = non-nucleated), from \citet{tre02a}, \citet{for11a}, and \citet{tol14a};
(3) $B-V$ colour excess, from \citet{sch98a} via NED;
(4) Optical depth at $1 \, \rm \mu m$, computed from $E(B-V)$ using an elliptical
SED at $z = 0$;
(5) Heliocentric velocity, from \citet{tol14a} and NED;
(6) [Fe/H] and uncertainty, from \citet{tol14a};
(7) Line-of-sight velocity dispersion and uncertainty, from \citet{for11a} and \citet{tol14a}, in $\rm km \, s^{-1}$;
(8): Central surface brightness in $K_s$ from sech fit, corrected for extinction and motion, in $\rm mag \, arcsec^{-2}$ ($\pm 0.12$);
(9): Scale length from sech fit, in arc seconds ($\pm 10\%$);
(10): Ratio of the length of the semi-minor to semi-major axis ($\pm 5\%$).


\end{TableNotes}

\setlength{\jot}{-0.0pt}

\begin{longtable}{lccccccccc}

\caption{
Observed Properties of Dwarf Elliptical Galaxies
\label{tab_obsde}
}
\\

\hline
\noalign{\smallskip}


Galaxy &
Nucleation &
$E$ &
$\tau_{1}$ &
$v_\odot$ &
[Fe/H] &
$\sigma_\textit{los}$ &
$\mu_{0}$ &
$r_{0}$ &
$q$
\\



(1) &
(2) &
(3) &
(4) &
(5) &
(6) &
(7) &
(8) &
(9) &
(10)
\\

\noalign{\smallskip}
\hline
\noalign{\smallskip}

\endfirsthead

\caption{cont'd.} \\

\hline
\noalign{\smallskip}


Galaxy &
Nucleation &
$E$ &
$\tau_{1}$ &
$v_\odot$ &
[Fe/H] &
$\sigma_\textit{los}$&
$\mu_{0}$ &
$r_{0}$ &
$q$
\\



(1) &
(2) &
(3) &
(4) &
(5) &
(6) &
(7) &
(8) &
(9) &
(10)
\\

\noalign{\smallskip}
\hline
\endhead

\noalign{\smallskip}
\hline

\endfoot
\hline
\noalign{\smallskip}
\insertTableNotes

\endlastfoot


PGC~32348	&	N	&	0.039	&	0.044	&	$	\phantom{-1}	541.0	$	&	$		\ldots		$	&	$	15.4	\pm	1.6	$	&	19.39	&	\phantom{~}	8.3	&	0.71	\\
VCC~0009	&	N	&	0.038	&	0.043	&	$	\phantom{-}	1674.0	$	&	$	-0.5	\pm	0.1	$	&	$	26.0	\pm	3.9	$	&	19.46	&		20.3	&	0.85	\\
VCC~0033	&	N	&	0.037	&	0.042	&	$	\phantom{-}	1179.3	$	&	$	-0.9	\pm	0.2	$	&	$	20.8	\pm	4.9	$	&	18.21	&	\phantom{~}	5.2	&	0.89	\\
VCC~0437	&	N	&	0.031	&	0.035	&	$	\phantom{-}	1412.3	$	&	$	-1.3	\pm	0.3	$	&	$	40.9	\pm	4.0	$	&	19.49	&		19.9	&	0.54	\\
VCC~0543	&	nN	&	0.031	&	0.035	&	$	\phantom{-1}	977.1	$	&	$	-0.7	\pm	0.1	$	&	$	35.1	\pm	1.4	$	&	19.19	&		14.2	&	0.57	\\
VCC~0634	&	N	&	0.027	&	0.031	&	$	\phantom{-1}	484.5	$	&	$	-0.4	\pm	0.0	$	&	$	31.3	\pm	1.6	$	&	19.18	&		14.5	&	0.78	\\
VCC~0750	&	N	&	0.027	&	0.031	&	$	\phantom{-}	1058.8	$	&	$	-0.3	\pm	0.1	$	&	$	43.5	\pm	2.9	$	&	19.59	&		10.5	&	0.72	\\
VCC~0846	&	N	&	0.028	&	0.032	&	$	\phantom{}	-730.0	$	&	$		\ldots		$	&	$	19.0	\pm	5.0	$	&	20.45	&	\phantom{~}	9.8	&	0.76	\\
VCC~0917	&	nN	&	0.032	&	0.036	&	$	\phantom{-}	1244.8	$	&	$	-0.6	\pm	0.1	$	&	$	28.4	\pm	1.4	$	&	18.43	&	\phantom{~}	6.1	&	0.52	\\
VCC~1087	&	N	&	0.026	&	0.030	&	$	\phantom{-1}	658.6	$	&	$	-0.4	\pm	0.1	$	&	$	42.0	\pm	1.5	$	&	18.59	&		12.2	&	0.68	\\
VCC~1261	&	N	&	0.028	&	0.032	&	$	\phantom{-}	1825.3	$	&	$	-0.6	\pm	0.2	$	&	$	44.8	\pm	1.4	$	&	18.25	&		14.2	&	0.57	\\
VCC~1355	&	N	&	0.034	&	0.039	&	$	\phantom{-}	1245.0	$	&	$	-0.7	\pm	0.1	$	&	$	20.3	\pm	4.7	$	&	20.47	&		17.5	&	0.96	\\
VCC~1407	&	N	&	0.031	&	0.035	&	$	\phantom{-}	1007.2	$	&	$	-1.1	\pm	0.1	$	&	$	31.9	\pm	2.1	$	&	18.70	&	\phantom{~}	7.1	&	0.83	\\
VCC~1431	&	N	&	0.054	&	0.061	&	$	\phantom{-}	1489.4	$	&	$	-0.3	\pm	0.3	$	&	$	52.4	\pm	1.6	$	&	17.66	&	\phantom{~}	6.0	&	0.96	\\
VCC~1453	&	N	&	0.035	&	0.040	&	$	\phantom{-}	1880.0	$	&	$	-0.2	\pm	0.0	$	&	$	35.6	\pm	1.4	$	&	19.22	&		13.0	&	0.82	\\
VCC~1528	&	nN	&	0.027	&	0.031	&	$	\phantom{-}	1615.4	$	&	$	-0.2	\pm	0.0	$	&	$	47.0	\pm	1.4	$	&	18.14	&	\phantom{~}	6.3	&	0.92	\\
VCC~1549	&	N	&	0.030	&	0.034	&	$	\phantom{-}	1389.3	$	&	$	-0.4	\pm	0.3	$	&	$	36.7	\pm	2.3	$	&	18.60	&	\phantom{~}	8.2	&	0.83	\\
VCC~1826	&	N	&	0.017	&	0.019	&	$	\phantom{-}	2033.0	$	&	$		\ldots		$	&	$	22.3	\pm	2.2	$	&	18.52	&	\phantom{~}	5.3	&	0.69	\\
VCC~1861	&	N	&	0.030	&	0.034	&	$	\phantom{-1}	629.7	$	&	$	-0.3	\pm	0.1	$	&	$	31.3	\pm	1.5	$	&	19.04	&		11.1	&	0.93	\\
VCC~1895	&	nN	&	0.018	&	0.020	&	$	\phantom{-1}	970.2	$	&	$	-0.9	\pm	0.1	$	&	$	23.8	\pm	3.0	$	&	18.40	&	\phantom{~}	8.2	&	0.47	\\
VCC~2083	&	N	&	0.018	&	0.020	&	$	\phantom{-1}	867.9	$	&	$	-0.7	\pm	0.4	$	&	$	28.4	\pm	2.4	$	&	19.61	&	\phantom{~}	7.9	&	0.85	\\

\noalign{\smallskip}

\end{longtable}

\end{ThreePartTable}

\end{center}

\clearpage
\onecolumn

\begin{center}
\begin{ThreePartTable}

\begin{TableNotes}

\item[]
(1) Name of galaxy;
(2) Absolute magnitude in $K_s$, in mag {\color{black} ($\pm 0.22$ for NGC 147, 185, and 205, and $\pm 0.43$ otherwise)};
(3) Logarithm of scale length of sech model, in pc ($\pm 0.02$);
(4) Logarithm of predicted potential, in $\mathcal{M}_\odot \, pc^{-1}$ ($\pm 0.12$);
(5) Logarithm of observed potential, in $\mathcal{M}_\odot \, pc^{-1}$ ($\pm 0.17$);
(6) Logarithm of stellar mass, in $\mathcal{M}_\odot$ ($\pm 0.17$);
(7) Logarithm of gas mass, in $\mathcal{M}_\odot$ ($\pm 0.12$);
(8) Logarithm of baryonic mass, in $\mathcal{M}_\odot$ ($\pm 0.12$);
(9) Logarithm of dynamical mass, in $\mathcal{M}_\odot$ ($\pm 0.14$);
(10) Escape velocity, in $\rm km \, s^{-1}$ ($\pm 15\%$).


\end{TableNotes}

\setlength{\jot}{-0.0pt}

\begin{longtable}{lccccccccc}

\caption{
Derived Properties of Dwarf Spheroidal Galaxies
\label{tab_comdsph}
}
\\

\hline
\noalign{\smallskip}


Galaxy &
$M_\textit{Ks}$ &
$\log r_{0}$ &
$\log P^\textit{\,pre}$ &
$\log P^\textit{\,obs}$ &
$\log \mathcal{M}_\textit{str}$ &
$\log \mathcal{M}_\textit{gas}$ &
$\log \mathcal{M}_\textit{bar}$  &
$\log \mathcal{M}_\textit{dyn}$ &
$v_\textit{esc}$ \\



(1) &
(2) &
(3) &
(4) &
(5) &
(6) &
(7) &
(8) &
(9) &
(10)
\\

\noalign{\smallskip}
\hline
\noalign{\smallskip}

\endfirsthead

\caption{cont'd.} \\

\hline
\noalign{\smallskip}


Galaxy &
$M_\textit{Ks}$ &
$\log r_{0}$ &
$\log P^\textit{\,pre}$ &
$\log P^\textit{\,obs}$ &
$\log \mathcal{M}_\textit{str}$ &
$\log \mathcal{M}_\textit{gas}$ &
$\log \mathcal{M}_\textit{bar}$  &
$\log \mathcal{M}_\textit{dyn}$ &
$v_\textit{esc}$ \\



(1) &
(2) &
(3) &
(4) &
(5) &
(6) &
(7) &
(8) &
(9) &
(10)
\\

\noalign{\smallskip}
\hline
\endhead

\noalign{\smallskip}
\hline

\endfoot
\hline
\noalign{\smallskip}
\insertTableNotes

\endlastfoot


And~I	&	$		-14.36	$	&	2.52	&	4.96	&	4.50	&	7.02	&	7.30	&	7.48	&	8.06	&	25	\\
And~II	&	$		-15.13	$	&	2.78	&	4.81	&	4.54	&	7.32	&	7.26	&	7.59	&	8.24	&	23	\\
And~III	&	$		-12.64	$	&	2.32	&	4.77	&	4.00	&	6.33	&	7.01	&	7.10	&	7.79	&	23	\\
And~V	&	$		-12.32	$	&	2.29	&	4.80	&	3.91	&	6.20	&	7.03	&	7.09	&	7.86	&	26	\\
And~VI	&	$		-14.32	$	&	2.35	&	5.22	&	4.65	&	7.00	&	7.43	&	7.57	&	8.06	&	30	\\
And~VII	&	$		-15.76	$	&	2.71	&	5.17	&	4.86	&	7.58	&	7.59	&	7.88	&	8.47	&	32	\\
Bootes~I	&	$	\phantom{}	-9.33	$	&	2.11	&	3.98	&	2.89	&	5.00	&	6.06	&	6.09	&	6.96	&	11	\\
Carina	&	$		-11.93	$	&	2.23	&	4.56	&	3.81	&	6.04	&	6.70	&	6.79	&	7.40	&	16	\\
Cetus	&	$		-13.57	$	&	2.46	&	4.77	&	4.24	&	6.70	&	7.07	&	7.23	&	7.82	&	20	\\
Coma~Ber	&	$	\phantom{}	-6.52	$	&	1.70	&	3.84	&	2.18	&	3.88	&	5.52	&	5.53	&	6.55	&	11	\\
CVn~I	&	$		-10.61	$	&	2.47	&	4.13	&	3.05	&	5.52	&	6.56	&	6.60	&	7.75	&	19	\\
CVn~II	&	$	\phantom{}	-8.51	$	&	1.66	&	4.27	&	3.02	&	4.68	&	5.90	&	5.93	&	6.51	&	11	\\
Draco	&	$		-11.22	$	&	2.07	&	4.73	&	3.69	&	5.76	&	6.76	&	6.80	&	7.51	&	22	\\
Fornax	&	$		-15.88	$	&	2.59	&	5.26	&	5.03	&	7.63	&	7.47	&	7.85	&	8.26	&	29	\\
Leo~I	&	$		-14.78	$	&	2.15	&	5.36	&	5.03	&	7.18	&	7.24	&	7.51	&	7.61	&	23	\\
Leo~II	&	$		-12.68	$	&	2.02	&	4.91	&	4.33	&	6.35	&	6.80	&	6.93	&	7.19	&	16	\\
Leo~IV	&	$	\phantom{}	-7.97	$	&	1.81	&	3.84	&	2.65	&	4.46	&	5.63	&	5.66	&	6.38	&	8	\\
NGC~147	&	$		-17.28	$	&	2.54	&	5.74	&	5.64	&	8.19	&	7.59	&	8.28	&	8.48	&	39	\\
NGC~185	&	$		-17.42	$	&	2.32	&	6.19	&	5.92	&	8.24	&	8.17	&	8.51	&	8.61	&	59	\\
NGC~205	&	$		-19.04	$	&	2.67	&	6.35	&	6.22	&	8.89	&	8.43	&	9.02	&	9.28	&	86	\\
Sculptor	&	$		-13.25	$	&	2.14	&	5.07	&	4.44	&	6.57	&	7.10	&	7.21	&	7.59	&	23	\\
Sextans	&	$		-12.07	$	&	2.59	&	4.32	&	3.51	&	6.10	&	6.83	&	6.91	&	7.91	&	19	\\
Ursa~Min	&	$		-11.28	$	&	2.27	&	4.57	&	3.51	&	5.78	&	6.80	&	6.84	&	7.75	&	23	\\

\noalign{\smallskip}

\end{longtable}

\end{ThreePartTable}

\end{center}

\clearpage
\onecolumn

\begin{center}
\begin{ThreePartTable}

\begin{TableNotes}

\item[]
(1) Name of galaxy;
(2) Absolute magnitude in $K_s$, in mag ($\pm 0.34$);
(3) Logarithm of scale length of sech model, in pc ($\pm 0.06$);
(4) Logarithm of predicted potential, in $\mathcal{M}_\odot \, pc^{-1}$ ($\pm 0.05$);
(5) Logarithm of observed potential, in $\mathcal{M}_\odot \, pc^{-1}$ ($\pm 0.08$);
(6) Logarithm of stellar mass, in $\mathcal{M}_\odot$ ($\pm 0.14$);
(7) Logarithm of dynamical mass, in $\mathcal{M}_\odot$ ($\pm 0.10$).


\end{TableNotes}

\setlength{\jot}{-0.0pt}

\begin{longtable}{lcccccc}

\caption{
Derived Properties of Dwarf Elliptical Galaxies
\label{tab_comde}
}
\\

\hline
\noalign{\smallskip}


Galaxy &
$M_\textit{Ks}$ &
$\log r_{0}$ &
$\log P^\textit{\,pre}$ &
$\log P^\textit{\,obs}$ &
$\log \mathcal{M}_\textit{str}$ &
$\log \mathcal{M}_\textit{dyn}$ \\



(1) &
(2) &
(3) &
(4) &
(5) &
(6) &
(7)
\\

\noalign{\smallskip}
\hline
\noalign{\smallskip}

\endfirsthead

\caption{cont'd.} \\

\hline
\noalign{\smallskip}


Galaxy &
$M_\textit{Ks}$ &
$\log r_{0}$ &
$\log P^\textit{\,pre}$ &
$\log P^\textit{\,obs}$ &
$\log \mathcal{M}_\textit{str}$ &
$\log \mathcal{M}_\textit{dyn}$ \\



(1) &
(2) &
(3) &
(4) &
(5) &
(6) &
(7)
\\

\noalign{\smallskip}
\hline
\endhead

\noalign{\smallskip}
\hline

\endfoot
\hline
\noalign{\smallskip}
\insertTableNotes

\endlastfoot

PGC~32348	&	$	-17.59	$	&	2.63	&	5.70	&	5.68	&	8.31	&	8.53	\\
VCC~0009	&	$	-20.51	$	&	3.19	&	5.99	&	6.29	&	9.48	&	9.54	\\
VCC~0033	&	$	-18.84	$	&	2.59	&	6.13	&	6.21	&	8.81	&	8.76	\\
VCC~0437	&	$	-19.95	$	&	3.18	&	6.10	&	6.07	&	9.25	&	9.93	\\
VCC~0543	&	$	-19.58	$	&	3.03	&	6.10	&	6.07	&	9.10	&	9.65	\\
VCC~0634	&	$	-19.99	$	&	3.04	&	6.11	&	6.22	&	9.27	&	9.56	\\
VCC~0750	&	$	-18.77	$	&	2.90	&	6.18	&	5.88	&	8.78	&	9.70	\\
VCC~0846	&	$	-17.84	$	&	2.87	&	5.61	&	5.54	&	8.41	&	8.96	\\
VCC~0917	&	$	-18.39	$	&	2.66	&	6.13	&	5.96	&	8.63	&	9.10	\\
VCC~1087	&	$	-20.05	$	&	2.97	&	6.35	&	6.32	&	9.29	&	9.74	\\
VCC~1261	&	$	-20.51	$	&	3.03	&	6.41	&	6.44	&	9.47	&	9.86	\\
VCC~1355	&	$	-19.31	$	&	3.12	&	5.69	&	5.87	&	8.99	&	9.26	\\
VCC~1407	&	$	-18.99	$	&	2.74	&	6.23	&	6.13	&	8.87	&	9.27	\\
VCC~1431	&	$	-19.79	$	&	2.66	&	6.71	&	6.53	&	9.19	&	9.62	\\
VCC~1453	&	$	-19.74	$	&	3.00	&	6.18	&	6.17	&	9.17	&	9.62	\\
VCC~1528	&	$	-19.37	$	&	2.68	&	6.55	&	6.34	&	9.02	&	9.55	\\
VCC~1549	&	$	-19.38	$	&	2.80	&	6.32	&	6.23	&	9.03	&	9.45	\\
VCC~1826	&	$	-18.29	$	&	2.61	&	6.05	&	5.98	&	8.59	&	8.83	\\
VCC~1861	&	$	-19.73	$	&	2.93	&	6.18	&	6.24	&	9.16	&	9.44	\\
VCC~1895	&	$	-18.98	$	&	2.80	&	6.02	&	6.07	&	8.86	&	9.08	\\
VCC~2083	&	$	-18.32	$	&	2.78	&	6.00	&	5.82	&	8.60	&	9.21	\\

\noalign{\smallskip}

\end{longtable}

\end{ThreePartTable}

\end{center}


\bsp	
\label{lastpage}
\end{document}